\documentclass[runningheads, twocolumn]{aastex63}

\usepackage[utf8]{inputenc}
\usepackage{savesym}
\savesymbol{tablenum}
\usepackage{siunitx}
\restoresymbol{SIX}{tablenum}
\usepackage{amsmath}
\savesymbol{iint}
\usepackage{txfonts}
\restoresymbol{TXF}{iint}
\usepackage{graphicx}
\usepackage[flushleft]{threeparttable}
\usepackage{multirow}
\usepackage{hyperref}
\hypersetup{    
     colorlinks = true,
     linkcolor = blue,
     anchorcolor = blue,
     citecolor = blue,
     filecolor = blue,
     urlcolor = cyan}
\usepackage{booktabs,tabularx}
\usepackage{natbib}
\bibpunct{(}{)}{;}{a}{}{,}
\received{\today}
\revised{}
\accepted{}
\submitjournal{ApJ}

\shorttitle{Binding energies of interstellar molecules}
\shortauthors{Ferrero et al.}

\begin{document} 

\title{Binding energies of interstellar molecules on crystalline and amorphous models of water ice by ab-initio calculations}
\correspondingauthor{C. Ceccarelli \& P. Ugliengo}
\email{cecilia.ceccarelli@univ-grenoble-alpes.fr, piero.ugliengo@unito.it}

\author[0000-0001-7819-7657]{Stefano Ferrero}
  \affil{Departament de Química, Universitat Autonoma de Barcelona, 08193 Bellaterra, Catalonia, Spain} \affil{Dipartimento di Chimica, Università degli Studi di Torino, via P. Giuria 7, 10125, Torino, Italy}
\author[0000-0003-0219-6150]{Lorenzo Zamirri}
  \affil{Dipartimento di Chimica, Università degli Studi di Torino, via P. Giuria 7, 10125, Torino, Italy}
  \affil{Nanostructured Interfaces and Surfaces (NIS) Centre, Università degli Studi di Torino, via P. Giuria 7, 10125, Torino, Italy}
\author[0000-0001-9664-6292]{Cecilia Ceccarelli}
\affiliation{Univ. Grenoble Alpes, CNRS, IPAG, 38000 Grenoble, France}
\author[0000-0003-0518-944X]{Arezu Witzel}
  \affil{Univ. Grenoble Alpes, CNRS, IPAG, 38000 Grenoble, France}
\author[0000-0002-9637-4554]{Albert Rimola}
  \affil{Departament de Química, Universitat Autonoma de Barcelona, 08193 Bellaterra, Catalonia, Spain}
\author[0000-0001-8886-9832]{Piero Ugliengo}
  \affil{Dipartimento di Chimica, Università degli Studi di Torino, via P. Giuria 7, 10125, Torino, Italy}

\begin{abstract} 
In the denser and colder ($\leq$20 K) regions of the interstellar medium (ISM), near-infrared observations have revealed the presence of sub-micron sized dust grains covered by several layers of H\textsubscript{2}O-dominated ices and “dirtied” by the presence of other volatile species. Whether a molecule is in the gas or solid-phase depends on its binding energy (\emph{BE}) on ice surfaces. Thus, \emph{BEs} are crucial parameters for the astrochemical models that aim to reproduce the observed evolution of the ISM chemistry. In general, \emph{BEs} can be inferred either from experimental techniques or by theoretical computations. In this work, we present a reliable computational methodology to evaluate the \emph{BEs} of a large set (21) of astrochemical relevant species. We considered different periodic surface models of both crystalline and amorphous nature to mimic the interstellar water ice mantles. Both models ensure that hydrogen bond cooperativity is fully taken into account at variance with the small ice cluster models. Density functional theory adopting both B3LYP-D3 and M06-2X functionals was used to predict the species/ice structure and their \emph{BEs}. As expected from the complexity of the ice surfaces, we found that each molecule can experience multiple \emph{BE} values, which depend on its structure and position at the ice surface. A comparison of our computed data with literature data shows agreement in some cases and (large) differences in others. We discuss some astrophysical implications that show the importance of calculating \emph{BEs} using more realistic interstellar ice surfaces to have reliable values for inclusion in the astrochemical models.

\end{abstract}


%-------------------------------------------------------------------
\section{Introduction} \label{sec:Introduction}

The presence of molecules in the extreme physical conditions of the interstellar medium (ISM) was considered impossible by astronomers, until the first diatomic species (CN, CH and CH$^+$) were detected in the ISM from optical and ultraviolet transitions \citep{swings1937considerations, mckellar1940evidence, douglas1942band}. 
Nowadays more than 200 gaseous molecular species (including radicals and ions) have been identified in the diffuse and dense regions of the ISM, thanks to their rotational and vibrational lines in the radio to Far-Infrared (FIR) wavelengths \citep[e.g. see the review by][]{mcguire20182018}.
In the coldest ($\leq$20--90 K) and densest ($\geq 10^3$ cm$^{-3}$) ISM, some of these molecules are also detected in the solid state via Near-Infrared (NIR) observations \citep[e.g. see the review by][]{boogert2015observations}.

We now know that the solid-state molecules are frozen species that envelop the sub-micron dust grains that permeate the ISM and whose refractory core is made of silicates and carbonaceous materials \citep[e.g.][]{jones2013heteroatom, jones2017global}. 
The grain iced mantles composition is governed by the adsorption of species from the gas phase and by chemical reactions occurring on the grain surfaces. 
For example, the most abundant component of the grain mantles is H\textsubscript{2}O, which is formed by the hydrogenation of O, O\textsubscript{2} and O\textsubscript{3} on the grain surfaces \citep[e.g.][]{hiraoka1998gas, dulieu2010experimental,oba2012water}. 

The water-rich ice is recognized from two specific NIR bands at about 3 and 6 $\mu$m which are associated with its O-H stretching and H-O-H bending modes, respectively \citep[e.g. see the review by][]{boogert2015observations}. 
In addition, species like CO, CO\textsubscript{2}, NH\textsubscript{3}, CH\textsubscript{4}, CH\textsubscript{3}OH and H\textsubscript{2}CO have also been identified as minor constituents of the ice mantles, which, for this reason, are sometimes referred to as “dirty ices” \citep{boogert2015observations}. 
Furthermore, the comparison between the astronomical spectroscopic observations and the laboratory spectra of interstellar ice sample analogous, principally based on the O-H stretching feature, has shown that the mantle ices very likely possess an amorphous-like structure resembling that of amorphous solid water (ASW) \citep[e.g.][]{oba2009formation,watanabe2008ice,boogert2015observations}.

Ice surfaces are known to have an important role in the interstellar chemistry because they can serve as catalysts for chemical reactions which cannot proceed in the gas phase, such as the formation of H\textsubscript{2}, the most abundant molecule in ISM \citep{hollenbach1971surface}. 
Ice surfaces can catalyze reactions either by behaving as: 
i) passive third body, this way absorbing part of the excess of energy released in the surface processes (adsorption and/or chemical reaction) \citep[e.g.][]{pantaleone2020chemical}; 
ii) chemical catalyst, this way directly participating in the reaction reducing the activation energies (e.g. \citep{rimola2018can, enrique2019reactivity, enrique2020revisiting}); 
and iii) reactant concentrator, this way retaining the reactants and keeping them in close proximity for subsequent reaction (e.g., CO adsorption and retention for subsequent hydrogenation to form H\textsubscript{2}CO and CH\textsubscript{3}OH \citep[e.g.][]{watanabe2002efficient, rimola2014combined, zamirri2019quantum}. 
All three processes depend on the binding energies (\emph{BEs}) of the molecules either directly (e.g. the adsorption of the species) or indirectly (e.g. because the diffusion of a particle on the grain surfaces is a fraction of its \emph{BE}) \citep[see][]{cuppen2017grain}.
In addition, molecules formed on the grain surfaces can be later transferred to the gas phase by various desorption processes, most of which depend, again, by the \emph{BE} of the species.
In practice, \emph{BEs} are crucial properties of the interstellar molecules and play a huge role in the resulting ISM chemical composition.
This key role of \emph{BEs} is very obvious in the astrochemical models that aim at reproducing the chemical evolution of interstellar objects, as clearly shown by two recent works by \cite{wakelam2017binding} and \cite{penteado2017sensitivity}, respectively.
   
Experimentally, the \emph{BEs} of astrochemical species are measured by temperature programmed desorption (TPD) experiments. 
These experiments measure the energy required to desorb a particular species from the substrate, namely a desorption enthalpy, which is equal to the \emph{BE} only if there are no activated processes \citep{he2016binding} and if thermal effects are neglected. 
A typical TPD experiment consists of two phases. In the first one, the substrate, maintained at a constant temperature, is exposed to the species that have to be adsorbed coming from the gas phase. In the second phase, the temperature is increased until desorption of the adsorbed species–collected and analyzed by a mass spectrometer–occurs. 
The \emph{BE} is then usually extracted by applying the direct inversion method on the Polanyi-Wigner equation \citep[e.g.][]{dohnalek2001physisorption, noble2012thermal}. 
The \emph{BE} values obtained in this way strongly depend on the chemical composition and morphology of the substrate and also on whether the experiment is conducted in the monolayer or multilayer regime \citep[e.g.][]{noble2012thermal, he2016binding, chaabouni2018thermal}. 
Another issue related to the TPD technique is that it cannot provide accurate B\emph{BEs} for radical species as they are very reactive. 
In literature, there are many works that have investigated the desorption processes by means of the TPD technique \citep[e.g.][]{collings2004laboratory, noble2012thermal, dulieu2013micron,fayolle2016n2, he2016binding, smith2016desorption} but they have been conducted for just a handful of important astrochemical species, whereas a typical network of an astrochemical model can contain up to five hundred species and very different substrates.  In a recent work, \citet{penteado2017sensitivity} collected the results of these experimental works, trying to be as homogeneous as possible in terms of different substrates, estimating the missing \emph{BE} values from the available data and performing a systematic analysis on the effect that the \emph{BE} uncertainties can have on astrochemical model simulations.
   
\emph{BE} values can also be obtained by means of computational approaches which, in some situations, can overcome the experimental limitations. 
Many computational works have so far focused on a few important astrochemical species like  H, H\textsubscript{2}, N, O, CO, CO\textsubscript{2}, in which \emph{BEs} are calculated on periodic/cluster models of crystalline/amorphous structural states using different computational techniques \citep[e.g.][]{al2007hydrogen, karssemeijer2014interactions, karssemeijer2014diffusion, asgeirsson2017long, senevirathne2017hydrogen, shimonishi2018adsorption, zamirri2019carbon}. 
In addition, other works have computed \emph{BE} in a larger number of species but with a very approximate model of the substrate. 
For example, in a recent work by \citet{wakelam2017binding} \emph{BE} values of more than 100 species are calculated by approximating the ASW surface with a single water molecule. 
The authors then fitted the most reliable \emph{BE} measurements (16 cases) against the corresponding computed ones, obtaining a good correlation between the two data sets. 
In this way, all the errors in the computational methods and limitations due to the adoption of a single water molecule are compensated by the fitting with the experimental values, in the view of the authors. 
The resulting parameters are then used to scale all the remaining computed \emph{BE} to improve their accuracy. 
This clever procedure does, however, consider the proposed scaling universal, leaving aside the complexity of the real ice surface and the specific features of the various adsorbates. 
In a similar work, \citet{das2018approach} have calculated the \emph{BEs} of 100 species by increasing the size of a water cluster from one to six molecules, noticing that the calculated \emph{BE} approaches the experimental value when the cluster size is increased. 
As we will show in the present work, these approaches, relying on an arbitrary and very limited number of water molecules, cannot, however, mimic a surface of icy grain. 
Furthermore, the strength of interaction between icy water molecules as well as with respect to the adsorbates depends on the hydrogen bond cooperativity, which is underestimated in small water clusters.   
  
In this work, we followed a different approach, focusing on extended periodic ice models, either crystalline or amorphous, adopting a robust computational methodology based on a quantum mechanical approach. 
We simulate the adsorption of a set of 21 interstellar molecules, whose 4 are radical species, on several specific exposed sites of the water surfaces of both extended models. 
\emph{BE} values have been calculated for more than one binding site (if present) to provide the spread of the \emph{BE} values that a same molecule can have depending on the position in the ice. 
Different approaches, with different computational cost, have been tested and compared, and the final computed \emph{BEs} have been compared with data from the computational approaches of \cite{wakelam2017binding} and \cite{das2018approach} and data from UMIST and KIDA databases as well as available experimental data \citep[e.g.][]{mcelroy2013umist, wakelam20152014}. 
One added value of this work is the definition of both a reliable, computationally cost-effective ab-initio procedure designed to arrive to accurate \emph{BE} values and an ice grain atomistic model, that can be applied to predict the \emph{BEs} of any species of astrochemical interest.

%%%%%%%%%%%%%%%%%%%%%%%%%%%%%%%%%%%%%%%%%%%%%%%%%%%%%%%%%%%%%%%%%%%%%%
%%%%%%%%%%%%%%%%%%%%%%%%%%%%%%%%%%%%%%%%%%%%%%%%%%%%%%%%%%%%%%%%%%%%%%
%%%%%%%%%%%%%%%%%%%%%%%%%%%%%%%%%%%%%%%%%%%%%%%%%%%%%%%%%%%%%%%%%%%%%%
%--------------------------------------------------------------------
\section{Computational details}
\label{sec:Computational details}
\subsection{Structure of the ice: periodic simulations}
\label{subsec:Structure of the ice: periodic simulations}
Water ice surfaces have been modelled enforcing periodic boundary conditions to define icy slabs of finite thickness either entirely crystalline or of amorphous nature. Adsorption is then carried out from the void region above the defined slabs. Periodic calculations have been performed with the ab initio CRYSTAL17 code \citep{dovesi2018quantum}. 
This software implements both the Hartree-Fock (HF) and Kohn-Sham self-consistent fields methods for the solution of the electronic Schrödinger equation, fully exploiting, if present, the crystalline or molecular symmetry of the system under investigation. 
CRYSTAL17 adopts localized Gaussian functions as basis sets, similar to the approach followed by molecular codes. This allows CRYSTAL17 to perform geometry optimizations and vibrational properties of both periodic (polymer, surfaces and crystals) and non-periodic (molecules) systems with the same level of accuracy. Furthermore, the definition of the surfaces through the slab model allows to avoid the 3D fake replica of the slab as forced when adopting plane waves basis set.

Computational parameters are set to values ensuring good accuracy in the results. 
The threshold parameters for the evaluation of the Coulomb and exchange bi-electronic integrals (TOLINTEG keyword in the CRYSTAL17 code; \cite{dovesi2018quantum}) have been set equal to 7, 7, 7, 7, 14. 
The needed density functional integration are carried out numerically over a grid of points, which is based on an atomic partition method developed by  \cite{becke1988multicenter}.
The standard pruned grid (XLGRID keyword in the CRYSTAL17 code; \cite{dovesi2018quantum}), composed by 75 radial points and a maximum of 974 angular points, was used. 
The sampling of the reciprocal space was conducted with a Pack-Monkhorst mesh \citep{pack1977special}, with a shrinking factor (SHRINK in the code CRYSTAL17; \cite{dovesi2018quantum}) of 2, which generates 4 \emph{k} points in the first Brillouin zone. 
The choice of the numerical values we assigned to these three computational parameters is fully justified in the Appendix \ref{app:comp-details}. 

Geometry optimizations have been carried out using the Broyden-Fletcher-Goldfarb-Shanno (BFGS) algorithm
\citep{broyden, fletcher1970new, goldfarb1970family, shanno1970conditioning}, relaxing both the atomic positions and the cell parameters. 
We adopted the default values for the parameters controlling the convergence, i.e., difference in energy between two subsequent steps, \num{1e-7} Hartree; and maximum components and root-mean-square of the components of the gradients and atomic displacements vectors, \num{4.5e-4} Hartree Bohr\textsuperscript{-1} and \num{3e-4} Hartree Bohr\textsuperscript{-1}, and \num{1.8e-3} Bohr and \num{1.2e-3} Bohr, respectively.
All periodic calculations were grounded on either the density functional theory (DFT) or the HF-3c method \citep{hohenberg1964inhomogeneous, sure2013corrected}. 
Within the DFT framework, different functionals were used to describe closed- and open-shell systems. 
For the former, we used the hybrid B3LYP method  \citep{becke1993, lee1988development}, which has been shown to provide a good level of accuracy for the interaction energies of non-covalent bound dimers \citep{kraus2018density}, added with the D3-BJ correction for the description of dispersive interactions \citep{grimme2010consistent, grimme2011effect}. 
For open-shell systems, treated with a spin-unrestricted formalism \citep{pople1995spin}, we used the hybrid M06-2X functional \citep{zhao2008m06}, which has been proved to give accurate results in estimating the interaction energy of non-covalent binary complexes involving a radical species and a polar molecule \citep{tentscher2013binding}. 
The choice of these two different functionals is justified by two previous works describing the accuracy on the energetic properties of molecular adducts \citep{kraus2018density, tentscher2013binding}. 
For all periodic DFT calculations we used the Ahlrichs’ triple-zeta quality VTZ basis set, supplemented with a double set of polarization functions \citep{schafer1992fully}. 
In the following, we will refer to this basis set as “A-VTZ*” (see Appendix \ref{basis set} for details of the adopted basis set).

The HF-3c method is a new method combining the Hartree-Fock Hamiltonian with the minimal basis set MINI-1 \citep{tatewaki1980systematic} and with three a posteriori corrections for: i) the basis set superposition error (BSSE), arising when localized Gaussian functions are used to expand the the basis set \citep{jansen1969non, liu1973accurate}; ii) the dispersive interactions; iii) short-ranged deficiencies due to the adopted minimal basis set \citep{sure2013corrected}.

Harmonic frequency calculations were carried out on the optimized geometries of both crystalline and amorphous ices to characterize the stationary points of each structure.
Vibrational frequencies have been calculated at the $\Gamma$ point by diagonalizing the mass-weighted Hessian matrix of second order energy derivatives with respect to atomic displacements \citep{pascale2004calculation, zicovich2004calculation}. 
The Hessian matrix elements have been evaluated numerically by a six-points formula (NUMDERIV=2 in the CRYSTAL17 code; \cite{dovesi2018quantum}), based on two displacements of $\pm0.003$ Å for each nuclear cartesian coordinates from the minimum structure.

To avoid computational burden, only a portion of the systems has been considered in the construction of the Hessian matrix, including the adsorbed species and the spatially closest interacting water molecules of the ice surface. 
This “fragment” strategy for the frequency calculation has already been tested by some of us in previous works and is fully justified by the non-covalent nature of the interacting systems where the coupling between the vibrational modes of bulk ice and adsorbate moieties is negligible \citep{tosoni2005quantum, rimola2008neutral, zamirri2017forsterite}. 

%%%%%%%%%%
%%%% PU
%%%%%%%%%%

From the set of frequencies resulting from the "fragment" calculations we worked out the zero point energy (\emph{ZPE}) for the free crystalline ice surface, the free adsorbate, the ice-surface/adsorbate complex to arrive to the corresponding correction $\Delta ZPE$, as reported in Appendix \ref{app:bssezpe}. From the $\Delta ZPE$ we corrected the electronic \emph{BE} for each adsorbates as: $BE(0) = BE - \Delta ZPE$ and found a good linear correlation \emph{BE(0)=0.854 BE}, as shown in Figure \ref{fig-app:ZPE_fit} of the Appendix \ref{app:bssezpe}.  While the "fragment frequency" strategy is fine for computing the $\Delta ZPE$ of the crystalline ice model due to the structural rigidity enforced by the system symmetry, the same does not hold for the amorphous ice. In that case, the large unit cell (60 water molecules) and their random organization renders the ice structure rather sensitive to the adsorbate interaction which causes large structural water molecules rearrangement. This, in turn, alter significantly the whole set of normal modes and the numerical value of the $\Delta ZPE$ becomes ill-defined. Nevertheless, considering that the kind of interactions operative for the crystal ice are of the same nature of those for the amorphous one, we adopted the same scaling factor 0.854 computed for the crystalline ice to correct the electronic \emph{BE} for the amorphous one.  In the following, we compared the experimental \emph{BE} usually measured for amorphous ices with the \emph{BE(0)} values. To discuss the internal comparison between adsorption features of different adsorbates on the crystalline ice, we still focused on the uncorrected \emph{BEs.
}
%%%%%%%%%
%%%% PU
%%%%%%%%%

%%%%%%%%%%
\subsection{Binding energies calculation and Counterpoise correction}
\label{subsec:Binding energies calculation and Counterpoise correction}

When Gaussian basis sets are used, a spurious contribution arises in the calculation of the molecule/surface interactions, called BSSE (basis set superposition error) \citep[e.g.][]{boys1970calculation}.
In this work, the BSSE for DFT calculations has been corrected making use of the a posteriori Counterpoise correction (CP) by Boys and Bernardi \citep{davidson1986basis}. 
The CP-corrected interaction $\Delta$E\textsuperscript{CP} energy has been calculated as
\begin{equation} \label{eq:1}
    \Delta E^{CP} = \Delta E^* + \delta E + \Delta E_L - BSSE
\end{equation}
where \emph{$\Delta$E\textsuperscript{*}} is the deformation free interaction energy, \emph{$\delta$E} is the total contribution to the deformation energy, and \emph{$\Delta$E\textsubscript{L}} is the lateral interaction (adsorbate-adsorbate interaction) energy contribution. 
Details on the calculation of each energetic term of Eq. \eqref{eq:1} can be found in the Appendix \ref{app:comp-details}.
By definition, $BE$ is the opposite of the CP-corrected interaction energy:
\begin{equation} \label{eq2}
    \Delta E^{CP} = - BE
\end{equation}
%

%%%%%%
\subsection{BE refinement with the embedded cluster method}
\label{subsec:BE refinement with the embedded cluster method}

With the aim of refining the periodic DFT \emph{BE} values for the crystalline ice model, single point energy calculations have been carried out on small clusters, cut out from the crystalline ice model, using a higher level of theory than the DFT methods with the Gaussian09 program \citep{frisch2009gaussian}. 
The adopted cluster models were derived from the periodic systems and are described in \S ~\ref{subsubsec:ONIOM2}.  
These refinements have been  performed through the ONIOM2 approach \citep{dapprich1999new}, dividing the systems in two parts that are described by two different levels of theory. 
The \emph{Model system} (i.e., a small moiety of the whole system including the adsorbate and the closest water molecules) was described by the \emph{High} level of theory represented by the single- and double-electronic excitations coupled-cluster method added with a perturbative description of triple excitations (CCSD(T)). 
The \emph{Real system} (i.e., the whole system) was described by the DFT level of theory adopted in the periodic calculations with the two different functionals for open- and closed-shell species. 
In the ONIOM2 methodology, the \emph{BE} can be written as: 

\begin{align} 
% \begin{split}
   BE(ONIOM2) = - BE(\emph{Low}, \emph{Real}) + \Delta BE \label{eq3} \\ 
   \Delta BE = BE(\emph{High}, \emph{Model}) - BE(\emph{Low}, \emph{Model}) \label{eq4}
% \end{split}
\end{align}

The final BE(ONIOM2) is also corrected by the BSSE following the same scheme described above. 
Our choices about the \emph{Model} and \emph{Real system} will be extensively justified in \S ~\ref{subsubsec:ONIOM2}.

%%%%%%%%%%%%%%%%%%%%%%%%%%%%%%%%%%%%%%%%%%%%%%%%%%%%%%%%%%%%%%%%%%%%%%%%%%
%%%%%%%%%%%%%%%%%%%%%%%%%%%%%%%%%%%%%%%%%%%%%%%%%%%%%%%%%%%%%%%%%%%%%%%%%%
%%%%%%%%%%%%%%%%%%%%%%%%%%%%%%%%%%%%%%%%%%%%%%%%%%%%%%%%%%%%%%%%%%%%%%%%%%
\section{Results}
\label{sec:Results}

%%%%%
\subsection{Ice surface models}\label{subsec:Ice ssurface models}

\subsubsection{Crystalline ice model}\label{subsubsec:crystalline ice}

\begin{figure}[ht]
   \centering
   \resizebox{\hsize}{!}{\includegraphics{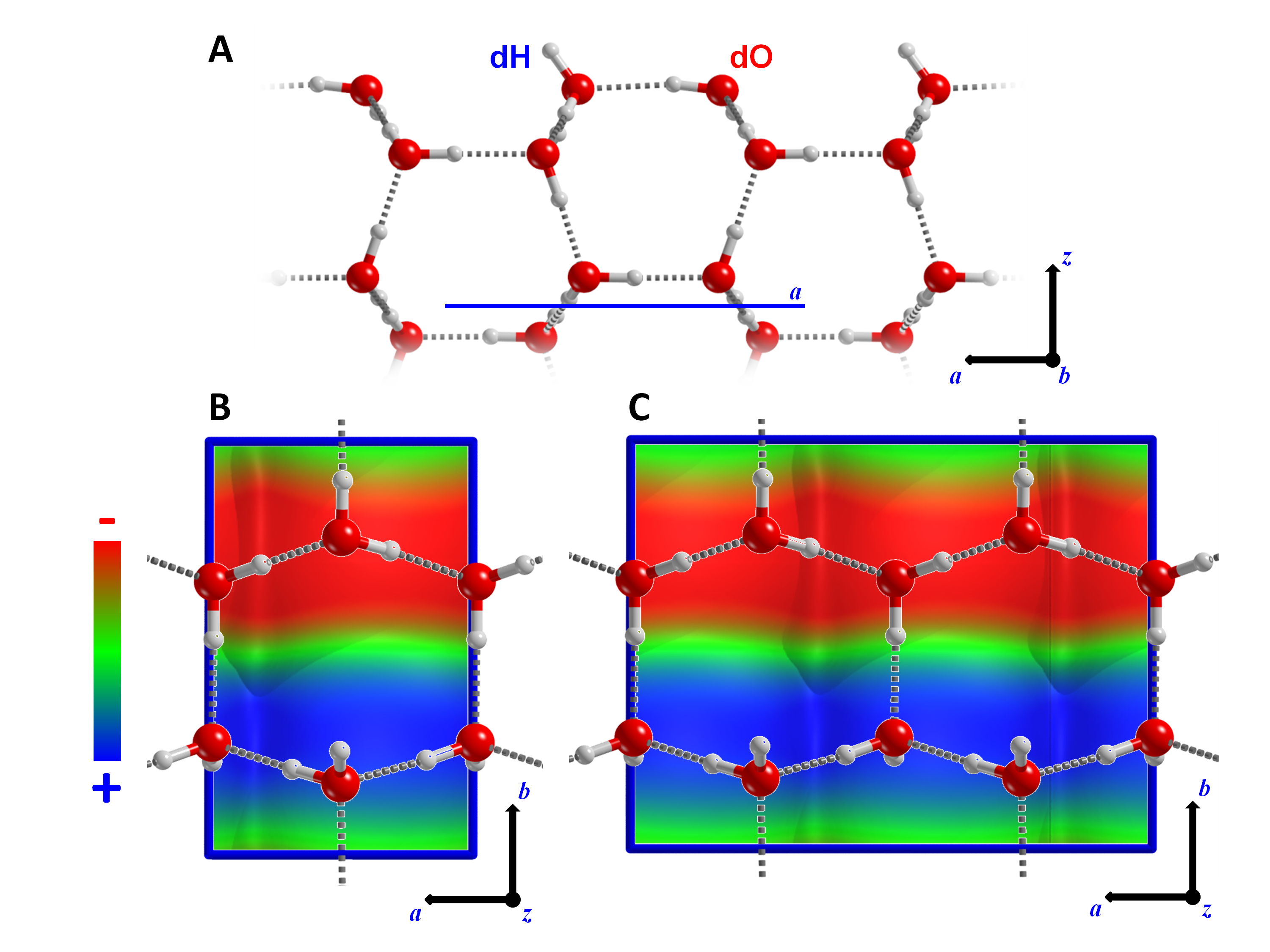}}
   \caption{(010) slab model of P-ice. A) Side view along the \textbf{\emph{b}} lattice vector. B) Top view of the 1x1 unit cell ($\vert$\textbf{\emph{a}}$\vert$ = 4.500 Å and $\vert$\textbf{\emph{b}}$\vert$ = 7.078 Å) superimposed to the electrostatic potential map, EPM. C) Top view of the 2x1 supercell ($\vert$\textbf{\emph{a}}$\vert$ = 8.980 Å, $\vert$\textbf{\emph{b}}$\vert$ = 7.081 Å) superimposed to the EPM. The iso-surface value for the electron density where the electrostatic potential is mapped is set equal to 10\textsuperscript{-6} au. Colour code: +0.02 au (blue, positive), 0.00 au (green, neutral) and -0.02 au (red, negative).}
    \label{fig:crystallinesurface}
    \end{figure}

Despite the amorphous and perhaps porous nature of the interstellar ice, we adopted, as a paradigmatic case, a proton-ordered crystalline bulk ice model usually known as \emph{P-ice} (\emph{Pna2\textsubscript{1}} space group) \citep{casassa1997proton}. 
From P-ice bulk, we cut out a slab model i.e. a 2D-periodic model representing a surface. 
Consequently, periodic boundary conditions are maintained only along the two directions defining the slab plane, while the third direction (z-axis) is non-periodic and defines the slab thickness. The slab model adopted in this work represents the P-ice (010) surface, in accordance with previous work \citep{zamirri2018ir}. 
This slab consists of twelve atomic layers, is stoichiometric and has a null electric dipole moment across the z-axis. 
This ensure an electronic stability of the model with the increase of the slab thickness \citep{tasker1979stability}. 
The slab structure has been fully optimized (unit cell and atomic fractional coordinates) at both B3LYP-D3/A-VTZ* and M06-2X/A-VTZ* DFT levels. 
As it can be seen from Figure \ref{fig:crystallinesurface} (panel A) the (010) P-ice unit cell is rather small, showing only one dangling hydrogen (dH) and oxygen (dO) as binding sites. 
For large molecules, to increase the number of adsorption sites and minimize the lateral interactions among replicas of the adsorbate we also considered a 2x1 supercell. 
The electrostatic potential maps (EPMs, see Fig. \ref{fig:crystallinesurface}, panels B and C), clearly reveal positive (blue EPM regions) and negative (red EPM regions) potentials around the dH and dO sites, respectively.

%-------------------------------------- Two column figure (place early!)
 
%%%
\subsubsection{Amorphous solid water (ASW)}\label{subsubsec:ASW}

\begin{figure}[ht]
   \centering
   \resizebox{\hsize}{!}{\includegraphics{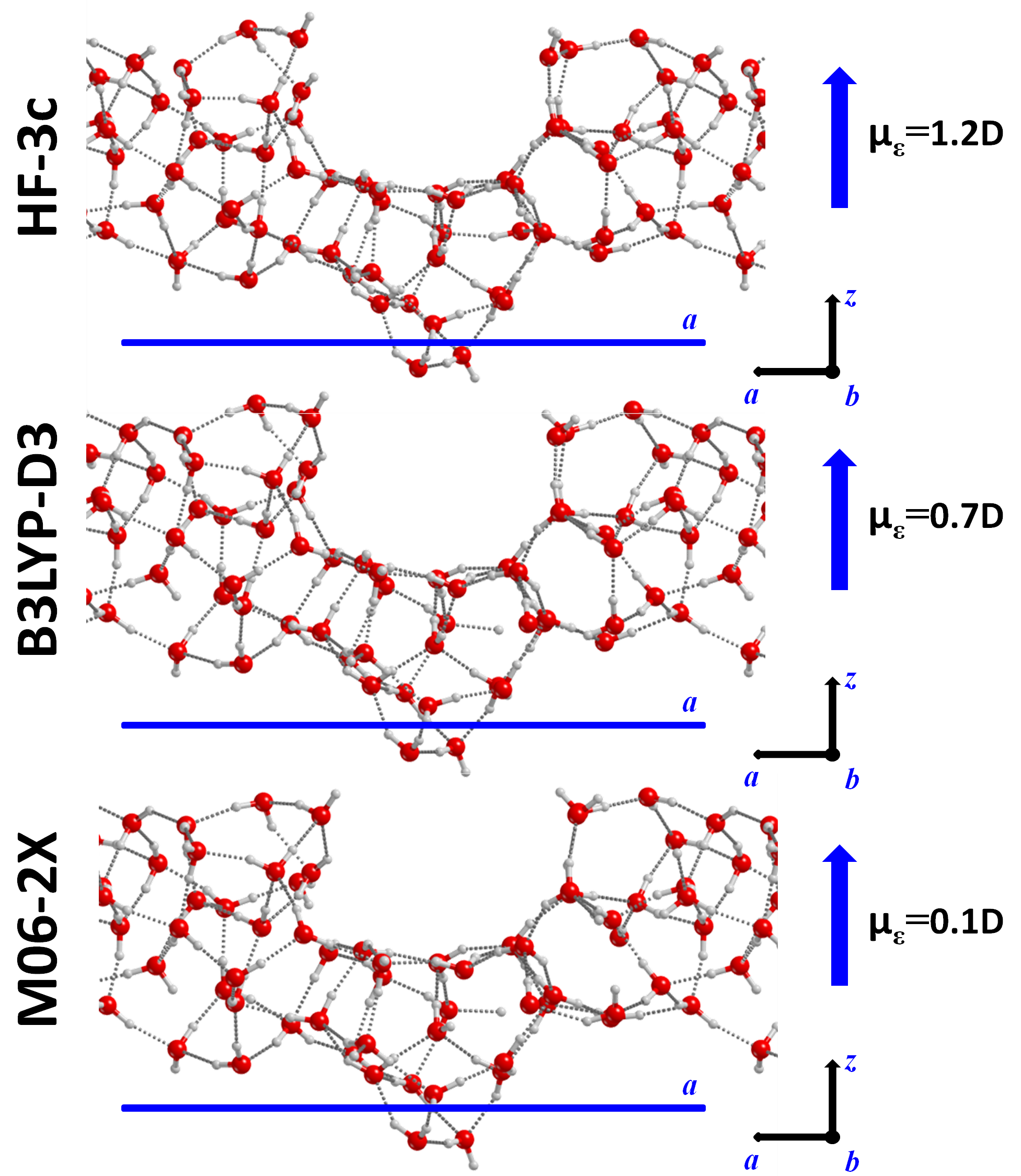}}
   \caption{Side view of the amorphous slab models. The cell parameter \textbf{\emph{a}} is highlighted as a blue line. Electric dipole moments \textbf{$\mu$\textsubscript{$\epsilon$}} along the \textbf{z} direction are shown on the right side.}
    \label{fig:amorfousurface}
\end{figure}

\begin{figure}[ht]
   \centering
   \resizebox{\hsize}{!}{\includegraphics{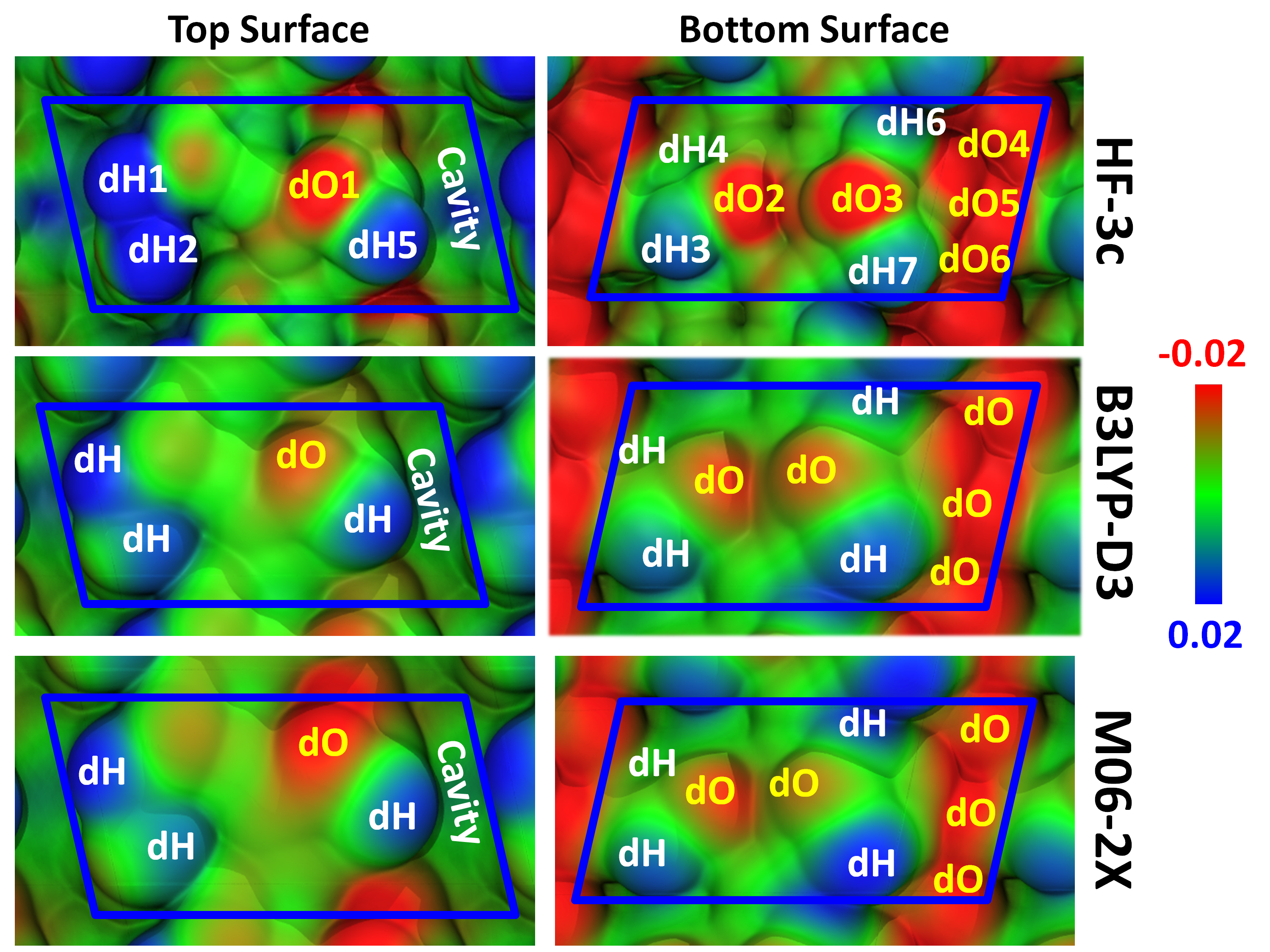}}
   \caption{Colour-coded electrostatic potential energy maps (EPMs) mapped to the electron density for the “top” and “bottom” surfaces of the amorphous slab (HF-3c, B3LYP-D3 and M06-2X optimized geometries). dO and dH sites are also labelled. The iso-surface value for the electron density is set equal to 10\textsuperscript{-6} au to which the electrostatic potential is mapped out. EPM color code: +0.02 au (blue, positive), 0.00 au (green, neutral) and -0.02 au (red, negative)..}
    \label{fig:amorfousurface2}
\end{figure}

As anticipated, the (010) P-ice surface might not be a physically sound model to represent actual interstellar ice surfaces, due to the evidence, from the spectroscopic feature of the interstellar ice, of its amorphous nature \citep{boogert2015observations}. 
The building up of amorphous surface models is a non-trivial and not unique procedure, because of the lack of a consistent and universally-accepted strategy. 
One common approach is to start from a crystalline model and heat it up to relatively high temperature by running molecular dynamics simulations (MDs) for few picoseconds. 
This step is followed by thermal annealing to freeze the ice in a glassy amorphous state. 
In this work, we adopted a different strategy. 
We refer to a recent work by \cite{shimonishi2018adsorption} in which the \emph{BEs} of a set of atomic species were computed on several water clusters, previously annealed with MD simulations. 
We re-optimized (at B3LYP-D3/A-VTZ* level only) the whole set of ice clusters and the three most stable clusters, composed by 20 water molecules each, were merged together to define a unit cell of an amorphous periodic ice. 
This procedure mimics somehow the collision of nanometric scale icy grains occurring in the molecular clouds. 
The merge of the three clusters was carried out by matching the dHs regions of one cluster with the dOs ones of the other. 
As a result, we ended up with a large 3D-periodic unit cell (with lattice parameters $\vert$\textbf{\emph{a}}$\vert$ = 21.11 Å, $\vert$\textbf{\emph{b}}$\vert$ = 11.8 Å and $\vert$\textbf{\emph{c}}$\vert$ = 11.6 Å) envisaging 60 water molecules. 
This initial bulk model was optimized at HF-3c level in order to fully relax the structure from the internal tensions of the initial guess. 
After this step, we cut out a 2D-periodic slab from the bulk structure. 
The amorphous slab is composed by 60 water molecules in the unit cell, and was further fully optimized (unit cell size and atomic coordinates)  at the HF-3c level, B3LYP-D3/A-VTZ* and M06-2X/A-VTZ* levels of theory. 
The three final structures show little differences in the positions of specific water molecules and, on the whole, the structures are very similar (Fig. \ref{fig:amorfousurface}). 
The computed electric dipole moment across the non-periodic direction (1.2, 0.7 and 0.1 Debyes for the HF-3c, B3LYP-D3 and M06-2X structures, respectively) showed a very good agreement between different models, also considering the dependence of the dipole value on the adopted quantum mechanical method.
These amorphous slab models show different structural features for the upper and lower surfaces which imparts the residual dipole moment across the slab, and, consequently, exhibit a variety of different binding sites for adsorbates. To characterize the electrostatic features of these sites, which in turn dictate the adsorption process, we resorted to the EPMs for the top/bottom surfaces of each optimized slab (Fig. \ref{fig:amorfousurface2}). 
The general characteristics are very similar for the three models: B3LYP-D3 and M06-2X giving the closest maps. 
HF-3c tends to enhance the differences between positive/negative regions due to overpolarization of the electron density caused by the minimal basis set. 
“Top” surfaces show a hydrophobic cavity (the central greenish region, Fig. \ref{fig:amorfousurface2}, absent in the P-ice slab, surrounded by dHs positive spots. 
“Bottom” surfaces shows several prominent negative regions (from five dOs) mixed with less prominent positive potentials (due to four buried dHs).

%%%%%
\subsection{BEs on crystalline ice}\label{subsec:BE-crystal}
\subsubsection{BE computed with DFT//DFT method}
\begin{figure}[ht]
   \centering
   \resizebox{\hsize}{!}{\includegraphics{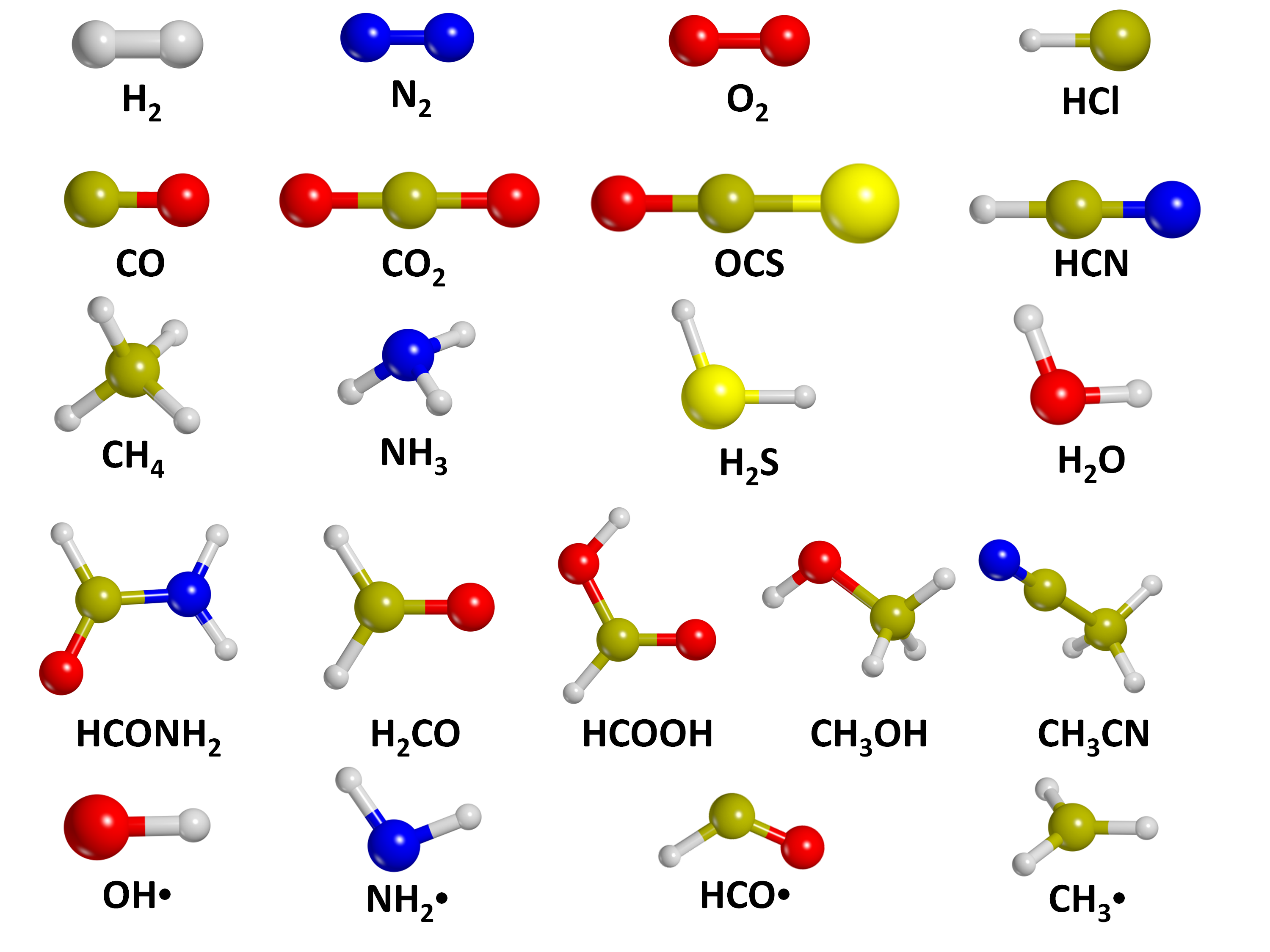}}
   \caption{Set of molecular and radical species adopted within this work for the calculation of binding energy on different ice models. O$_2$ is an open-shell (spin-triplet) species \citep{borden2017dioxygen}.}
    \label{fig:species}
\end{figure}

In this work, we simulated the adsorption of 17 closed-shell species and 4 radicals, shown in Fig. \ref{fig:species}. 
For each molecule/surface complex, geometry optimizations (unit cell plus all atomic coordinates without constraints) were performed. Initial structures were guessed by manually setting the maximum number of H-bonds between the two partners. 
The pure role of dispersion is estimated by extracting the D3 contribution from the total energy at B3LYP-D3 level of theory. 
The energetics of the adsorption processes were then computed according to Eq. \eqref{eq:1}.

As it can be seen from the results of Table \ref{tab:crystBE}, a range of interactions of different strength is established between the adsorbed species and the crystalline P-ice surface. 
Some molecules do not possess a net electric dipole moment, while exhibiting relevant electric quadrupole moments (i.e., H\textsubscript{2}, N\textsubscript{2}  and O\textsubscript{2}) or multipoles moments of higher order (i.e., CH\textsubscript{4}, see their EPMs in \ref{fig:ch4}). 
For these cases, only weak interactions are established so that \emph{BEs} are lower than 1800 K (see \emph{BE} disp values in Table \ref{tab:crystBE}). 
Interestingly, for the N\textsubscript{2}, O\textsubscript{2} and CH\textsubscript{4} cases, interactions are almost repulsive if dispersive contributions are not accounted for in the total \emph{BE} (compare \emph{BE} \emph{disp} with \emph{BE} \emph{no disp} values of Table \ref{tab:crystBE}). 
Therefore, the adsorption is dictated by dispersive forces, which counterbalance the repulsive electrostatic interactions. 
For the H\textsubscript{2} case, electrostatic interactions are attractive mainly because of the synergic effect of both the surface dH and the dO on the negative and positive parts of the H\textsubscript{2} quadrupole, respectively (see Fig. \ref{fig:h2}).

CO, OCS and CO\textsubscript{2} also exhibit a quadrupole moment, but due to the presence of heteroatoms in the structure, they can also establish H-bonds with the dH site. Consequently, \emph{BEs} are larger than the previous set of molecules (i.e., $>$ 2400 K, see Table \ref{tab:crystBE}). For these three cases, pure electrostatic interactions are attractive, but the dispersion contribution is the most dominant one over the total \emph{BE} values (compare \emph{BE} \emph{disp} with \emph{BE} \emph{no disp} values of Table \ref{tab:crystBE}). CO, in addition to a net quadrupole, also possesses a weak electric dipole, with the negative end at the carbon atom (see its EPM in \ref{fig:co}) \citep{zamirri2017forsterite}. Thus, although the two negative poles (C and O atoms) of the quadrupole, both can interact with the positive dH site, the interaction involving the C atom is energetically slightly favoured over the O atom \citep{zamirri2017forsterite, zamirri2019carbon}. Accordingly, we only considered the C-down case, the computed \emph{BE} being in good agreement with previous works \citep{zamirri2017forsterite, zamirri2018ir}. OCS also possesses a dipole and can interact with the surface through either its S- or O-ends, through dO or dH sites. However, due to the softer basic character of S compared to O, the interaction through oxygen is preferred and only considered here.

NH\textsubscript{3}, H\textsubscript{2}O, HCl, HCN and H$_2$S are all amphiprotic molecules that can both serve as acceptors and donors of H-bonds from/to the dH and dO sites. The relative strong H-bonds with the surface result in total \emph{BE} values that are almost twice higher than the values of the previous set of molecules (i.e., CO, OCS and COS). Although also in these cases dispersive forces play an important contribution to the BE, the dominant role is dictated by the H-bonding contribution.

For the adsorption of CH\textsubscript{3}OH, CH\textsubscript{3}CN and the three carbonyl-containing compounds, i.e., H\textsubscript{2}CO, HCONH\textsubscript{2} and HCOOH, all characterized by large molecular sizes, we adopted the 2x1 supercell (shown in Fig. \ref{fig:crystallinesurface}) to minimize the lateral interactions between adsorbates. Consequently, two dHs and two dOs are available for adsorption. Therefore, for some of these species (i.e., the carbonyl-containing ones), we started from more than one initial geometry to improve a better sampling of the adsorption features on the (010) P-ice surface (the different cases on the supercell are labeled as SC1 and SC2 in Table \ref{tab:crystBE} and the geometries are reported in \ref{fig:hconh2-sc2}). The \emph{BE} values of these species are among the highest ones, due to the formation of multiple H-bonds with the slab (and therefore increasing the electrostatic contribution to the interactions), and a large dispersion contribution due to the larger sizes of these molecules with respect to the other species.

The adsorption study has also been extended to four radicals (i.e., OH• NH$_2$• CH$_3$• HCO•), since they are of high interest due to their role in the formation of interstellar compounds \citep{sorrell2001origin, bennett2007formation}. OH• and NH$_2$• form strong H-bonds with the dH and dO sites of the slab, at variance with CH$_3$• and HCO• cases, as shown by the higher \emph{BE} values. Because of the nature of the M06-2X functional, we cannot separate the dispersion contributions to the total \emph{BEs}. Interestingly, in all cases, we did not detect transfer of the electron spin density from the radicals to the ice surface, i.e., the unpaired electron remains localized on the radical species upon adsorption.

%%%%%
\subsubsection{The ONIOM2 correction and the accuracy of the DFT//DFT \emph{BE} values}
\label{subsubsec:ONIOM2}
As described in the Computational Details section, the ONIOM2 methodology has been employed to check the accuracy of the B3LYP-D3/A-VTZ* and M06-2X/A-VTZ* theory levels, both representing the \emph{Low} level of calculation. For this specific case, to reduce the computational burden, we only considered 15 species, leaving aside N\textsubscript{2}, O\textsubscript{2}, H\textsubscript{2}O, CH\textsubscript{4}, CH\textsubscript{3}CN and CH\textsubscript{3}• radical. Here, the \emph{Real system} is the periodic P-ice slab model without adsorbed species. Therefore, the BE(\emph{Low},\emph{Real}) term in Eq. \eqref{eq3} corresponds to the \emph{BEs} at the DFT theory levels, hereafter referred to as BE(\emph{DFT}, \emph{Ice}). The \emph{Model system} is carved from the optimized geometry of the periodic system: it is composed by the adsorbed molecule plus $n$ ($n = 2,6$; the latter only for the H\textsubscript{2} case) closest water molecules of the ice surface to the adsorbates. For the \emph{Model systems}, two single point energy calculations have been carried out: one at the \emph{High} level of theory, i.e., CCSD(T), calculated with Gaussian09, and the other at the \emph{Low} level of theory, employing the same DFT methods as in the periodic calculations, calculated with CRYSTAL17. For the sake of clarity, we renamed the two terms BE(\emph{High},\emph{Model}) and BE(\emph{Low},\emph{Model}) in Eq. \eqref{eq3} for any molecular species $\mu$, as BE(CCSD(T), $\mu$-$n$H\textsubscript{2}O) and BE(DFT, $\mu$-$n$H\textsubscript{2}O), respectively.

\begin{figure}[ht]
   \centering
   \resizebox{\hsize}{!}{\includegraphics{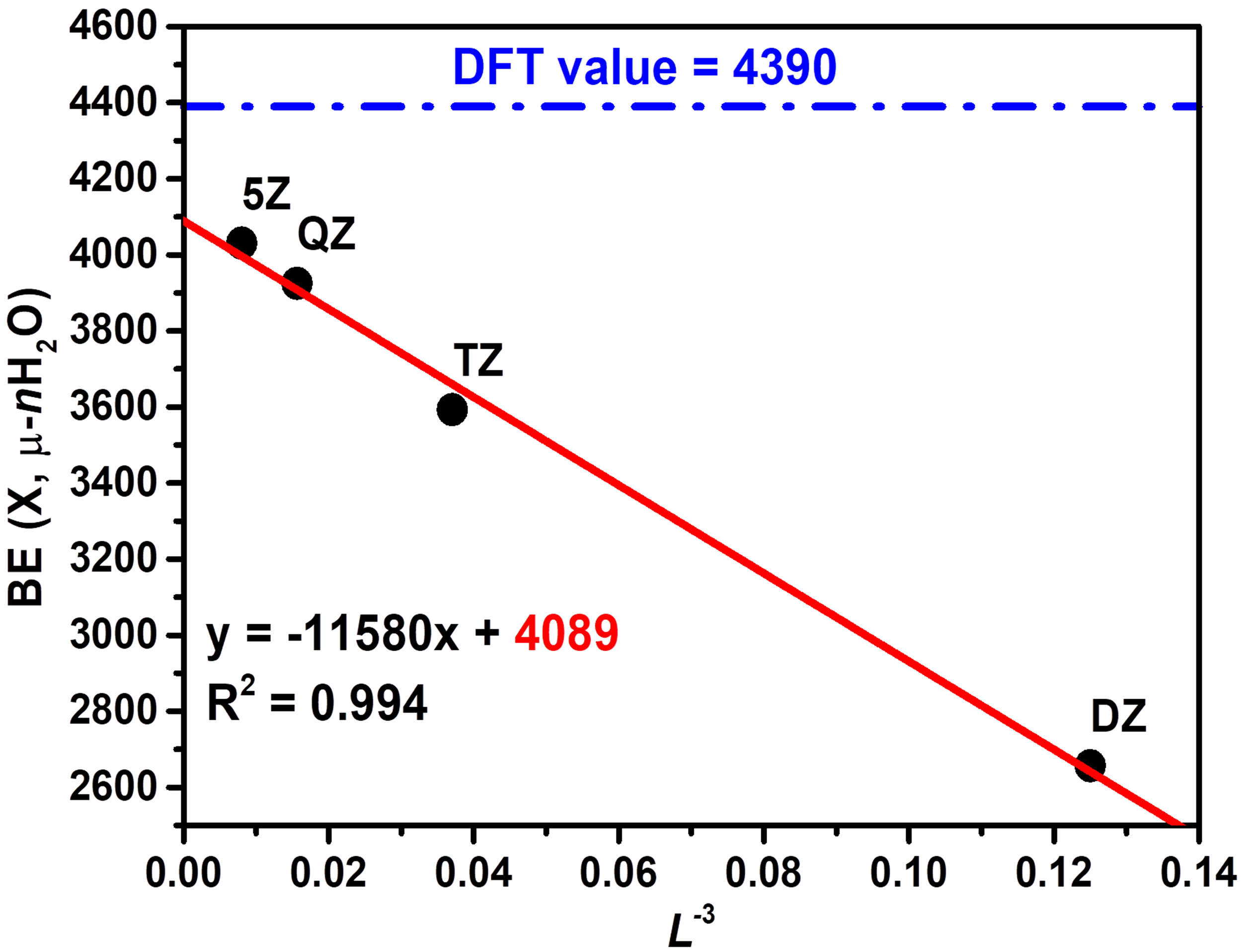}}
   \caption{BE(X, $\mu$-$n$H\textsubscript{2}O) extrapolated value at infinite basis set for the case of NH\textsubscript{3}. The dashed-dot blue line represents the \emph{BE} computed for the \emph{BE}(DFT, $\mu$-$n$H\textsubscript{2}O) at DFT//A-VTZ* level (4390 K). Solid red line represents the linear fit of the BE(CCSD(T), $\mu$-$n$H\textsubscript{2}O) values (red squares) calculated with DZ, TZ, QZ and 5Z basis sets. The extrapolated BE(X, $\mu$-$n$H\textsubscript{2}O) at infinite basis set is highlighted in red in the fitting equation (4089 K).}
    \label{fig:CCSD(T)extrap}
    \end{figure}

As CCSD(T) is a wavefunction-based method, the associated energy strongly depend on the quality of the adopted basis set \citep{cramer2002essentials}. Consequently, accurate results are achieved only when complete basis set extrapolation is carried out \citep{cramer2002essentials}; accordingly, we adopted correlation consistent basis sets \citep{dunning1989gaussian}, here named as cc-pV\emph{N}Z, where “cc” stands for correlation consistent and \emph{N} stands for double (D), triple (T), quadruple (Q), etc... Therefore, we performed different calculations improving the quality of the basis set from \textbf{Jun}-cc-pVDZ to \textbf{Jun}-cc-pVQZ (and even \textbf{Jun}-cc-pV5Z when feasible) \citep{bartlett2007coupled, papajak2011perspectives}, extrapolating the BE(CCSD(T), $\mu$-$n$H\textsubscript{2}O) values for \emph{N} $\to\infty$.  
Figure \ref{fig:CCSD(T)extrap} shows, using NH\textsubscript{3} as illustrative example, the plot of the BE(CCSD(T), $\mu$-$n$H\textsubscript{2}O) values as a function of $1/L^3$ where \emph{L} is the cardinal number corresponding to the \emph{N} value for each correlation-consistent basis set. For all other species, we observed similar trends. This procedure was used in the past to extrapolate the \emph{BE} value of CO adsorbed at the Mg(001) surface \citep{ugliengo2002dispersive}.

\begin{figure}[ht]
   \centering
   \resizebox{\hsize}{!}{\includegraphics{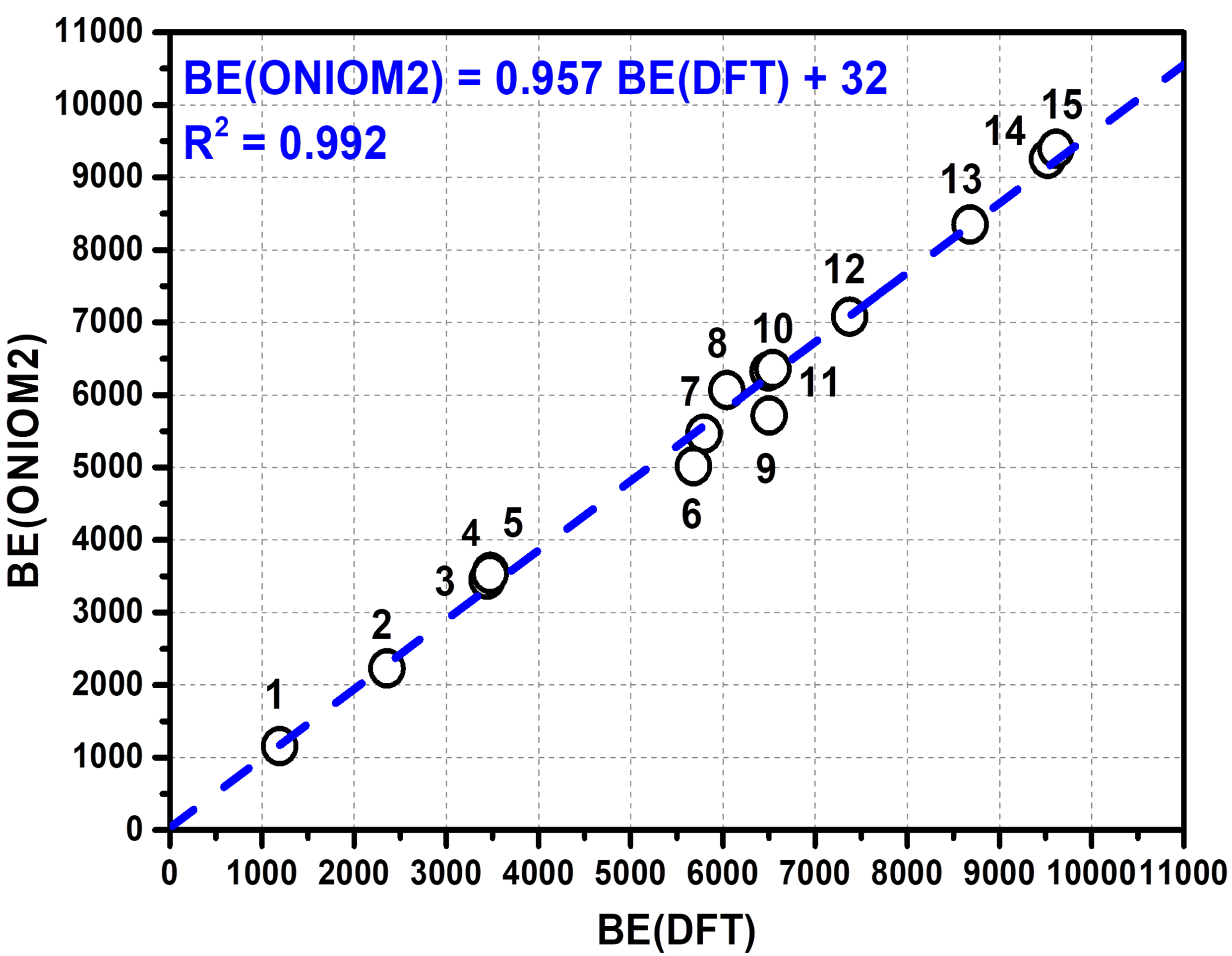}}
   \caption{Linear fit between periodic DFT/A-VTZ* \emph{BE} values (\emph{BE(DFT)}) and the basis set extrapolated ONIOM2 \emph{BE} values (\emph{BE(ONIOM2)}). All values are in K. Fit parameters are also reported. Legend: 1:H\textsubscript{2}; 2:CO; 3:CO\textsubscript{2}; 4:HCO•; 5:OCS; 6:H\textsubscript{2}S; 7:HCN; 8:NH\textsubscript{2}•; 9:H\textsubscript{2}CO; 10:HCl; 11:OH•; 12:NH\textsubscript{3}; 13:CH\textsubscript{3}OH; 14:HCOOH; 15:HCONH\textsubscript{2}.}
    \label{fig:CCSD(T)linfit}
    \end{figure}

The procedure gives for the extrapolated BE(CCSD(T), $\mu$-$n$H\textsubscript{2}O) a value of 4089 K in excellent agreement with the value computed by the plain B3LYP-D3/A-VTZ* at periodic level of 4390 K (see Fig. \ref{fig:CCSD(T)extrap}). Very similar agreement was computed for all considered species as shown in Fig. \ref{fig:CCSD(T)linfit}, in which a very good linear correlation is seen between BE(ONIOM2) and BE(DFT). Therefore, we can confidently assume the periodic B3LYP-D3/A-VTZ* (closed shell molecules) or the M06-2X/A-VTZ* (radical species) plain \emph{BE} values as reliable and accurate enough and are those actually used in this work.
 
%%%%%    
\subsubsection{BE computed with composite DFT/HF-3c method}\label{subsubsec:3.2.3}

In the previous Section we proved the DFT/A-VTZ* as a reliable and accurate method to compute the \emph{BEs} of molecules and radicals on the crystalline (010) P-ice ice slab. However, this approach can become very computational costly when moving from crystalline to amorphous model of the interstellar ice, as larger unit cells are needed to enforce the needed randomness in the water structure. Therefore, we tested the efficiency and accuracy of the cost-effective computational HF-3c method (see \S ~\ref{sec:Computational details}).

\begin{figure}[ht]
   \centering
   \resizebox{\hsize}{!}{\includegraphics{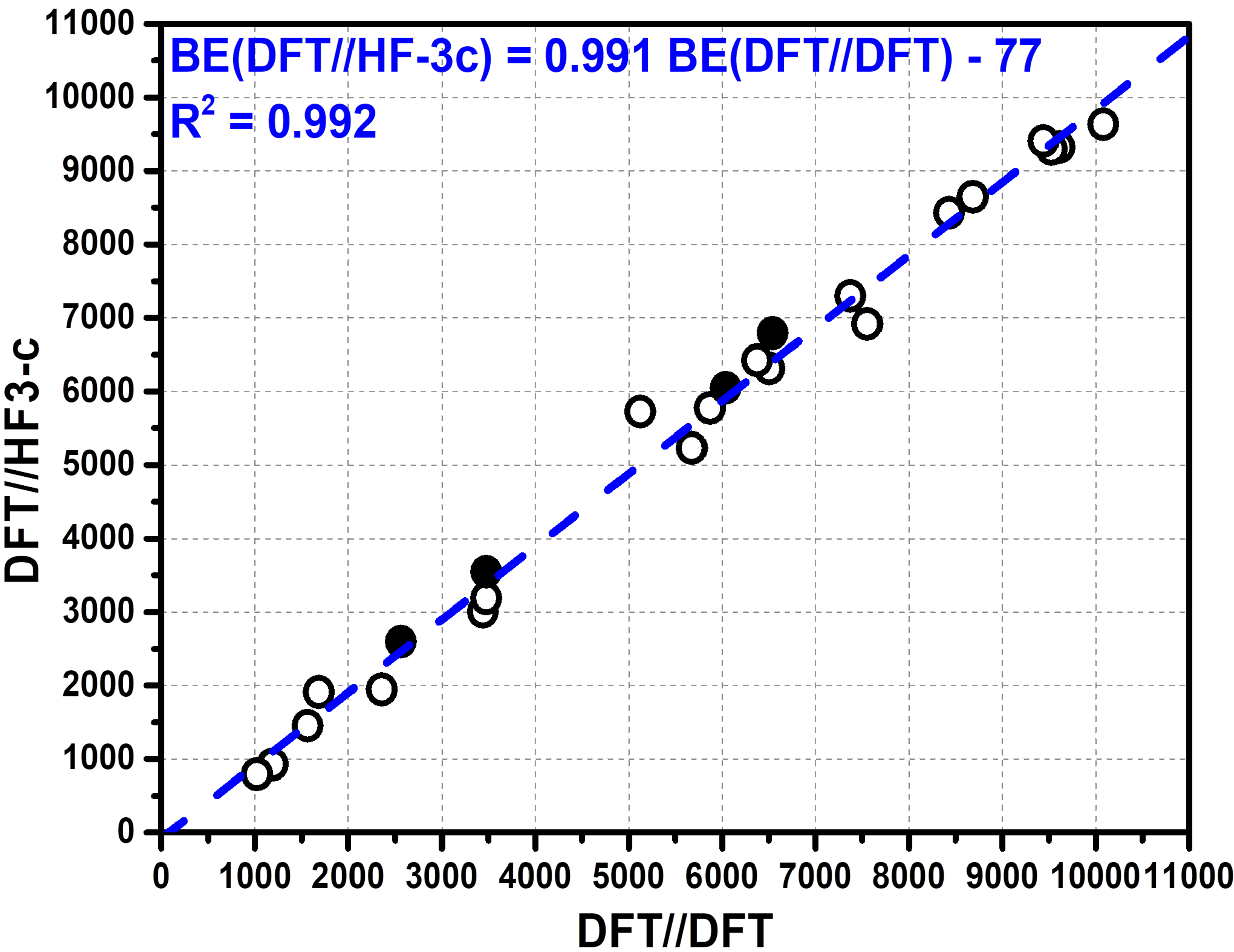}}
   \caption{Linear fit between the \emph{BE} values calculated with the full DFT computational scheme and the \emph{BE} values calculated with the composite DFT//HF-3c computational scheme for the crystalline ice model (all values in K). Black-filled and empty triangles stand for open-shell and closed-shell species, respectively.}
    \label{fig:DFTHF3clinfit}
    \end{figure}

To this end, we adopted a composite procedure which has been recently assessed and extensively tested in the previous work by some of us on the structural and energetic features of molecular crystals, zeolites and biomolecules \citep{cutini2016assessment, cutini2017method, cutini2019cost}. We started from the DFT/A-VTZ* optimized structure just discussed for the crystalline ice. We re-optimize each structure at HF-3c level to check the changes in the structures resulting from the more approximated method. Then we run a single point energy calculations at the DFT/A-VTZ* (B3LYP-D3 and M06-2X) levels to evaluate the final \emph{BE} values. 
The results obtained are summarized in Fig. \ref{fig:DFTHF3clinfit}, showing a very good linear correlation between the \emph{BE} values computed as described.

The largest percentage differences are found for the smallest BEs, that is, those dominated by dispersion interactions or very weak quadrupolar interactions (i.e., N\textsubscript{2}, O\textsubscript{2}, H\textsubscript{2} and CH\textsubscript{4}) in which the deficiencies of the minimal basis set encoded in the HF-3c cannot be entirely recovered by the internal corrections. For higher \emph{BE} values, the match significantly improves, in some cases being almost perfect. Even for radicals, the composite approach gives good results. It is worth mentioning that HF-3c optimized geometries are very similar to the DFT-optimized ones (only slight geometry alterations occurred), indicating that the adducts are well-defined minima in both potential energy surfaces.
This successful procedure calibrated on crystalline ice is therefore adopted to model the adsorption of all 21 species on the proposed amorphous slab model, a task which would have been very expensive at the full DFT/A-VTZ* level.

%%%%%%%%%%%%%%%%%%
\subsection{BEs on amorphous solid water (ASW)}\label{subsec:BE-ASW}                     
\begin{figure}[ht]
   \centering
   \resizebox{\hsize}{!}{\includegraphics{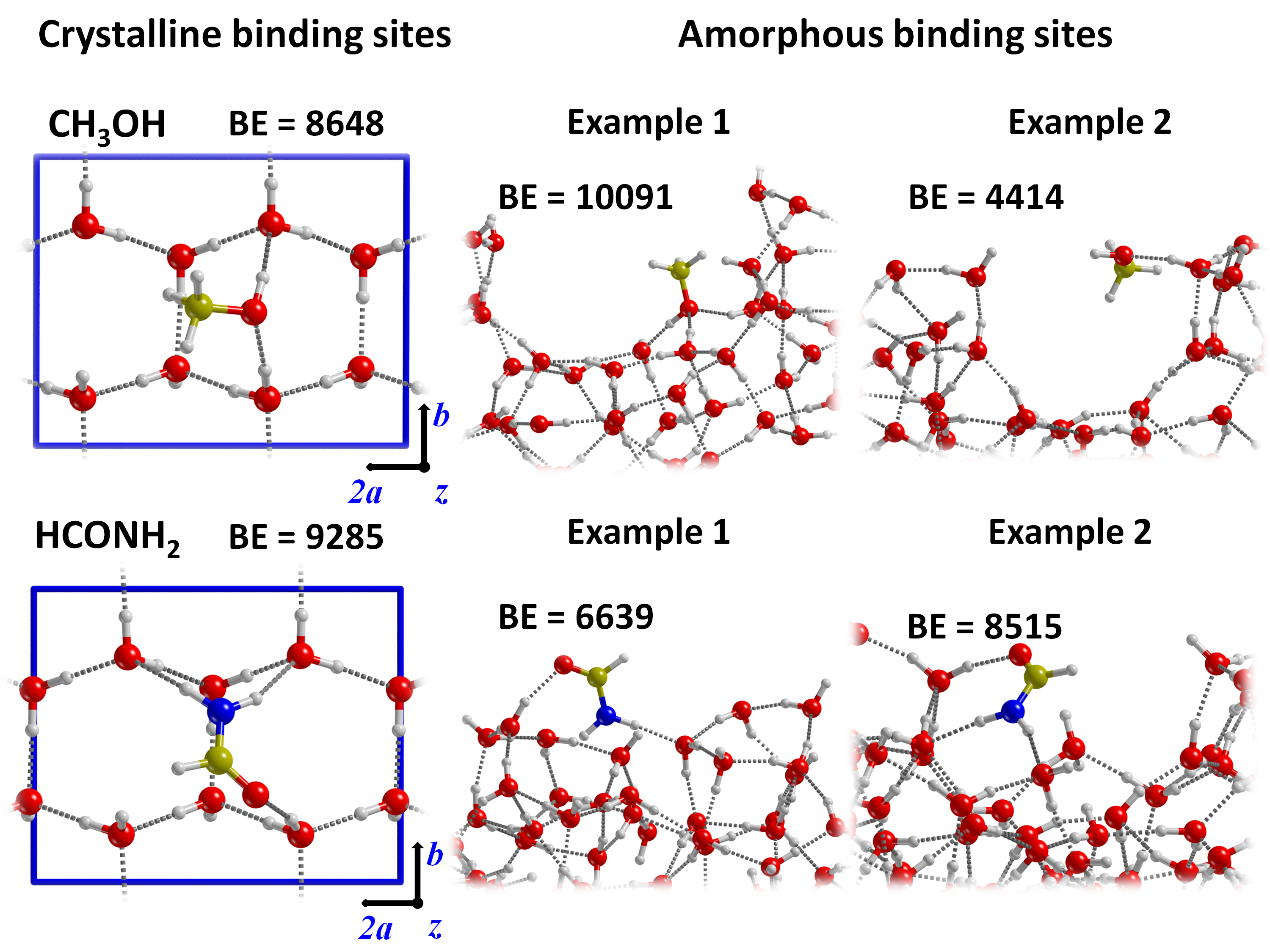}}
   \caption{Comparison of the final optimized geometries for CH\textsubscript{3}OH and HCONH\textsubscript{2} (as illustrative examples) on the crystalline ice (\S ~\ref{subsubsec:crystalline ice}) and on the ASW (\S ~\ref{subsubsec:ASW}). The \emph{BE} values (in Kelvin) are reported in each plot.}
    \label{fig:methanolformcomparison}
    \end{figure}

On the ASW model, due to the presence of different binding sites, a single \emph{BE} value is not representative of the whole adsorption processes as it is the case for almost all adsorbates on the crystalline surface. Therefore, we computed the \emph{BE} with the composite DFT//HF-3c procedure (see \S ~\ref{subsubsec:3.2.3}) by sampling different adsorption sites at both the “top” and “bottom” surfaces of the amorphous slab. The starting initial structures of each adsorbate were setup by hand, following the maximum electrostatic complementarity between the EPMs (see Fig. \ref{fig:amorfousurface2}) of the ice surface and that of a given adsorbate. For each molecule, at least four \emph{BE} values have been computed on different surface sites. Figure \ref{fig:methanolformcomparison} reports the examples of methanol and formamide: for each molecule, we show the geometry on the crystalline ice and in two different sites of ASW, as well as the \emph{BE} associated with each geometry. For methanol, the \emph{BE} is 8648 K in the crystalline ice, whereas it is 4414 and 10091 K in the two shown ASW sites. Similarly, for formamide, \emph{BE} is 9285 K on the crystalline ice and 6639 and 8515 K on the ASW. These two examples show that \emph{BEs} on ASW can differ more than a twice depending on the site and that the value on the crystalline ice can also be substantially different to that on the ASW.

\begin{table*}[t]
\begin{threeparttable}
\centering
      \caption{Summary of the \emph{BE} values (in Kelvin) obtained for the crystalline P-ice (010) slab with DFT//DFT and DFT//HF-3c methods. Legend: “\emph{BE disp}” = \emph{BE} value including the D3 contribution, “\emph{BE no disp}” = \emph{BE} values not including D3 contribution; “-disp(\%)” = absolute (percentage) contribution of dispersive forces to the total \emph{BE disp}.}
         \label{tab:crystBE}
          \begingroup
\setlength{\tabcolsep}{6pt} % Default value: 6pt
\renewcommand{\arraystretch}{1.1} % Default value: 1
  \begin{tabular}{@{} l c c c c c c c@{}} \toprule
    \multirow{2}{*}{\textbf{Species}} & \multicolumn{3}{c}{\textbf{(010) P-ice crystalline slab DFT//DFT}} & \phantom{abc} & \multicolumn{3}{c}{\textbf{(010) P-ice crystalline slab DFT//HF-3c}} \\
     \cmidrule{2-4} \cmidrule{6-8}
       & \emph{BE disp} & \emph{BE no disp} & -disp(\%) && \emph{BE disp} &  \emph{BE no disp} & -disp(\%) \\
      \midrule
H\textsubscript{2}          & 1191 & 565 & 625(53) && 926 & 241 & 686(74)   \\
O\textsubscript{2}          & 1022 &-373 & 1034(137) && 794 &-84  & 878(110)   \\
N\textsubscript{2}          & 1564 &-72  & 1636(104) && 1455&-180 & 1636(160)   \\
CH\textsubscript{4}         &     1684    & -229 &    1912(113)  && 1912 &-349&	2261(118)\\
CO                          &     2357    &  698&  1660(71)  &&	1948    & 60 &  1888 (97) \\
CO\textsubscript{2}         &     3440    &   1540& 1900(55) &&   3007    & 938&  2069(69)\\
OCS                         &     3476    &    120& 3356(97) &&   3187	& 265&  2923(92)  \\
HCl                         &     6507    &   4402&	2093(32)&&	6314	&3488&	2237(39) \\
HCN                         &     5124    &   3067&	2057(29)&&	5725	&3271&	3043(48) \\
H\textsubscript{2}O         &     8431    &   6844&	1588(19)&&	8431	&6808&	1612(19) \\
H\textsubscript{2}S         &     5677    &   3380&	2297(40)&&	5232	&3199&	2105(40) \\
NH\textsubscript{3}         &     7373    &   5533&	1852(25)&&	7301	&5484&	1816(25) \\
CH\textsubscript{3}CN       &     7553    &   4450&	3103(41)&&	6916	&3259&	2598(44) \\
CH\textsubscript{3}OH       &     8684    &   6014&	2670(31)&&	8648	&6026&	2237(27)  \\
H\textsubscript{2}CO-SC1    &     5869    &   3885&	1985(34)&&	5773	&4053&	2369(37)  \\
H\textsubscript{2}CO-SC2    &     6375    &   3692&	2682(42)&&	6423	&3716&	2057(36)  \\
HCONH\textsubscript{2}-SC1  &     9610    &   6459&	3151(33)&&	9321	&6158&	3163(34)\\
HCONH\textsubscript{2}-SC2  &     10079    &  6483&	3608(36)&&	9634	&6074&	3560(37)  \\
HCOOH                       &     9526    &   7325&	2189(23)&&	9297	&7168&	2117(23) \\
HCOOH-SC                    &     9442    &   7301&	2021(21)&&	9405	&7541&	1864(20)  \\
OH•                         &     6543*    &   &  &&   6795*   &  &       \\
HCO•                        &     3476*    &   &  &&   3548*   &  &      \\
CH\textsubscript{3}•        &     2562*    &   &  &&   2598*   &  &     \\
NH\textsubscript{2}•        &     6038*    &   &  &&   6050*   &  &     \\

    \bottomrule    
  \end{tabular}
  \endgroup
\begin{tablenotes}
      \small
      \item *Notes: For radical species (energy at M06-2X level) we cannot discern between \emph{disp} and \emph{no disp} data. 
    \end{tablenotes}
  \end{threeparttable}
   \end{table*}

\begin{figure}[ht]
   \centering
   \resizebox{\hsize}{!}{\includegraphics{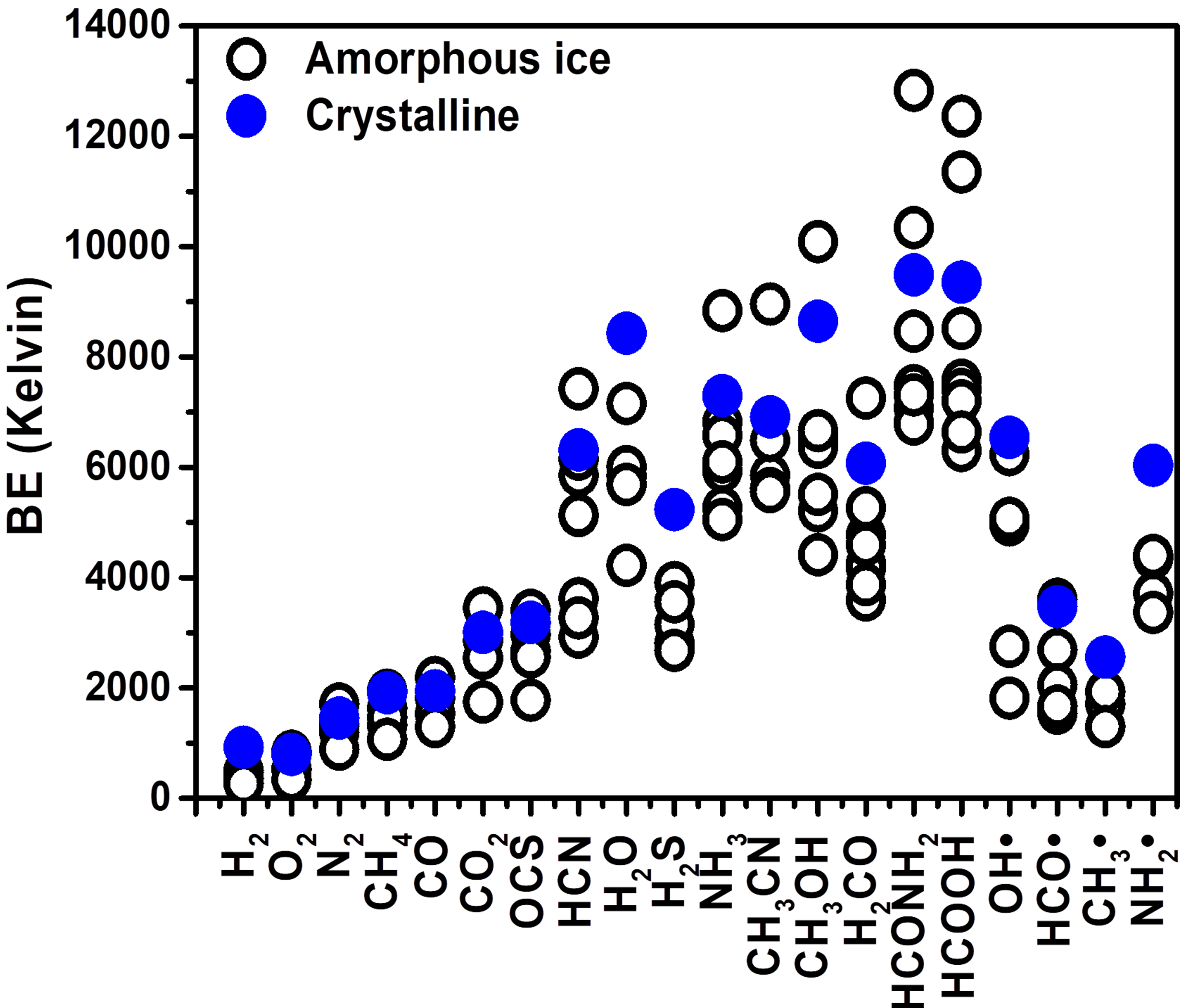}}
   \caption{Comparison between the DFT//HF-3c \emph{BEs} (in Kelvin) computed on the crystalline ice (filled blue circles) and ASW (open circles), respectively, for 20 species studied here: HCl is missing as it dissociates on the ASW (see text).}
    \label{fig:cryst vs asw}
    \end{figure}

Figure \ref{fig:cryst vs asw} shows the computed BEs, on crystalline ice and ASW, for the studied species. The list of all computed \emph{BE} values on ASW is reported in Table \ref{tab:amorphousvalues}, while Table \ref{tab:2} reports the computed minimum and maximum \emph{BE} values on ASW and the \emph{BEs} on the crystalline ice for all the studied species. As already mentioned when presenting the methanol and formamide examples, the amorphous nature of the ice can yield large differences in the calculated \emph{BEs} with respect to the crystalline values. Figure \ref{fig:cryst vs asw} shows that while the \emph{BEs} for crystalline \emph{versus} amorphous ices are very close to each other for H\textsubscript{2}, O\textsubscript{2}, N\textsubscript{2}, CH\textsubscript{4}, CO, CO\textsubscript{2}, OCS, the ones computed for the remaining molecules for the crystalline ice fall in the highest range of the distribution of the amorphous \emph{BE} values. This behavior can be explained considering the smaller distortion energy cost upon adsorption for the crystalline ice compared to the amorphous one. The different local environment provided by crystalline \emph{versus} amorphous ices is also the reason for HCl being molecularly adsorbed at the crystalline ice while becomes dissociated at the amorphous one. 
Further details about the case of HCl are reported in section \ref{app:hcl} of the Appendix.
Probably this will not occur for HF, not considered here, which is expected to be molecularly adsorbed on both ices due to its higher bond strength compared to HCl. Nevertheless, as we did not explore exhaustively all possible configurations of the adsorbates at the amorphous surfaces, we cannot exclude that some even more/less energetic binding cases remain to be discovered.

\begin{figure}[ht]
   \centering
   \resizebox{\hsize}{!}{\includegraphics{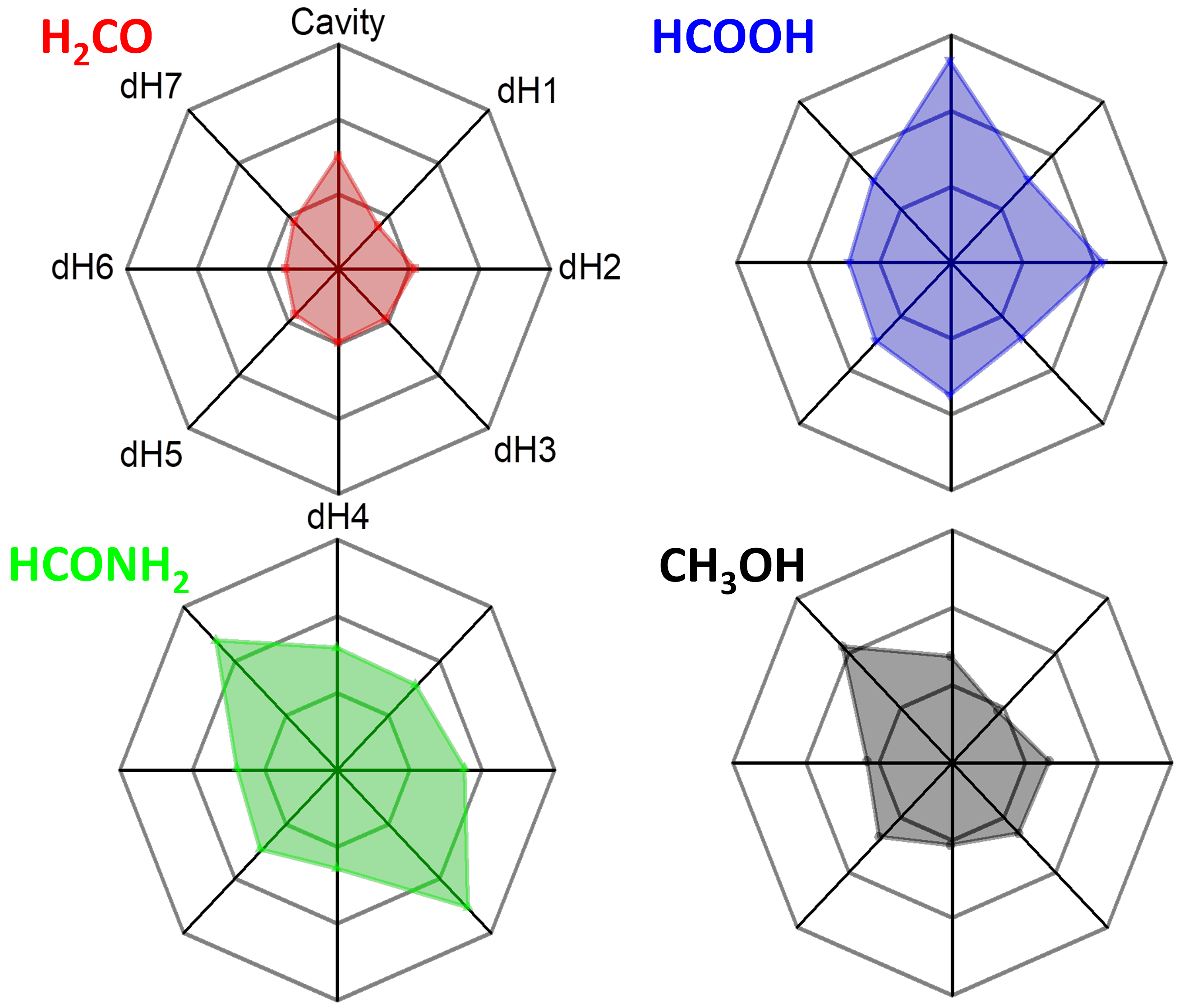}}
   \caption{Spider graphs of the DFT//HF-3c \emph{BE} values (in Kelvin) calculated on the same adsorption eight sites of the ASW for the H\textsubscript{2}CO (red), CH\textsubscript{3}OH (black), HCONH\textsubscript{2} (green) and HCOOH (blue) molecules. The \emph{BE} values scale goes from 0 Kelvin (center of the graph) to 14433 Kelvin (vertices of the polygon) in steps of 4811 Kelvin. Labeling of dH and dO sites is referring to Fig. \ref{fig:amorfousurface2}}
    \label{fig:spider graphs}
    \end{figure}

Some adsorbates show similar trends in the BEs, despite their different chemical nature. This is shown in Fig. \ref{fig:spider graphs}, in which we plot the \emph{BE} values on ASW for four molecules that have been adsorbed at the same adsorption sites: formaldehyde, formic acid, formamide and methanol. The \emph{BE} distributions for the H\textsubscript{2}CO and HCOOH are very similar (in their relative values), and those for CH\textsubscript{3}OH and HCONH\textsubscript{2} show some similarities, despite the large difference in the chemical functionality.

\begin{table*}[ht]
  \centering
  \setlength{\tabcolsep}{6pt} % Default value: 6pt
\renewcommand{\arraystretch}{1.1} % Default value: 1
  \caption{\emph{BE} values ($\mathrm{K}$) calculated with DFT//HF-3c method for every case on the amorphous slab model, \textbf{where the ZPE correction has not been added.}}
    \begin{tabular}{@{}l c c c c c c c c@{}}
    \toprule
     & \multicolumn{8}{c}{\textbf{Amorphous ice \emph{BE} values}} \\
    \toprule
    \textbf{Species} & {\textbf{Case 1}} & {\textbf{Case 2}} & {\textbf{Case 3}} & {\textbf{Case 4}} & {\textbf{Case 5}} & {\textbf{Case 6}} & {\textbf{Case 7}} & {\textbf{Case 8}} \\
    \midrule
    H\textsubscript{2}    & 469   & 505   & 277   & 361   & 265   &       &       &  \\
    O\textsubscript{2}    & 818   & 854   & 854   & 529   & 337   &       &       &  \\
    N\textsubscript{2}    & 1347  & 1708  & 1311  & 1191  & 890   &       &       &  \\
    CH\textsubscript{4}   & 1323  & 1960  & 1636  & 1467  & 1070  &       &       &  \\
    CO    & 1816  & 2189  & 1540  & 1527  & 1299  &       &       &  \\
    CO\textsubscript{2}   & 2863  & 3452  & 2538  & 2550  & 1744  &       &       &  \\
    OCS   & 3404  & 2971  & 2670  & 2562  & 1780  &       &       &  \\
    HCN   & 2923  & 5124  & 3620  & 5136  & 3271  & 5857  & 6146  & 7421 \\
    H\textsubscript{2}O   & 7156  & 5845  & 6014  & 5689  & 4222  &       &       &  \\
    H\textsubscript{2}S   & 2814  & 3909  & 3151  & 3560  & 2682  &       &       &  \\
    NH\textsubscript{3}   & 8840  & 5268  & 6820  & 5930  & 6579  & 6098  & 5052  &  \\
    CH\textsubscript{3}CN & 8960  & 5857  & 5617  & 5557  & 6483  &       &       &  \\
    CH\textsubscript{3}OH & 6531  & 4414  & 6519  & 6362  & 5208  & 6663  & 5509  & 10091 \\
    H\textsubscript{2}CO  & 3596  & 4258  & 4174  & 4775  & 5268  & 4594  & 3873  & 7253 \\
    HCONH\textsubscript{2} & 12833 & 7481  & 10344 & 6820  & 8467  & 7072  & 6783  & 7313 \\
    HCOOH & 7577  & 7409  & 8515  & 12364 & 6302  & 7204  & 6639  & 11354 \\
    OH•   & 6230  & 1816  & 4955  & 2754  & 5076  &       &       &  \\
    HCO•  & 2694  & 2057  & 1540  & 3608  & 1672  &       &       &  \\
    CH\textsubscript{3}•  & 1708  & 1936  & 1299  &       &       &       &       &  \\
    NH\textsubscript{2}•  & 4354  & 3716  & 4402  & 3368  &       &       &       &  \\
    \bottomrule
    \end{tabular}
  \label{tab:amorphousvalues}
\end{table*}

 \begin{table*}[t]
  \begin{threeparttable}
  \centering
   \caption{Summary of our computed \emph{BEs} and comparison with data from the literature. The first column reports the species, the columns 2 to 4 the \emph{BEs} computed in the present work {and corrected for the ZPE}, columns 5 and 6 the values obtained via calculations from other authors, column 7 and 8 the values in the two astrochemical databases KIDA and UMIST (see text), and the last columns the values measured in different experiments. Units are in K and the references are listed in the notes below.}
\begingroup
\setlength{\tabcolsep}{3pt} % Default value: 6pt
\renewcommand{\arraystretch}{1.2} % Default value: 1
%\begin{tabular}{@{} l c c c c c c c c c c @{}} 
\begin{tabular}{l c c c c c c c c c c} 
     \toprule
         & \multicolumn{3}{c}{\textbf{BEs from this work}} & \phantom{abc} & \multicolumn{6}{c}{\textbf{BEs from literature}} \\
     \cmidrule{2-4} \cmidrule{6-11}
       & \textbf{Crystalline ice} & \multicolumn{2}{c}{\textbf{ASW}} && \multicolumn{2}{c}{\textbf{Computed}}  & \multicolumn{2}{c}{\textbf{Databases}}  & \multicolumn{2}{c}{\textbf{Experiments}} \\
      
     Species & BE\textbf{(0)} \emph{disp} & Min & Max && Das\textsuperscript{(a)} & Wakelam\textsuperscript{(b)} & UMIST\textsuperscript{(c)} & KIDA\textsuperscript{(d)} & Penteado\textsuperscript{(e)} & Others \\
      \midrule
H\textsubscript{2}      & {790}  & {226} & {431} && 545 & 800 & 430 & 440 & 480 $\pm$ 10 & 322-505\textsuperscript{(f)} \\
O\textsubscript{2}      &{677}  & {287} & {729} &&1352 &	1000&1000 &	1200& 914-1161     & 920-1520\textsuperscript{(f)} \\
{N\textsubscript{2}} & {1242}& {760} & {1458} && 1161 &	1100& 790 &	1100& 1200  &	790-1320\textsuperscript{(f1)}\\
 & & & & && & & & & 900-1800\textsuperscript{(f1)} \\
CH\textsubscript{4} & {1633} & {914} & {1674} && 2321 & 800& 1090 &960& 1370& {960-1947\textsuperscript{(f,g,h)}}  \\
CO &{1663} & {1109} & {1869} && 1292 & 1300 & 1150 & 1300 & 863-1420 &  870-1600\textsuperscript{(f)}  \\
& & & & && & & & & 980-1940\textsuperscript{(f1)} \\
CO\textsubscript{2}     & {2568} & {1489}&	{2948}&&	2352&	3100&	2990&	2600&	2236-2346  \\
OCS  & {2722} & {1520}&	{2907}&&	1808&	2100&	2888&	2400&  {2325}\textsuperscript{**} & {2430}\textsuperscript{(i)}            \\
HCl   & {5557} & (l) & (l)  && 4104 &  {4800}     &       & 5172   &     & {5172\textsuperscript{(m)}}     \\
HCN  & {6392} & {2496} &	{6337}&&	2352&	3500&	2050&	3700 &     &          \\
H\textsubscript{2}O     & {7200} & {3605}&	{6111}&&	4166&	4600&	4800&	5600&	4815-5930&   \\
H\textsubscript{2}S     & {4468} & {2291}&	{3338}&&	3232&	{2500-2900}&	2743&	2700&   {2296}\textsuperscript{**}   &         \\
NH\textsubscript{3}     & {6235} & {4314}&	{7549}&&	5163&	5600&	5534&	5500&    {2715}\textsuperscript{**}  &        \\
CH\textsubscript{3}CN   & {5906} & {4745}&	{7652}&&	3786&	4300&	4680&	4680&   {3790}\textsuperscript{**}   & \\
CH\textsubscript{3}OH   & {7385} & {3770}&	{8618}&&	4511&	{4500-5100}&	4930&	5000&  {3820}\textsuperscript{**}    & {3700-5410\textsuperscript{(n,o,p)}}\\
H\textsubscript{2}CO    & {5187} & {3071}&	{6194}&&	3242&	5100&	2050&	4500&	3260$\pm$60 & \\
HCONH\textsubscript{2}  & {8104} & {5793}& {10960}&&	    &   6300&	5556&	    &	         &7460-9380\textsuperscript{(q)}\\
HCOOH                   & {7991} & {5382}&	{10559}&&	3483&       &	5000&	5570&  {4532}\textsuperscript{**}          & \\
OH•                     & {5588} & {1551}&	{5321}&&	3183&	{3300-5300}&	2850&	4600&	1656-4760&  \\
HCO•                    & {2968} & {1315}&	{3081}&&	1857&	{2300-2700} &	1600&	2400&            & \\
CH\textsubscript{3}•    & {2188} & {1109}&	{1654}&&	1322&	2500&	1175&	1600&            &  \\
NH\textsubscript{2}•    & {5156} & {2876}&	{4459}&&	3240&	{2800-4500}&	3956&	3200&           &       \\
        \bottomrule
\end{tabular}
\endgroup
   \label{tab:2}
      \begin{tablenotes}
      \small
      \item *Notes: \textsuperscript{(a)}\citep{das2018approach}; \textsuperscript{(b)}\cite{wakelam2017binding}; \textsuperscript{(c)}\cite{mcelroy2013umist}; \textsuperscript{(d)}\cite{wakelam20152014}; \textsuperscript{(e)}\cite{penteado2017sensitivity}; \textsuperscript{(f)}\cite{he2016binding}, note that \textsuperscript{(f1)} refers to porous ice;   
       \textsuperscript{(g)}\cite{raut2007characterization};
       \textsuperscript{(h)}\cite{smith2016desorption};
       \textsuperscript{(i)}\cite{ward2012thermal};
       \textsuperscript{(l)}HCl molecule dissociate;
       \textsuperscript{(m)}\cite{olanrewaju2011probing}
       \textsuperscript{(n)}\cite{minissale2016dust}
       \textsuperscript{(o)}\cite{martin2014thermal}
       \textsuperscript{(p)}\cite{bahr2008interaction}
       \textsuperscript{(q)}\cite{chaabouni2018thermal} note that the \emph{BE} refers to the silicate substrate because it is larger than that of water ice;
         \textsuperscript{**} Results estimated from the work of \cite{collings2004laboratory}, reported in Table 2 of \cite{penteado2017sensitivity}      
    \end{tablenotes}
  \end{threeparttable}
 
\end{table*}

%%%%%%%%%%%%%%%%%%%%%%%%%%%%%%%%%%%%%%%%%%%%%%%%%%%%%%%%%%%%%%%%%%%%%%%%%%%
%%%%%%%%%%%%%%%%%%%%%%%%%%%%%%%%%%%%%%%%%%%%%%%%%%%%%%%%%%%%%%%%%%%%%%%%%%%
%%%%%%%%%%%%%%%%%%%%%%%%%%%%%%%%%%%%%%%%%%%%%%%%%%%%%%%%%%%%%%%%%%%%%%%%%%%
\section{Discussion}
A first rather expected result of our computations is that the \emph{BE} of a species on the ASW is not a single value: depending on the species and the site where it lands, the \emph{BE} can largely differ, even by more than a factor two (Table \ref{tab:2}). This has already discussed in the literature, for instance for H adsorption on both crystalline and amorphous ice models  \citep{asgeirsson2017long}.
This has important consequences both when comparing the newly computed \emph{BEs(0)} with those in the literature \S ~\ref{subsec:literature} and for the astrophysical implications \S ~\ref{subsec:astroimpl}. 
We will discuss these two aspects separately in the next two sections.

\begin{figure*}
    \centering
   \resizebox{\hsize}{!}{\includegraphics{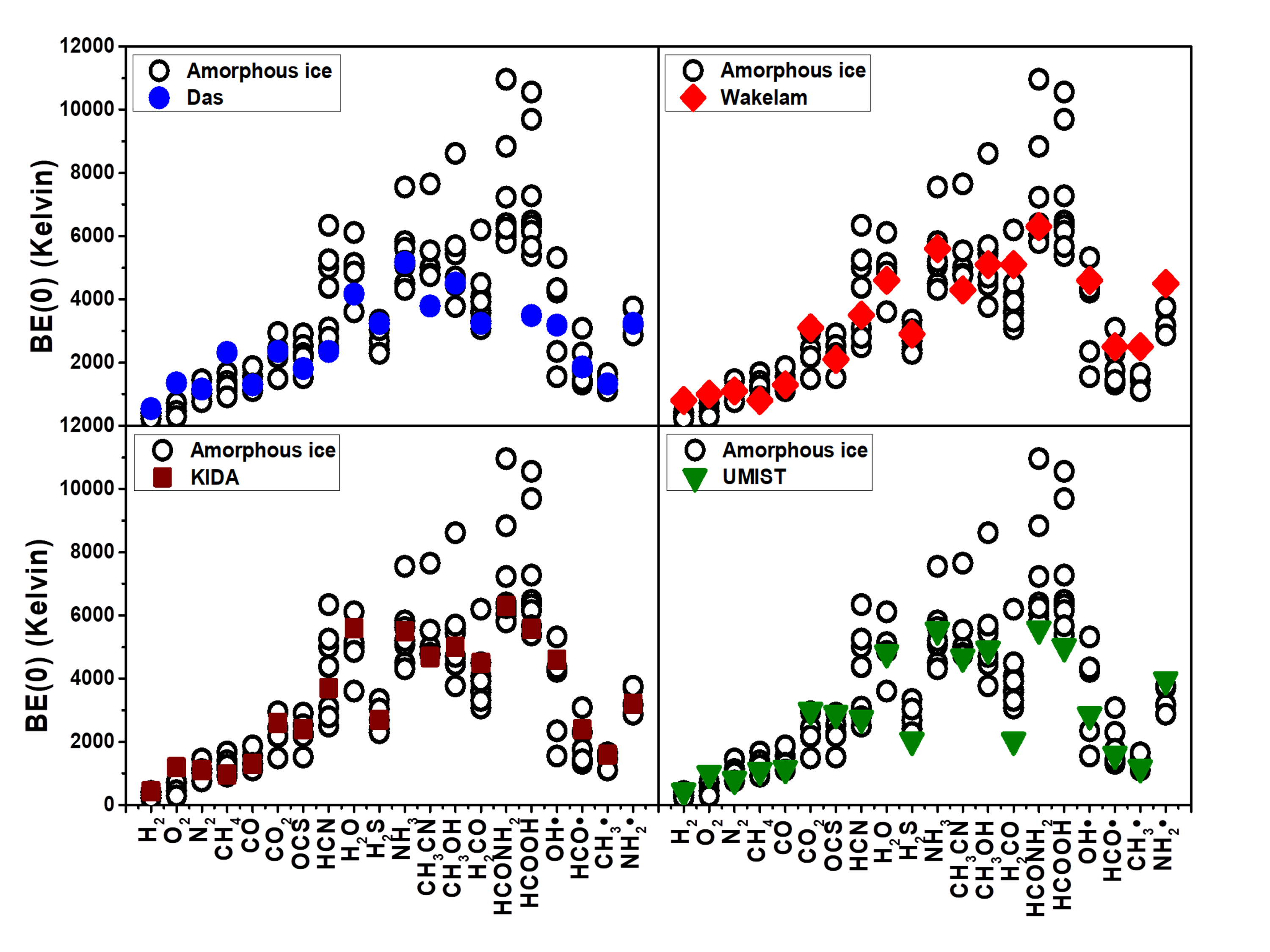}}
   \caption{Comparison of the computed zero point energy corrected \emph{BE(0)s} for the amorphous ice model with respect to those by Das et al., Wakelam et al. and reported in the KIDA and UMIST databases \citep{mcelroy2013umist, wakelam20152014, wakelam2017binding, das2018approach}.}
    \label{fig:bigcomparison}
    \end{figure*}

%%%%%
\subsection{Comparison \emph{BE} values in literature}
\label{subsec:literature}
Being such a critical parameter, \emph{BEs} have been studied from an experimental and theoretical point of view. In this section, we will compare our newly computed values with those in the literature, separating the discussion for the experimental and theoretical values, respectively. We will then also comment on the values available in the databases that are used in many astrochemical models.

%%%
\subsubsection{Comparison with experimental values}
\label{subsubsec:experimental}
%%%%%%%%%%%
%%%%% PU
%%%%%%%%%%
In the present computer simulation we have computed the binding energy  (\emph{BE}) released when a species is adsorbed on the surfaces of the ice models (either crystalline or amorphous) at very low adsorbate coverage $\theta$ ($\theta \longrightarrow 0$). The correct comparison with experiments would therefore be with microcalorimetric measurements at the zero-limit adsorbate coverage. In astrochemical laboratories, Temperature Programmed Desorption (TPD) is, instead, the method of choice and is related to the desorption activation energy (\emph{DAE}). \emph{DAE} derives indirectly from the TPD peaks through the Readhead’s method \citep{redhead}, or more sophisticated techniques. TPD usually start from an ice surface hosting a whole monolayer of the adsorbate and therefore depends also on $\theta$ \citep{he2016binding}, rendering the comparison with the theoretical \emph{BE} not straightforward \citep{king}. Ice restructuring processes may also affect the final \emph{DAE}. Sometimes, TPD experiments only provide desorption temperature peaks $T_{des}$, without working out the \emph{DAE}.
This is the case of the fundamental work by Collings et al \citep{collings2004laboratory}. Therefore, \emph{BE} reported in the review by Penteado et al \citep{penteado2017sensitivity} relative to the Collings et al data (see Table \ref{tab:2}) were computed through the approximate formula: $BE(X)=[T_{des}(X)/T_{des}(H_2O)] BE(H_2O)$, in which $T_{des}(X)$ is the desorption temperature of the $X$ species contrasted with that of water $T_{des}(H_2O)$ to arrive to the corresponding \emph{BE(X)} by assuming that of water to be 4800 K.
For the above reasons, a one-by-one comparison between experiment and modeling is outside the scope of the present paper.

Following the above warnings we can now analyze Table \ref{tab:2} reporting the recent compilation by \citet{penteado2017sensitivity} (\emph{vide supra}) \emph{plus} the values appeared in the literature after that compilation. 
We start with the cases of two measurements carried out by \citet{he2016binding} on porous amorphous ice surfaces, for N$_2$ and CO. 
Table \ref{tab:2} shows two values, reported by He et al. for the two extreme cases of when the ice is completely covered by the species (the smaller value) and when, on the contrary, it is less than a monolayer (the largest value), which is the one to compare with our computed values. 
Our \emph{BE} on amorphous ice models are in reasonable agreement with that measured by He et al. for CO (1109-1869 K versus 1940 K), and on the lower side for N$_2$ (760-1455 K versus 1800 K). It is worth-noting that the comparison is much better when referring to the non-porous amorphous ice measurements by the same authors: for CO and N$_2$ the measured \emph{BE} values are 1600 K and 1320 K, indeed well bracketed by our \emph{BE(0)}.
%%%%%%
%%% PU
%%%%%%

Data from  Penteado et al \citep{penteado2017sensitivity} extracted  from the TPD of Collings et al \citep{collings2004laboratory} for NH$_3$ adsorbed on ice layer gave a \emph{BE} of 2715 K. This value is, however, identical to that from the TPD of NH$_3$ adsorbed on gold surface (no water ice), proving that that \emph{BE} is relative to the NH$_3$/NH$_3$ lateral interaction within the adsorbed NH$_3$ multilayer and not due to the interaction with the ice surface. For reasons explained in Collings et al \citep{collings2004laboratory} the \emph{BE} of NH$_3$ on ice is assumed of the same order of that of water, i.e. around 4800 K, in better agreement with the UMIST value of 5500 K \citep{mcelroy2013umist}. Indeed, our data (see Table \ref{tab:2}, ZPE corrected) of 4300-7500 K brackets the experimental ones. The computed highest values emphasize the H-bond acceptor capability of NH$_3$ occurring on few specific sites characterized by very high electrostatic potential, only important for very low NH$_3$ coverage, not easily accessible in the TPD experiments.  
For the H$_2$O case, the computed \emph{BE(0)s} (Table \ref{tab:2}) for the amorphous ice are in the 3605-6111 K range, reasonably bracketing the experimental one of 4815-5930 K.

In general, the comparison of our \emph{BE} values computed on the ASW with those measured by the various experiments reported in Table \ref{tab:2} shows an excellent agreement, when considering the ranges in our values and the ranges in the values of the experiments. 
Only one species seems to have a relatively different computed and measured BEs: O$_2$. 
For O$_2$, experiments tend to provide larger values with respect to what we computed (our largest value is 729 K while the lowest measured value is 914 K). 
For many other species, except H$_2$, our computed lowest \emph{BEs(0)} are within the range of the measured ones, but we predict sites where \emph{BEs(0)} are larger, which may have important astrophysical implications (\S ~\ref{subsec:astroimpl}). 
Finally, for H$_2$ we predict sites where the \emph{BE(0)} is (slightly) lower than the measured ones.

%%%
\subsubsection{Comparison with computed values}
\label{subsubsec:computed}
In the literature, there are two works that reported computations of \emph{BEs} for a large set of molecules, those by \cite{wakelam2017binding} and by \cite{das2018approach}. 
The former carried out computations considering only one water molecule whereas the latter considered a cluster of up to six water molecules. 

The first aspect to notice is, therefore, that none of these two studies can, by definition, reproduce the strong adsorption sites that we have in our ASW model. 
Indeed, only the adoption of more realistic and periodically extended ice models allows to fully considering the hydrogen bond cooperativity, which will enhance the strength of the interaction with adsorbates at the terminal dangling hydrogen atoms exposed at the surface. 
This important effect is entirely missed by the two above mentioned works. 
It is not surprising, then, that our crystalline and ASW \emph{BEs} differ, sometimes substantially, from the Wakelam et al. and Das et al. values (as, by the way, they differ between themselves as well). 
This is clearly shown in Fig. \ref{fig:bigcomparison} where we report the comparison of our computed values with those by Wakelam et al. and Das et al., respectively. 
In general, both work values tend to lay in the low end of our computed \emph{BEs}. 
As extreme examples, our ASW \emph{BEs} are larger for CH$_3$CN and HCOOH. 
The inverse effect is observed for the smallest studied species: our \emph{BEs} are smaller than those computed by Wakelam et al. and Das et al. for H$_2$ and O$_2$.

%%%
\subsubsection{Comparison with values in astrochemical databases}
Two databases list the \emph{BE} of the species used by the astrochemical models: the KIDA (Kinetic Database for Astrochemistry, \url{http://kida.astrophy.u-bordeaux.fr/}; \citet{wakelam20152014}) and UMIST (\url{http://udfa.ajmarkwick.net/index.php}; \citet{mcelroy2013umist}). 
The comparison between our newly computed values and those reported in the two databases is shown in Fig. \ref{fig:bigcomparison}. 
The general remarks that we wrote for the comparison with the literature experimental and theoretical values (\S ~\ref{subsubsec:experimental} and ~\ref{subsubsec:computed}) roughly apply here: the databases quote \emph{BE} values in the low end of ours. 
This is not surprising, as the databases are compiled based on the experimental and theoretical values in the literature. 
We just want to emphasize here, once again, that the sites with large binding energies are lacking and this may have important consequences in the astrochemical model predictions.

%%%%%
\subsection{Astrophysical implications}
\label{subsec:astroimpl}
\emph{BEs} enter in two hugely important ways in the chemical composition of interstellar objects/clouds: (i) they determine at what dust temperature the frozen species sublimate, and (ii) at what rate the species can diffuse in the ice, as the diffusion energy is a fraction of the \emph{BE} species. 
Both processes are mathematically expressed by an exponential containing the \emph{BE}. 
Therefore, even relatively small variations of the \emph{BE} can cause huge differences in the species abundances in the gas phase and on the grain surfaces, where they can react with other species.

In this context, probably the most important astrophysical implication of the present study is that in our ASW model (which is likely the best description to represent the interstellar amorphous ice so far available in the context of the \emph{BE} estimates) a species does not have a single value, but a range of values that depend on the species itself and the site where it is bound. 
The range can spread by more than a factor two: this obviously can have a huge impact on the modeling and, consequently, our understanding of the interstellar chemical evolution.

%%%
\subsubsection{Impact of multiple \emph{BEs} in astrochemical modeling}
\label{subsubsec:multiple}
To give a practical example of the impact on the gaseous abundance, we built a toy model for the interstellar ice and simulated the desorption rate of the ice as a function of the temperature. Our scope here is not to compare the toy model predictions with astronomical observations or laboratory experiments: we only mean to show how multiple \emph{BEs} (we used the electronic \emph{BEs}) would lead, in principle, to a different behavior of the ice sublimation process. Therefore, we developed a toy model that does not contain diffusion or reaction processes on the ice surface or rearrangement of the ice during the ice heating, but only a layered structure with two species, specifically water and methanol, where molecules have the range of \emph{BEs} calculated in section \ref{sec:Results}. We then show how the multiple \emph{BEs} affect the temperature at which peaks of desorption appear, considering that only species at the surface of the ice, namely exposed to the void, can sublimate (and not the entire bulk).
In this toy model, we considered ten layers for an icy grain mantle, where the bottom five are made entirely of water and each of the top five layers contains 80\% of water and 20\% of methanol. 
The methanol molecules with different \emph{BEs} are distributed randomly on each layer, with a same proportion of \emph{BE} sites. 
Looking at the \emph{BE} values computed with the DFT//HF-3c method on the ASW model, methanol has eight \emph{BE} values (4414, 5208, 5509, 6362, 6519, 6531,6663 and 10091 K) and, at each layer, there will be 12.5\% of methanol molecules with each of the eight \emph{BEs}. 
The same applies for water molecules, for which we have computed five possible \emph{BEs} (4222, 5689, 5845, 6014 and 7156 K), with 20\% molecule with each \emph{BE}. 
In this model, only molecules of the layers in contact with the void can evaporate: for example, methanol molecules can be trapped if they have water molecules with larger \emph{BEs} on top of them.

\begin{figure}
    \centering
   \resizebox{\hsize}{!}{\includegraphics{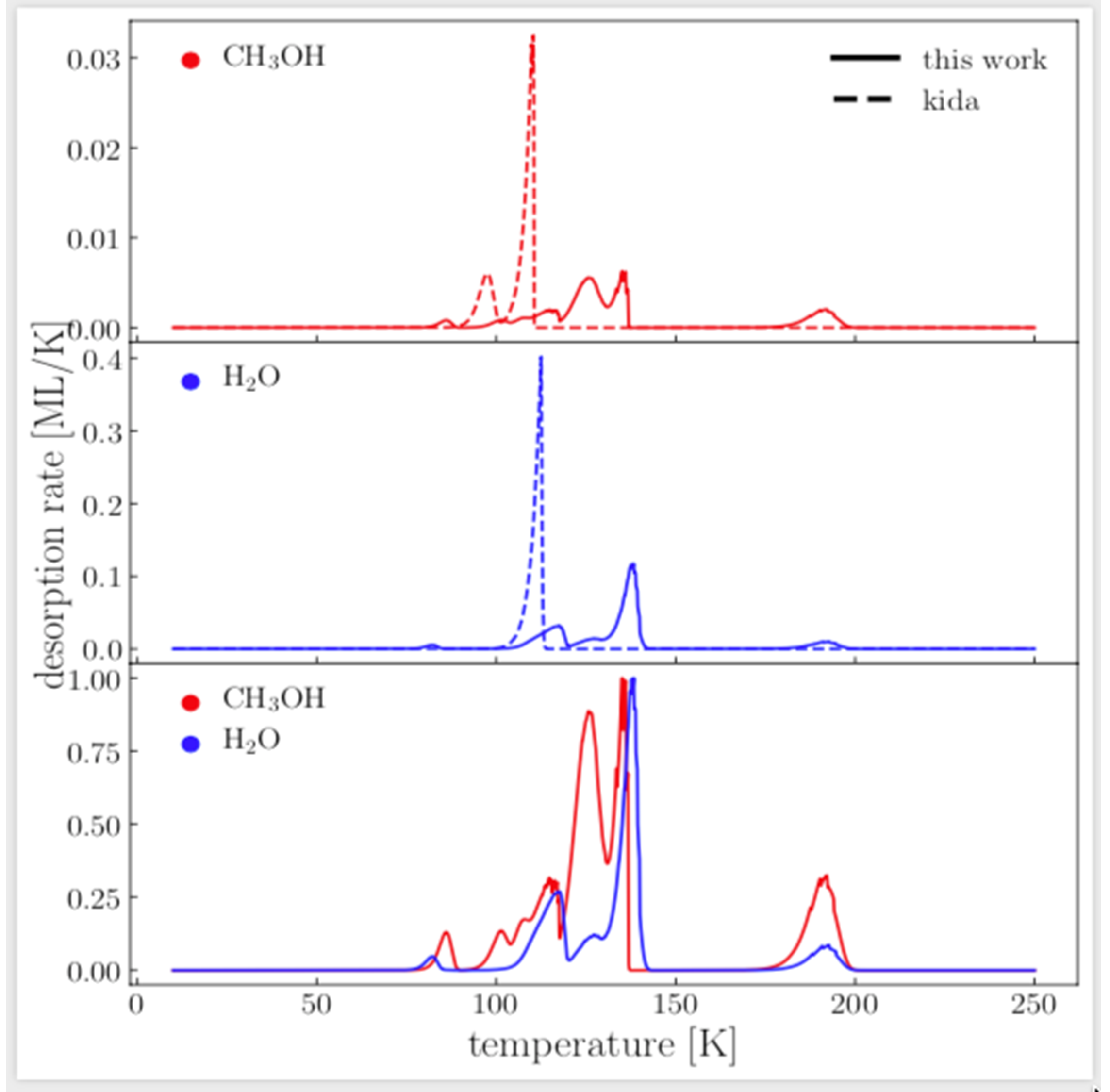}}
   \caption{Desorption rate of methanol (red curves) and water (blue curves) as a function of the temperature. In these computations, the ice is assumed to be composed by ten layers: the bottom five layers contain water while the top fives contain 80\% of water and 20\% of methanol. The ice is assumed to be at 10 K at the beginning of the simulation and it reaches 400 K in $10^5$ yr. The curves refer to the case when a single \emph{BE} (from the KIDA database) is considered for water and methanol, respectively (dashed line), and when the multiple \emph{BEs} of this work are considered (solid line). The bottom panel shows the methanol and water desorption rates normalised to one.}
    \label{fig:arezu}
    \end{figure}

We start with an ice temperature of 10 K and at the end it reaches 400 K in $10^5$ yr, to simulate the heating of a collapsing Solar-like protostar. 
The plot of the desorption rates is shown in Fig. \ref{fig:arezu}, where we also show them assuming the \emph{BE} values from the KIDA database for methanol and water, respectively.
First, when only the KIDA values for \emph{BE} are assumed, water molecules desorb at about 110 K; methanol has two peaks of desorption rates, the first at about 95 K, corresponding to the desorption of the methanol molecules not trapped by the water molecules, and the second peak at about 110 K, when all water molecules desorb so that no methanol molecules are trapped.
Note that the desorption of the water molecules of the bottom layers arrives at a slightly larger temperature. 

Not surprising, the introduction of multiple \emph{BEs} produces multiple peaks of desorption, both for water and methanol. 
Figure \ref{fig:arezu} shows that the water desorption rate has a small peak at $\sim$75 K, a larger one at $\sim$120 K, then another at $\sim$140 K and, finally, a last peak at $\sim$190 K. 
Methanol starts to desorb at 80 K, the bulk is desorbed between 120 and 140 K, and a last peak is seen around 190 K.
We emphasize again that this is a toy model meant to show the potential impact of the new \emph{BEs} on the astrochemical modeling. 
The details will depend on the real structure of the water ice and how molecules are distributed on the icy mantles. 
They will determine how many sites have a certain \emph{BE} value and how molecules are trapped in the ice.
As a very general remark, we can conclude that species can be in the gas phase at lower and higher dust temperatures than if one only considers a single \emph{BE}.

%%%
\subsubsection{Looking forward: implementation of multiple \emph{BEs} in astrochemical models}
Our toy model introduced in the previous subsection shows the importance of considering multiple \emph{BEs} for each species in the astrochemical models to have more realistic predictions. 
In this work, we provide the possible \emph{BEs} for 21 species (Table \ref{tab:2}). 
Very likely, they cover most of the possibilities, as they span a large range of H-bonds within the water molecules of the ASW. However, from a computational standpoint, such an adsorption variability has to be fully explored, in which plotting the different calculated \emph{BEs} in histograms is useful to provide insights on the shape of the \emph{BE} distribution \citep{Song_2017}. 
Moreover, in order to build a reliable astrochemical model one would also need to know the relative probability for each \emph{BE} and our present study is unable to provide sensible numbers. 
For that, a statistical study on an ASW model that is much larger than the one used here is necessary. This is a step that we indeed plan to take in future studies. 
Meanwhile, we adopted a distribution in which we assign an equal fraction of molecules to each \emph{BE}. 
If one looks carefully at the distribution of the \emph{BEs} for each molecule, they are not uniformly distributed but peak around some values: for example, methanol has a peak around 6000 K and an extreme value of 10091 K only, so that, very likely, this site will be less populated than the sites around 6000 K, as shown by Figure \ref{fig:spider graphs}. 
Yet, considering even a smaller fraction of these extreme values may have important consequences, for example in the so-called snow lines of protoplanetary disks, or even on the observed abundances towards hot cores and hot corinos, or, finally, towards prestellar cores.

%%%
\subsubsection{Comments on N$_2$, CO and HCl}
Finally, we would like to comment on three species of the studied list, N$_2$, CO and the hydrogen chloride (HCl).

\textit{N$_2$ and CO}: 
Our computations show that the \emph{BE} of CO is definitively larger than that of N$_2$, against the values that are in the astrochemical databases (see Table \ref{tab:2}): on average, our computed \emph{BEs} differ by about 400 K whereas in the data bases the difference is 200 and 360 K in KIDA and UMIST, respectively. 
This difference very likely can explain why observations detect N$_2$H$^+$, which is formed in the gas phase from N$_2$, where CO is already frozen on the grain mantles \citep[e.g.][]{bergin_2002, tafalla2004A&A...416..191T,radaelli2019A&A...629A..15R}, a debate which has been going on for almost two decades \citep[e.g.][]{oberg2009A&A...496..281O, pagani2012A&A...548L...4P}.
%%%%%%
%%%% PU
%%%%%%

In order to quantify the effect a specifically focused modeling will be necessary, which is out of the scope of the present work. Here we want just to alert that the new \emph{BEs} might explain some long-standing mysteries.
Another comment regards the difference in the \emph{BEs} on crystalline surfaces and ASW.
Again the CO binding energy is about 500 K larger than that of N$_2$, and both are larger by about 300 K than those on the ASW, a difference that also has an impact on the snow lines of these two species in protoplanetary disks, where crystalline water ices have been detected \citep{Terada_2012}.

\textit{HCl}:
Astrochemical models predict that HCl is the reservoir of Cl in molecular gas \citep[e.g.][]{schilke1995hydrogen,neufeld2009chemistry, acharyya2017gas}. 
However, all the observations so far carried out have found that only a tiny fraction of Cl is in the gaseous HCl, even in sources where all the grain mantles are supposed to be completely sublimated \citep{peng2010comprehensive, codella2011first, kama2015depletion}. 
One possible explanation is that HCl, once formed in the gas phase, is adsorbed on the grain icy mantles and dissociates, as shown by our calculations on the ASW model and also by previous calculations on the crystalline P-ice model \citep{casassa2000ab} and for more sophisticated proton-disordered crystalline ice models \citep{svanberg}. 
It is a matter to be studied whether the sublimation of the water, when the dust reaches about 100-120 K, would also provide a reactive channel transforming the Cl anion in the neutral atom, this latter obviously unobservable. This would help solving the mystery of HCl not being observed in gas-phase. Furthermore, if that would be the case, the population of the chemically reactive atomic Cl will be increased, with important role in the gas-phase chemistry \citep[see e.g.][]{balucani2015, skouteris2018genealogical}.

%%%%%%%%%%%%%%%%%%%%%%%%%%%%%%%%%%%%%%%%%%%%%%%%%%%%%%%%%%%%%%%%%%%%%%%%%%
%%%%%%%%%%%%%%%%%%%%%%%%%%%%%%%%%%%%%%%%%%%%%%%%%%%%%%%%%%%%%%%%%%%%%%%%%%
%%%%%%%%%%%%%%%%%%%%%%%%%%%%%%%%%%%%%%%%%%%%%%%%%%%%%%%%%%%%%%%%%%%%%%%%%%
\section{Conclusions}
In this work, we present both a new computational approach and realistic models for crystalline and amorphous water ice to be used to address an important topic in Astrochemistry: the binding energies (\emph{BEs}) of molecules on interstellar ice surfaces. 
We simulated such surfaces by means of two (antipodal) models, in both cases adopting a periodic approach: a crystalline and an amorphous 2D slab models. 
We relied on density functional theory (DFT) calculations, using the B3LYP-D3 and M06-2X widely used functionals. This approach was further validated by an ONIOM-like correction at CCSD(T) level. Results from this combined procedure confirm the validity of the \emph{BEs}  computed with the adopted DFT functionals.
The reliability of a cost-effective HF-3c method adopted to optimize the structures at the amorphous ice surface sites was proved by comparing the binding energies computed at the crystalline ice surface at DFT//DFT and  DFT//HF-3c level, which were found in very good agreement.

On both ice surface models, we simulated the structure and adsorption energetic features of 21 molecules, including 4 radicals, representative of the most abundant species of the dense ISM.
A main conclusion is that the crystalline surfaces only show very limited variability in the adsorption sites, whereas the amorphous surfaces provide a wide variety of adsorption binding sites, resulting in a distribution of the computed \emph{BE}. Furthermore, \emph{BE} values at crystalline ice surface are in general higher than those computed at the amorphous ice surfaces. 
This is largely due to the smaller geometry relaxation cost upon adsorption compared to the amorphous cases, imposed by the tighter network of interactions of the denser crystalline ice over the amorphous ice.

Finally, the \emph{BEs} obtained by the present computations were compared with literature data, both from experimental and computational works, as well as those on the public astrochemical databases KIDA and UMIST.
In general, our \emph{BEs} agree relatively well with those measured in laboratory, with the exception of O$_2$ and, at less extent, H$_2$.
On the contrary, previous computations of \emph{BEs}, which considered a very small ($\leq 6$) water molecules, provide generally lower values with respect to our new computations and, with no surprise, miss the fact that \emph{BEs} have a spread of values which depend on the position of the molecule on the ice.
Since the two astrochemical databases mentioned above are based on the literature data, our \emph{BEs} differ, sometimes substantially, from those quoted and do not report multiple \emph{BEs} values.

We discussed some astrophysical implications, showing that the multiple computed \emph{BEs} give rise to a complex process of interstellar iced mantles desorption, with multiple peaks as a function of the temperature that depends (also) on the ice structure. 
Our new computations do not allow to estimate how the \emph{BEs} are distributed for each molecule as only a statistical study is necessary for that.
The new (multiple) \emph{BEs} of N$_2$ and CO might explain why N$_2$H$^+$ deplete later than CO in prestellar cores, while the relatively low abundance of HCl, observed in protostellar sources, could be due to the fact that it dissociates into the water ice, as shown by our calculations.

Finally, the present study shows the importance of theoretical calculations of \emph{BEs} on as realistic as possible ice surfaces. 
This first study of 21 molecules needs to be extended to the hundreds molecules that are included in the astrochemical models to have a better understanding of the astrochemical evolution of the ISM.

%%%%%%%%%%%%%%%%%%%%%%%%%%%%%%%%%%%%%%%%%%%%%%%%%%%%%%%
\acknowledgements
A part of the computational results were from the SF Master thesis "\emph{Ab initio} quantum mechanical study of the interaction of astrochemical relevant molecules with interstellar ice models", Dipartimento di Chimica, University of Torino, Torino, 2018. SF, LZ and PU acknowledge financial support from the Italian MIUR (Ministero dell’Istruzione, dell’Università e della Ricerca) and from Scuola Normale Superiore (project PRIN 2015, STARS in the CAOS - Simulation Tools for Astrochemical Reactivity and Spectroscopy in the Cyberinfrastructure for Astrochemical Organic Species, cod. 2015F59J3R). 
The Italian CINECA consortium is also acknowledged for the provision of supercomputing time for part of this project. 
AR is indebted to the “Ramón y Cajal” program. 
MINECO (project CTQ2017-89132-P) and DIUE (project 2017SGR1323) are acknowledged. 
This project has received funding from the European Union’s Horizon 2020 research and innovation programme under the Marie Sklodowska-Curie grant agreement No 811312 for the project "Astro-Chemical Origins” (ACO) and from the European Research Council (ERC) under the European Union's Horizon 2020 research and innovation programme, for the Project “The Dawn of Organic Chemistry” (DOC), grant agreement No 741002.
Finally, we wish to acknowledge the extremely useful discussions with Prof. Gretobape.

\bibliography{biblio}{}
\bibliographystyle{aasjournal}

%%%%%%%%%%%%%%%%%%%%%%%%%%%%%%%%%%%%%%%%%%%%%%%%%%%%%%%%%%%%%%
%%%%%%%%%%%%%%%%%%%%%%%%%%%%%%%%%%%%%%%%%%%%%%%%%%%%%%%%%%%%%%
%%%%%%%%%%%%%%%%%%%%%%%%%%%%%%%%%%%%%%%%%%%%%%%%%%%%%%%%%%%%%%
\appendix

%%%%%%%%%%%%%%%%%%%%%%%%%%%%%%%%%%%%%%%%%%%%%%%%%%%%%%%%%%%%%%
\section{Computational details}\label{app:comp-details}
In CRYSTAL17, the multielectron wavefunction is built as a Slater determinant of crystalline/molecular orbitals which are linear combinations of localized functions on the different atoms of the structure which are called Atomic Orbitals (AOs). In a similar manner, the AOs are constructed by linear combinations of localized Gaussian functions which form a basis set. The basis set employed for this work is an Ahlrichts-TVZ \citep{schafer1992fully}, added with polarization functions.

%%%%%%%%%%%%%
%%%% PU
%%%%%%%%%%%%%%%%%
\subsection{Binding energies, Counterpoise and zero point energy corrections}\label{app:bssezpe}
 In a periodic treatment of surface adsorption phenomena one of the most relevant energy value, useful to describe the interacting system, is the binding energy \emph{BEs} which is related to the interaction energy $\Delta E$, so that:

\begin{equation}
    BE = -\Delta E
\label{eq:A1}
\end{equation}

The binding energy per unit cell per adsorbate molecule \emph{BE} is a positive quantity (for a bounded adsorbate) defined as:

\begin{equation}
    BE = [E_m(M//M) + E(S//S)] - E(SM//SM)
\label{eq:A2}
\end{equation}

where $E(SM//SM)$ is the energy of a fully relaxed unitary cell containing the surface slab $S$ in interaction with the adsorbate molecules $M$, $E(S//S)$ is the energy of a fully relaxed unitary cell containing the slab alone and $E\textsubscript{m}(M//M)$ is the molecular energy of the free fully optimized adsorbate molecule (the symbol following the double slash identifies the geometry at which the energy h

\begin{align}
BE &= BE^* - \delta E_S- \delta E_M	\label{eq:A3}\\
\delta E_S &= E(S//SM) - E(S//S)	\label{eq:A4}\\
\delta E_M &= E(M//SM) - E\textsubscript{m} (M//M)	\label{eq:A5}\\
BE^* &= [E(S//SM) + E(M//SM)] - E(SM//SM) \label{eq:A6}
\end{align}

in which $\delta E_S$ is the deformation energy of the surface ($\delta E_S>0$) whereas $\delta E_M$ (=$\Delta E_M + \Delta E_L)$ counts both the deformation energy of the molecule ($\Delta E_M$) and the lateral intermolecular interactions ($\Delta E_L$) between the infinite molecule images in the same spatial configuration occurring in the $SM$ periodic system. The purely molecule’s deformation energy can be computed as:

\begin{equation}
\Delta E_M = E_m(M//SM) - E_m(M//M)	
\label{eq:A7}
\end{equation}

in which $E_m(M//SM)$ is the molecular energy of the molecule frozen at the geometry occurring on the surface and $E_m(M//M)$ is the molecular energy of a fully optimized free molecule, so that $\Delta E_M>0$. The lateral intermolecular interactions, $\Delta E_L$ are defined as:

\begin{equation}
\Delta E_L = E(M//SM)-E_m(M//SM)
\label{eq:A8}
\end{equation}

and can be either positive (repulsion) or negative (attraction). With those positions, the $BE^*$ interaction energy is then deformation and lateral interactions free, being the result of energy differences between periodic calculations carried out at the geometry of the $SM$ system. For instance, $E(M//SM)$ is the energy of the unit cell of a crystal containing only a molecule in the same geometry assumed in the $SM$ system. 
The above \emph{BE} definition can be easily recast to include the basis set superposition error (\emph{BSSE}) correction, using the same counterpoise method adopted for intermolecular complexes \citep{boys1970calculation, davidson1986basis}.  
The definition of the \emph{BSSE} corrected interaction energy $BE^C$ is then:

\begin{align}
BE^C &= BE^{*C}-\delta E_S - \Delta E_M - \Delta E_L^C	\label{eq:A9}\\
BE^{*C} &= [E(S[M]//SM) + E([S]M//SM)] - E(SM//SM)	\label{eq:A10}\\
BSSE &= BE - BE^C	\label{eq:A11}
\end{align}

in which $E(S[M]//SM)$ and $E([S]M//SM)$ are the energy of the slab plus the ghost functions of the molecules and the energy of the infinite replica of molecules with the ghost functions of the underneath slab, respectively. Because the variational theorem ensures that $BE^{*C} < BE^*$ it immediately results that \emph{BSSE} $>$ 0.\\

Each of the terms of equation \ref{eq:A2} can be corrected by the zero point vibrational contribution (in the harmonic approximation), $ZPE$, therefore arriving to the definition of the zero point correct binding energy $BE(0)$ as:

\begin{align}
BE(0) &= [E_m(M//M) + E(S//S)] - E(SM//SM) + [ZPE(M) + ZPE(S) - ZPE(SM)] \label{eq:A12}\\
BE(0) &= BE - \Delta ZPE  \label{eq:A13}\\
\Delta ZPE &= ZPE(SM) - ZPE(M) - ZPE(S)
\end{align}

In this work the $\Delta ZPE$ has been evaluated for the crystalline ice cases only and the same correction was adopted also to correct the $BE$ for the the amorphous ice model for the reason described in the Computational Details section.

\begin{figure}[h!]
    \centering
   \resizebox{9cm}{!}{\includegraphics{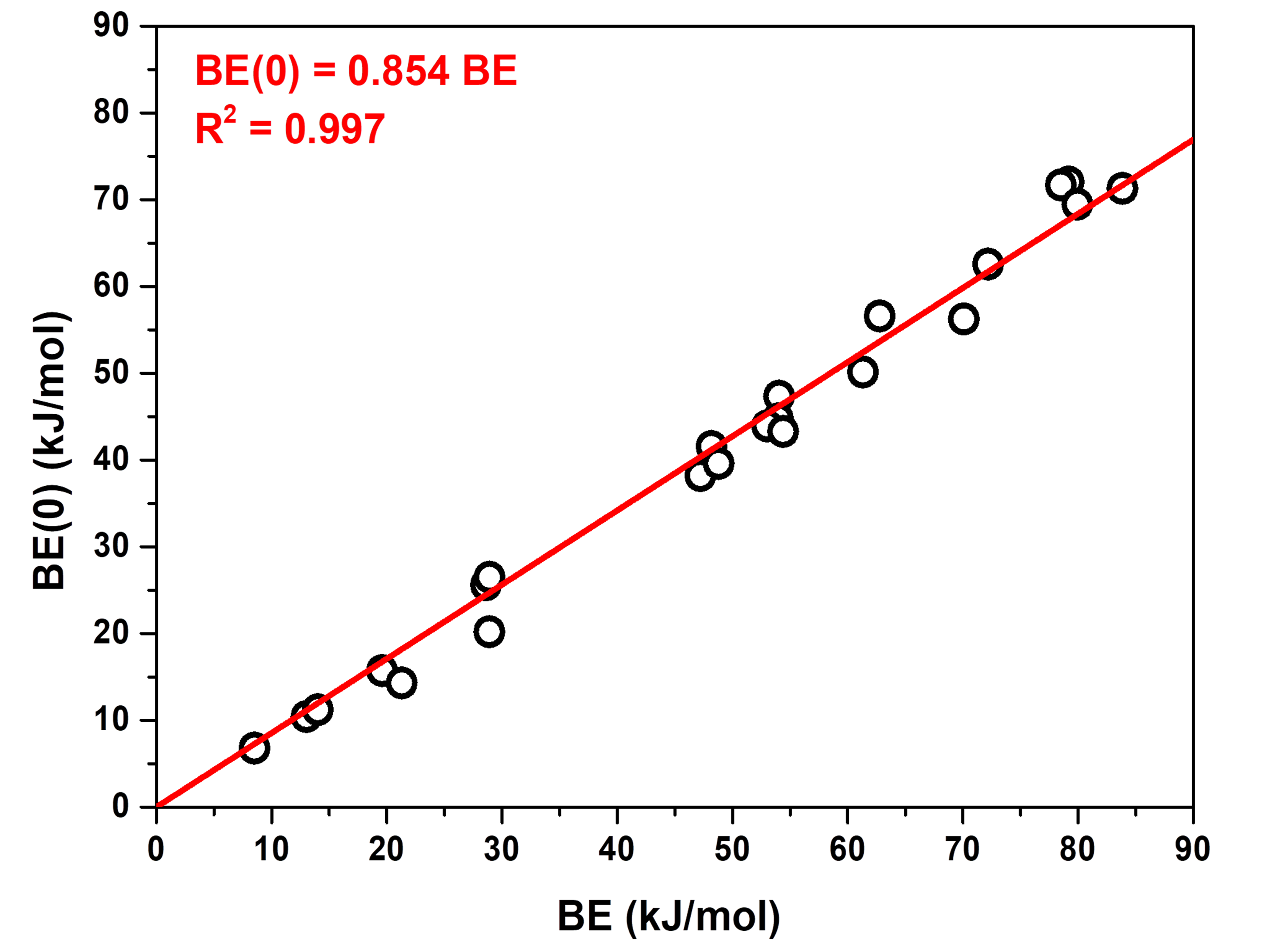}}
   \caption{Linear regression between \emph{BE} and \emph{BE(0)} (zero point corrected)
values computed for the considered adsorbates on the crystalline ice}
    \label{fig-app:ZPE_fit}
    \end{figure}

%%%%%%%%%%%%%%%
%%%% PU
%%%%%%%%%%%%%%

%%%%%%%

\subsection{Details on HCl adsorbed on crystalline and amorphous ices}\label{app:hcl}
\begin{figure}[h!]
    \centering
   \resizebox{11cm}{!}{\includegraphics{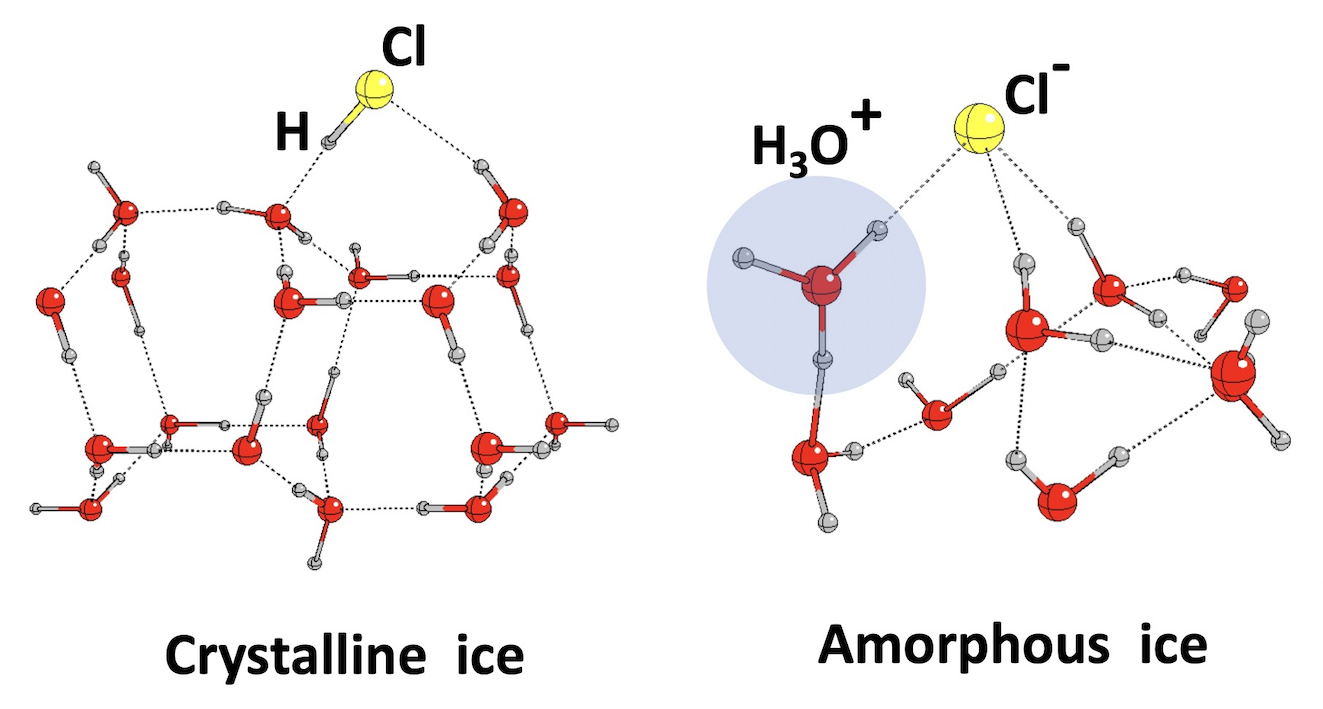}}
   \caption{Enlarged views of HCl adsorbed on a crystalline and amorphous ice models}
    \label{fig-app:hcl}
    \end{figure}

Our calculations showed that HCl remains molecularly adsorbed at the crystalline ice surface. At the amorphous one it spontaneously deprotonates making a locally stable ion-pair (Cl$^-$/H$_3$O$^+$) (see the figure below). The reason is that for the surface selected to represent the crystalline ice, the ion pair cannot be stabilized by a large enough network of H-bonds due to the rigidity of the structure. On the contrary, at amorphous surface, Cl$^-$ is engaged in three H-bonds while the hydronium ion remains embedded in the H-bonds provided by the ice surface. A good solvation of the ion-pair is the key determining the final fate of HCl at the ice surfaces (molecular or ion pair) as already pointed out many years ago by Novoa and Sosa for a small water cluster hosting HCl \citep{sosa}. While deprotonation of HCl cannot be excluded for specific crystalline ice surface sites as simulated by Svanberg \emph{et al} \citep{svanberg} when enough dangling hydrogen can stabilize the anion, deprotonation would be much more common at amorphous ice surface or locally distorted crystalline ones, due to the presence of favorable local environment (see Figure \ref{fig-app:hcl}). Dissociative adsorption was early found by Horn \emph{et al} \citep{horn} through RAIRS spectroscopy of DCl adsorption on thin D$_2$O film, and by Olanrewaju \emph{et al} \citep{orlando} through thermal and electron-stimulated desorption experiments.  
\pagebreak

\subsection{CRYSTAL17 computational parameters}
In order to optimize the values of the shrinking factor, the tolerances on integrals  of the integration grid (SHRINK, TOLINTEG and GRID parameters \citep{dovesi2018quantum} in the code) as described in \S \ref{sec:Computational details} of the article, NH\textsubscript{3} has been adopted as a test case. Geometry optimizations for the adsorption process on the crystalline slab model have been run with the previously described computational scheme, varying one-by-one the aforementioned parameters, with convergence on the pure $\Delta E$ as define by Equation \eqref{eq:A1} being the threshold. For this benchmark calculations, $BSSE$, distortion and lateral interaction contributions have not been computed. Results are resumed in what follows.

\begin{table}[h!]
    \centering
    \caption{Optimization of the SHRINK parameter. Tolerances of integrals (TOLINTEG) and integration grid (XLGRID) are fixed to default values.}   \label{tab-app:SHRINK}
    \begin{tabular}{c c c}
    \toprule
     SHRINK & \emph{k} points in BZ & $\Delta E$ ($\mathrm{K}$)  \\
     \toprule
     2 2  &  4 &  7873 \\
     3 3  &  5 &  7889 \\
     4 4  &  10 &  7889 \\
     5 5  &  13&  7890 \\
     6 6  &  20&  7890 \\
     7 7  &  25 & 7890 \\
     8 8  &  34 & 7890 \\
    \bottomrule 
   \end{tabular}
\end{table}

From Table \ref{tab-app:SHRINK} it is clear that $\Delta E$ is practically unaffected by the SHRINK value, thus SHRINK = (2 2) has been used for all calculations, allowing the saving of computational time.

\begin{table}[h!]
    \centering
    \caption{Optimization of the TOLINTEG parameter (SHRINK = 2 2 and XLGRID).}   \label{tab-app:TOLINTEG}
    \begin{tabular}{c c}
    \toprule
     TOLINTEG & $\Delta E$ ($\mathrm{K}$)  \\
     \toprule
     6 6 6 6 12  &  7889 \\
     7 7 7 7 14  &  7906 \\
     8 8 8 8 16  &  8063 \\
     9 9 9 9 18  &  7981 \\
     10 10 10 10 20 &  7890 \\
     \bottomrule 
   \end{tabular}
\end{table}

As for the SHRINK parameter, the variations of TOLINTEG values does not practically affect the final $\Delta E$ value. Consequently, its values have been set equal to 7, 7, 7, 7, and 14.

\newpage
\begin{table}[h!]
    \centering
    \caption{Optimization of the integration grid parameter (SHRINK = 2 2 and TOLINTEG = 7 7 7 7 14).}   \label{tab-app:GRID}
    \begin{tabular}{c c}
    \toprule
     GRID & $\Delta E$ ($\mathrm{K}$)  \\
     \toprule
     XLGRID  &  7951 \\
     XXLGRID  &  7951 \\
  \bottomrule 
   \end{tabular}
\end{table}

Again, the $\Delta E$ is practically unaffected by the changing the adopted grid, therefore XLGRID was selected for all the remaining calculations. Resuming, every other calculation has been carried out with SHRINK = (2 2), TOLINTEG = (7 7 7 7 14) and XLGRID as computational parameters.

\subsection{Description of dispersive forces}
\begin{figure}[h!]
    \centering
   \resizebox{9cm}{!}{\includegraphics{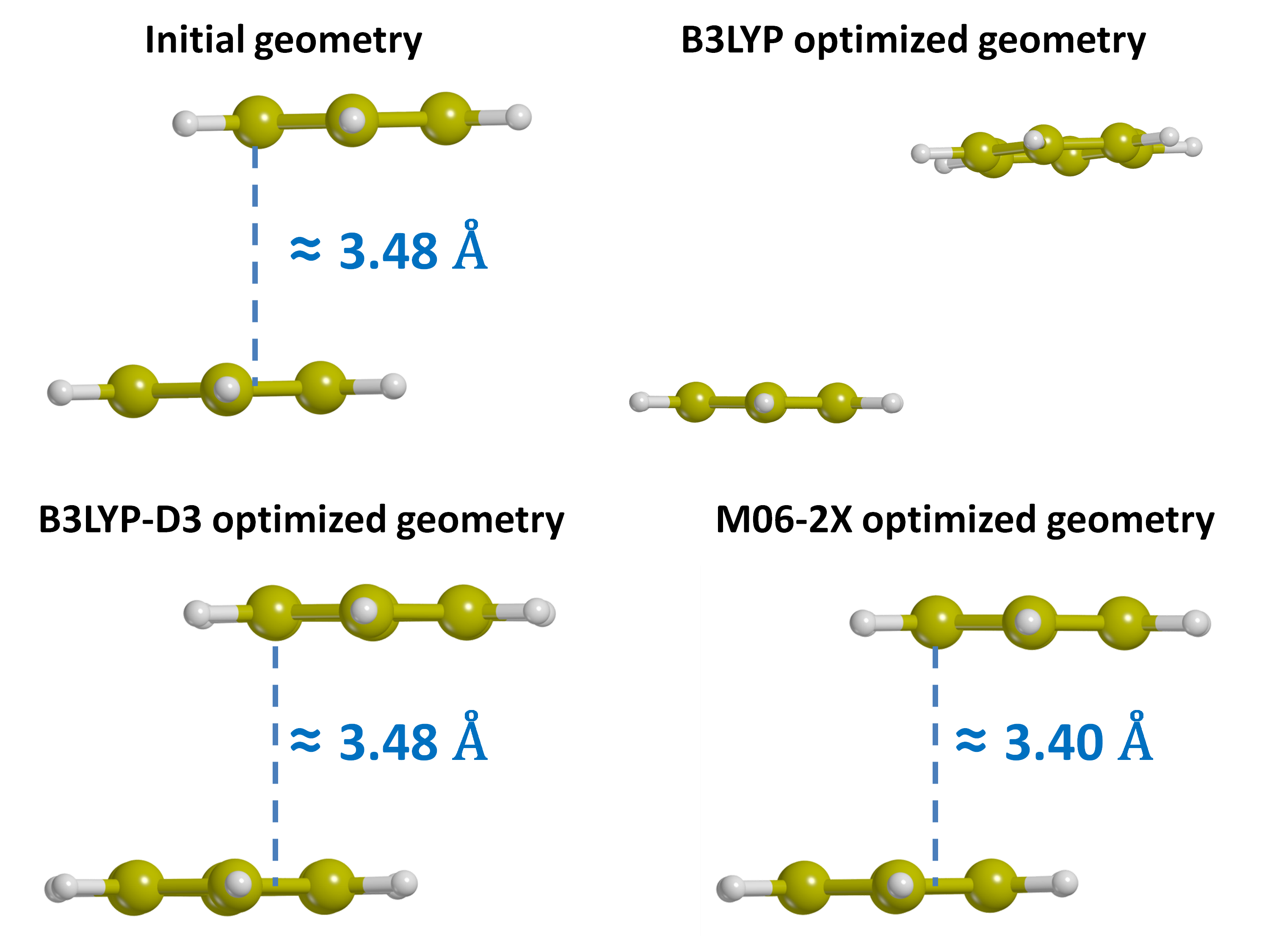}}
   \caption{Initial and optimized geometry of two parallel benzene rings obtained with B3LYP (without dispersion corrections), B3LYP-D3 and M06-2X functionals. Distances between the planes of the two rings are reported in Angstrom.}
    \label{fig-app:dispersionA1}
    \end{figure}

We optimized the geometry of two parallel benzene rings which interact with each other just because of dispersive forces given the apolar nature of the molecule. The results of Figure \ref{fig-app:dispersionA1} clearly show that the B3LYP functional needs some a posteriori correction (like the Grimme's D3) for dispersive forces in order to describe correctly the interaction \citep{grimme2010consistent}, whereas the M06-2X functional is able to describe it correctly without any a posteriori correction \citep{zhao2008m06}.

%%%%%%    
\subsection{ONIOM2 correction}
Hereafter we summarize the single components (namely \emph{BE}(CCSD(T), $\mu$-2H\textsubscript{2}O) extrapolated, \emph{BE}(DFT, $\mu$-2H\textsubscript{2}O), \emph{BE}(DFT, ice) and finally the corrected \emph{BE}(ONIOM2)) of the ONIOM2 correction done with the extrapolation procedure already described in \S ~\ref{sec:Computational details}.

\begin{table*}[htbp]
  \centering
  \caption{Summary of the components used for the calculation of the ONIOM2 correction to the \emph{BEs}. The last column shows the absolute value of the difference between the \emph{BE}(DFT, $\mu$-2H\textsubscript{2}O) value and the corrected \emph{BE}(ONIOM2).}  \label{tab-app:summary-components}%
    \begin{tabular}{lccccc}
    \toprule
    \textbf{Species} & \textbf{\emph{BE}(CCSD(T), $\mu$-nH\textsubscript{2}O) extrapolated} & \textbf{\emph{BE}( DFT, $\mu$-nH\textsubscript{2}O)} & \textbf{\emph{BE}(DFT, ice)} & \textbf{\emph{BE}(ONIOM2)} & \textbf{$\vert$Diff$\vert$} \\
    \toprule
    H\textsubscript{2} & 843   & 796   & 1192  & 1240  & 47 \\
    CO & 1127  & 1165  & 2356  & 2318  & 38 \\
    CO\textsubscript{2} & 1939  & 1817  & 3442  & 3564  & 121 \\
    OCS & 1354  & 1138  & 3478  & 3695  & 216 \\
    H\textsubscript{2}S & 2443  & 2875  & 5679  & 5248  & 432 \\
    HCN & 3300  & 3453  & 5795  & 5642  & 153 \\
    H\textsubscript{2}CO & 3199  & 3114  & 6490  & 6575  & 85 \\
    HCl & 3993  & 4475  & 6501  & 6019  & 482 \\
    NH\textsubscript{3} & 4246  & 4389  & 7376  & 7233  & 143 \\
    CH\textsubscript{3}OH & 5196  & 5275  & 8681  & 8602  & 79 \\
    HCOOH & 6263  & 6223  & 9522  & 9562  & 40 \\
    HCONH\textsubscript{2} & 5080  & 5056  & 9614  & 9638  & 24 \\
    OH• & 4492  & 4502  & 6542  & 6532  & 10 \\
    NH\textsubscript{2}• & 3021  & 2788  & 6043  & 6276  & 233 \\
    HCO• & 2339  & 2088  & 3474  & 3725  & 251 \\
    \bottomrule
    \end{tabular}%
\end{table*}%

\newpage

%%%%%%%%%%%%%%%%%%%%%%%%%%%%%%%%%%%%%%%%%%%%%%%%%%%%%%%%%%%%%%
%%%%%%%%%%%%%%%%%%%%%%%%%%%%%%%%%%%%%%%%%%%%%%%%%%%%%%%%%%%%%%
%%%%%%%%%%%%%%%%%%%%%%%%%%%%%%%%%%%%%%%%%%%%%%%%%%%%%%%%%%%%%%
\section{Crystalline adsorption geometries}\label{app-cryst-ads-geo}
The final optimized structures, together with \emph{BEs} and structural properties for every molecule in our set are presented in this section. For some notable cases, more than one initial structure has been modeled. All energetic quantities are in Kelvin ($\mathrm{K}$), while distances are in \AA. For every molecule, the gas phase optimized geometry together with its ESP map are also present. In the latter, the isovalue for the electron density is set equal to \num{1e-6} au, while the values for the ESP (again in au) vary case by case and thus are reported close to the RGB scale legend.

\figsetstart
\figsetnum{16}
\figsettitle{Adsorption on crystalline ice}

"The complete figure set (24 images) is available in the online journal".

\begin{figure}[h!]
    \centering
   \resizebox{8cm}{!}{\includegraphics{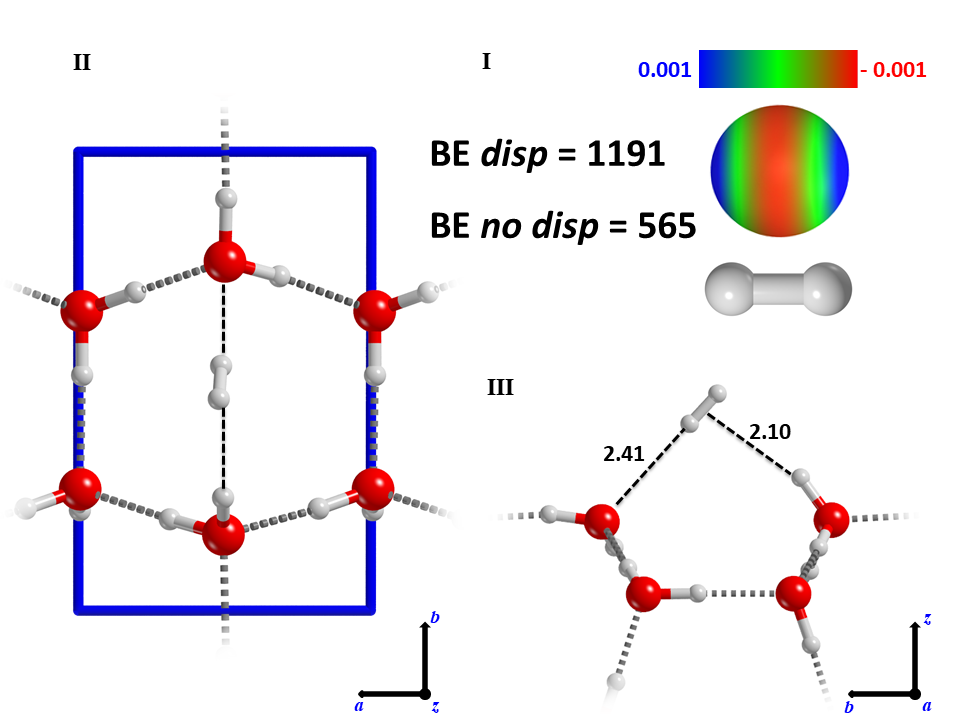}}
   \caption{I) Representation of the hydrogen (H\textsubscript{2}) molecule along with its ESP surface, BSSE-corrected binding energies (\emph{BE}) with (\emph{disp}) and without (\emph{no disp}) dispersion. II) Top view of H\textsubscript{2}-(010) P-ice surface interaction (unit cell highlighted in blue). III) Detail of the side view.}
    \label{fig:h2}
    \end{figure}
\figsetgrpend

\figsetgrpstart
\figsetgrpnum{16.1}
\figsetgrptitle{$O_2$}
\figsetplot{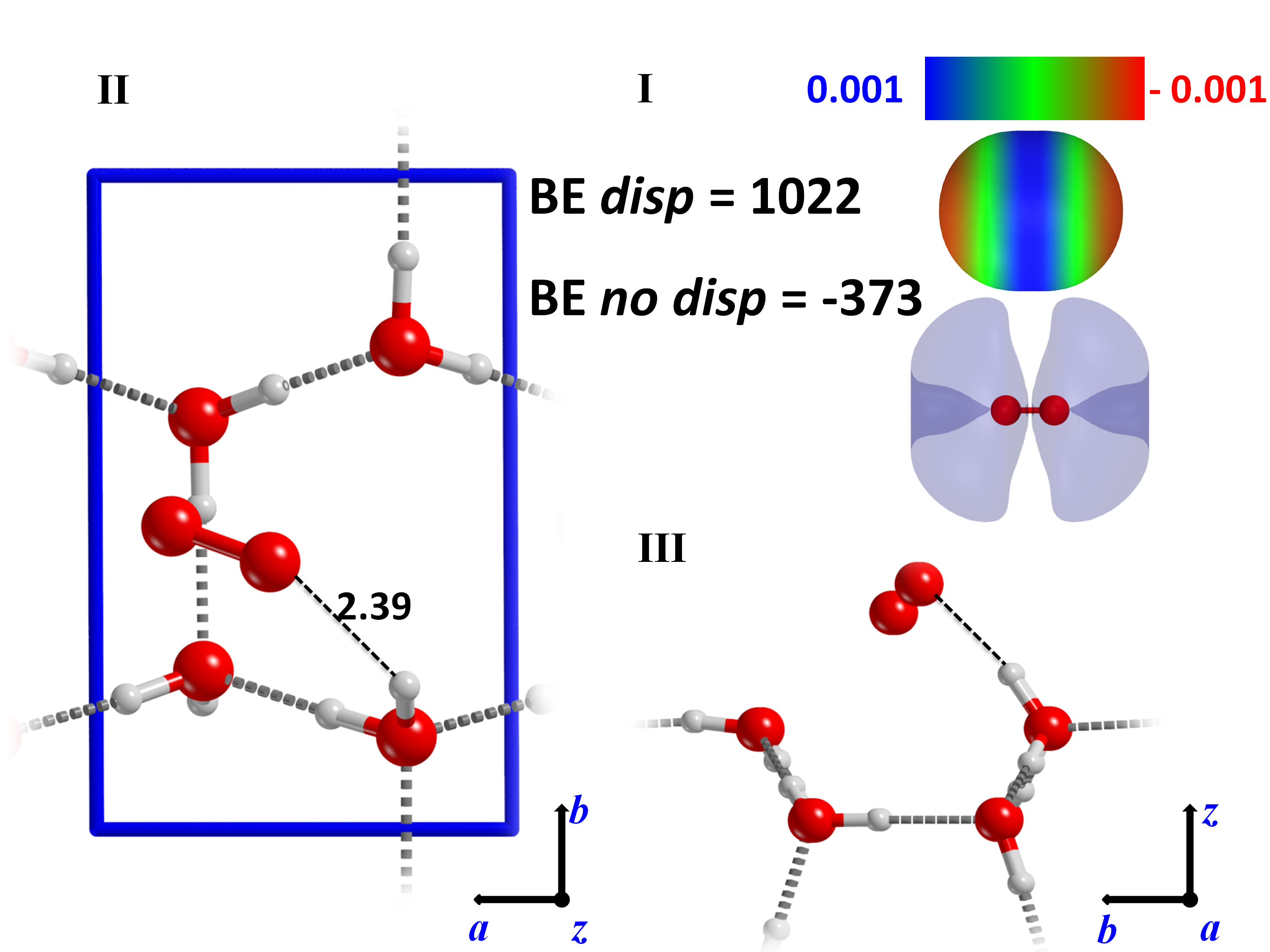}
\figsetgrpnote{I) Representation of the oxygen (O\textsubscript{2}) molecule along with its ESP surface, spin density map, BSSE-corrected binding energies (\emph{BE}) with (\emph{disp}) and without (\emph{no disp}) dispersion. II) Top view of O\textsubscript{2}-(010) P-ice surface interaction (unit cell highlighted in blue). III) Detail of the side view.}
\label{fig:o2}
\figsetgrpend

\figsetgrpstart
\figsetgrpnum{16.2}
\figsetgrptitle{$N_2$}
\figsetplot{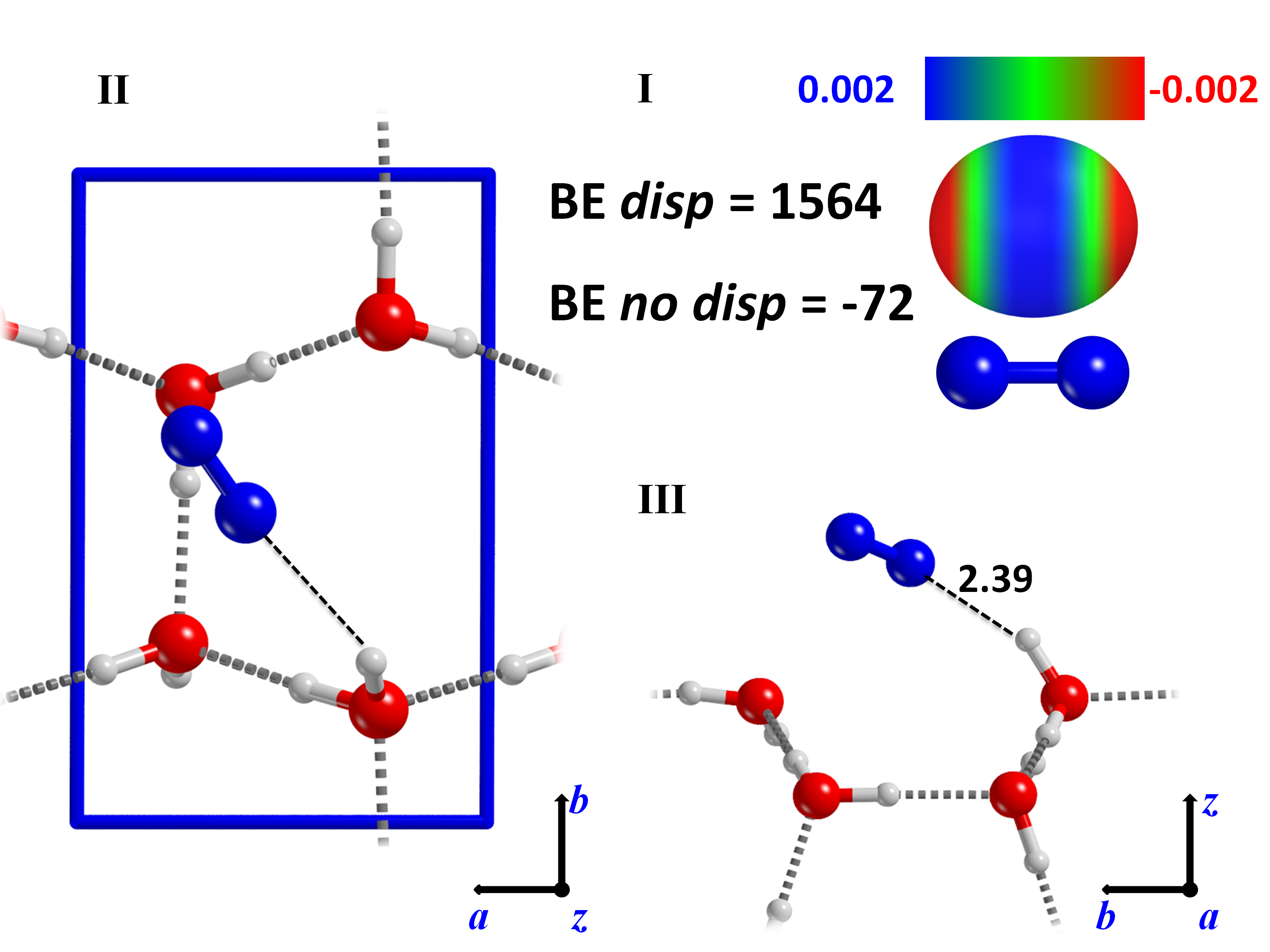}
\figsetgrpnote{I) Representation of the nitrogen (N\textsubscript{2}) molecule along with its ESP surface, BSSE-corrected binding energies (\emph{BE}) with (\emph{disp}) and without (\emph{no disp}) dispersion. II) Top view of N\textsubscript{2}-(010) P-ice surface interaction (unit cell highlighted in blue). III) Detail of the side view.}        \label{fig:n2}
\figsetgrpend

\figsetgrpstart
\figsetgrpnum{16.3}
\figsetgrptitle{$CH_4$}
\figsetplot{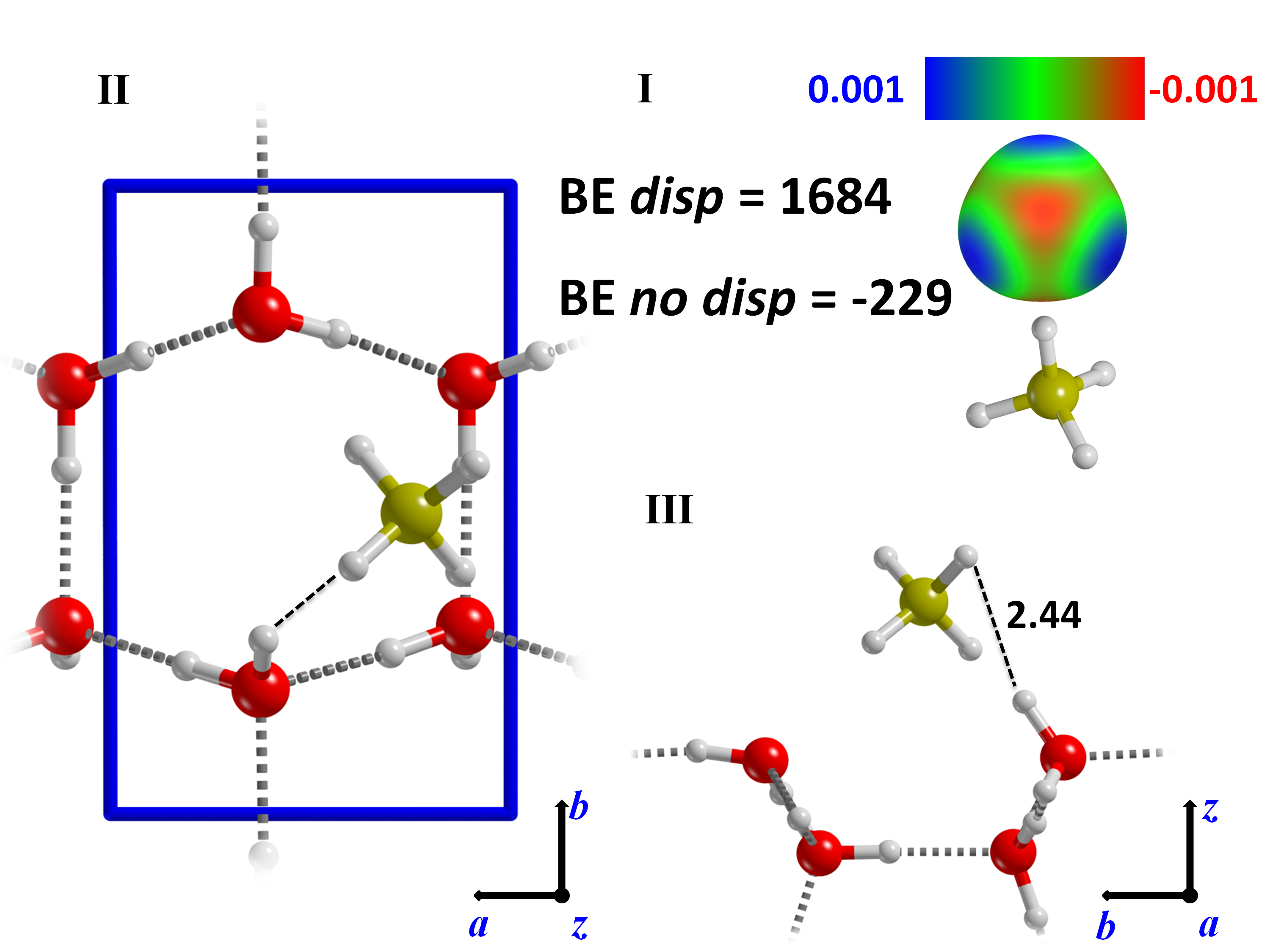}
\figsetgrpnote{I) Representation of the methane (CH\textsubscript{4}) molecule along with its ESP surface, BSSE-corrected binding energies (\emph{BE}) with (\emph{disp}) and without (\emph{no disp}) dispersion. II) Top view of N\textsubscript{2}-(010) P-ice surface interaction (unit cell highlighted in blue). III) Detail of the side view.}        \label{fig:ch4}
\figsetgrpend

\figsetgrpstart
\figsetgrpnum{16.4}
\figsetgrptitle{CO}
\figsetplot{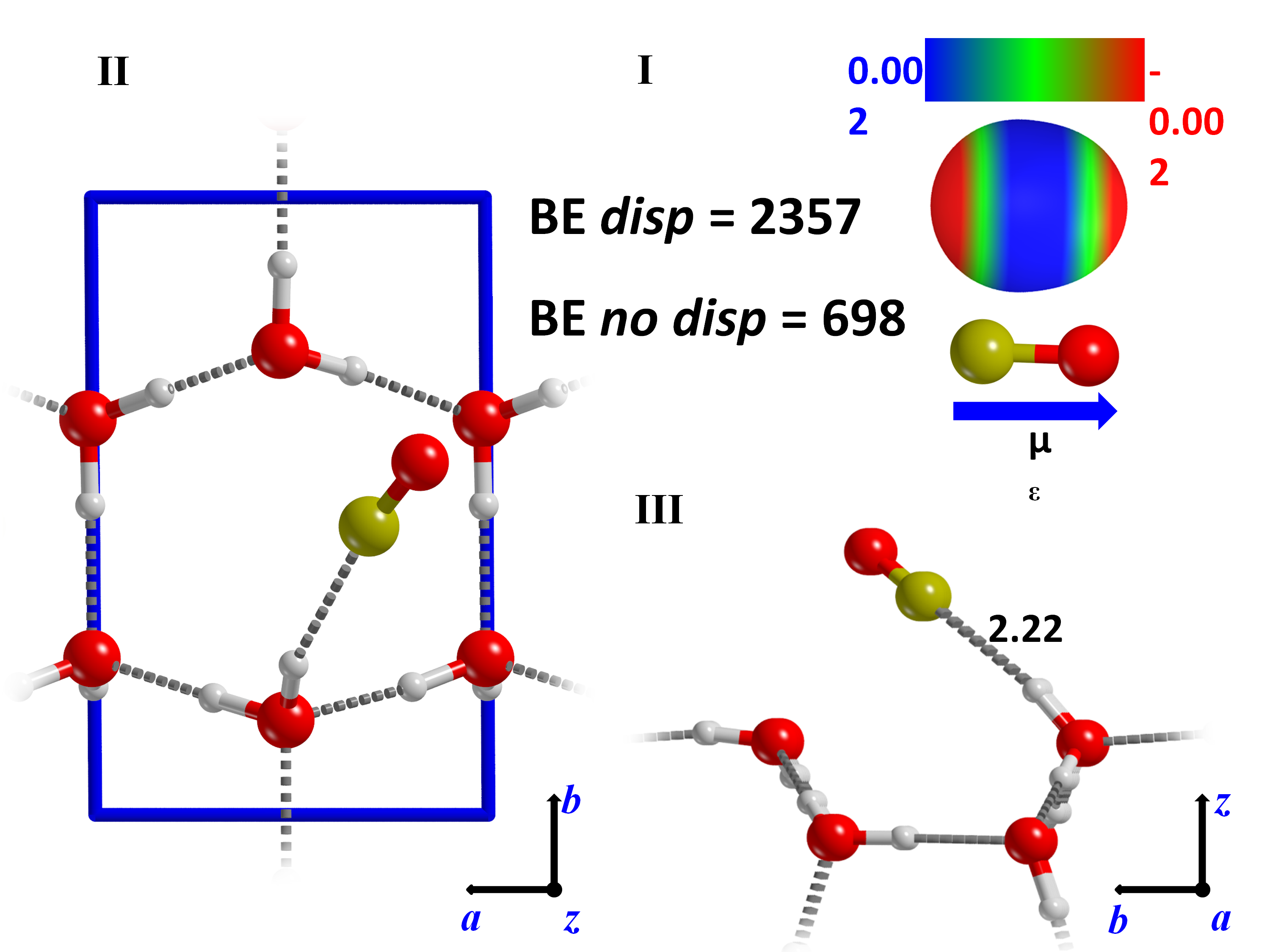}
\figsetgrpnote{I) Representation of the carbon monoxide (CO) molecule along with its ESP surface, BSSE-corrected binding energies (\emph{BE}) with (\emph{disp}) and without (\emph{no disp}) dispersion. II) Top view of CO-(010) P-ice surface interaction (unit cell highlighted in blue). III) Detail of the side view.}
\label{fig:co}
\figsetgrpend

\figsetgrpstart
\figsetgrpnum{16.5}
\figsetgrptitle{$CO_2$}
\figsetplot{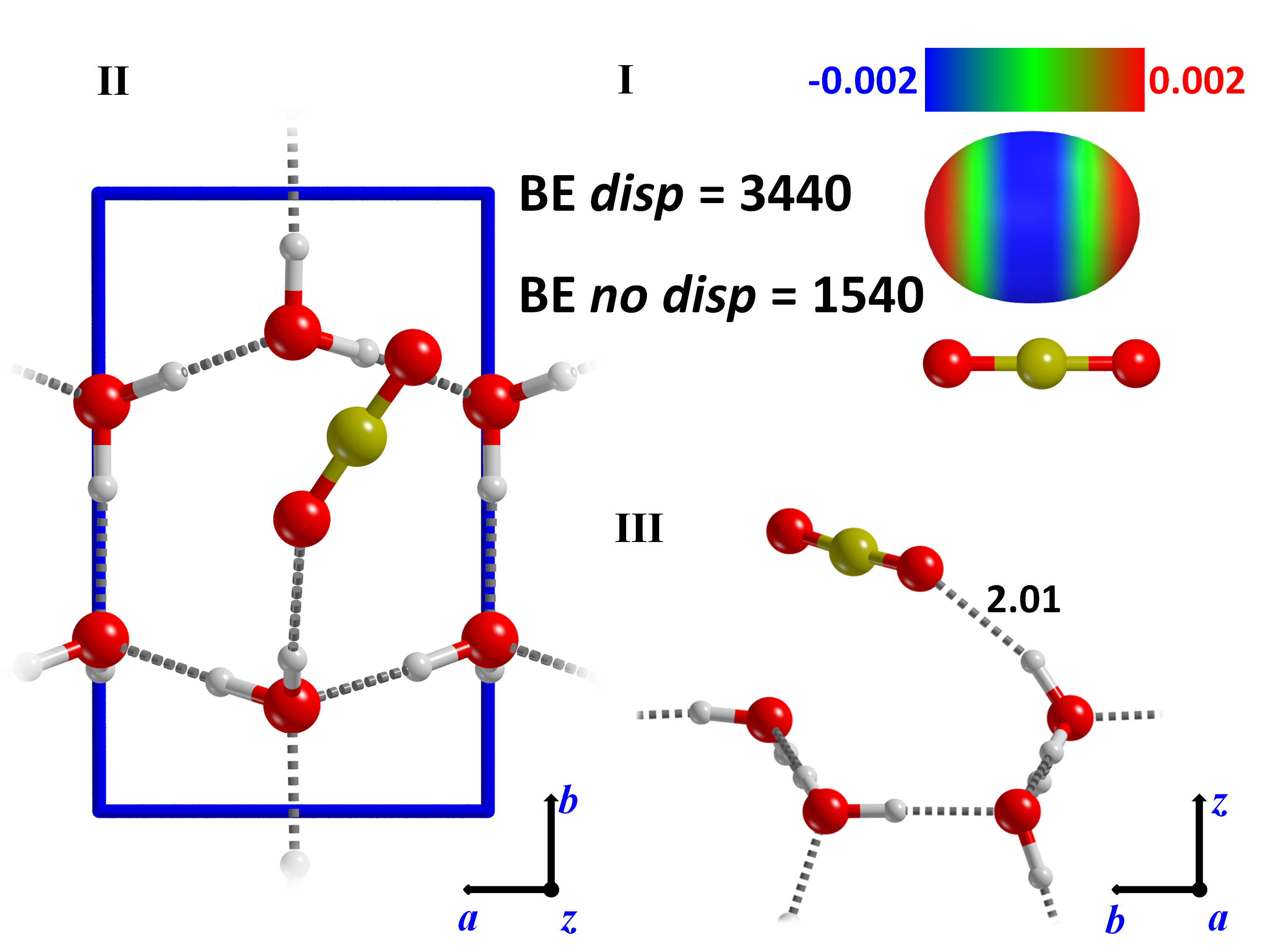}
\figsetgrpnote{I) Representation of the carbon dioxide (CO\textsubscript{2}) molecule along with its ESP surface, BSSE-corrected binding energies (\emph{BE}) with (\emph{disp}) and without (\emph{no disp}) dispersion. II) Top view of CO\textsubscript{2}-(010) P-ice surface interaction (unit cell highlighted in blue). III) Detail of the side view.}
\label{fig:co2}
\figsetgrpend

\figsetgrpstart
\figsetgrpnum{16.6}
\figsetgrptitle{OCS}
\figsetplot{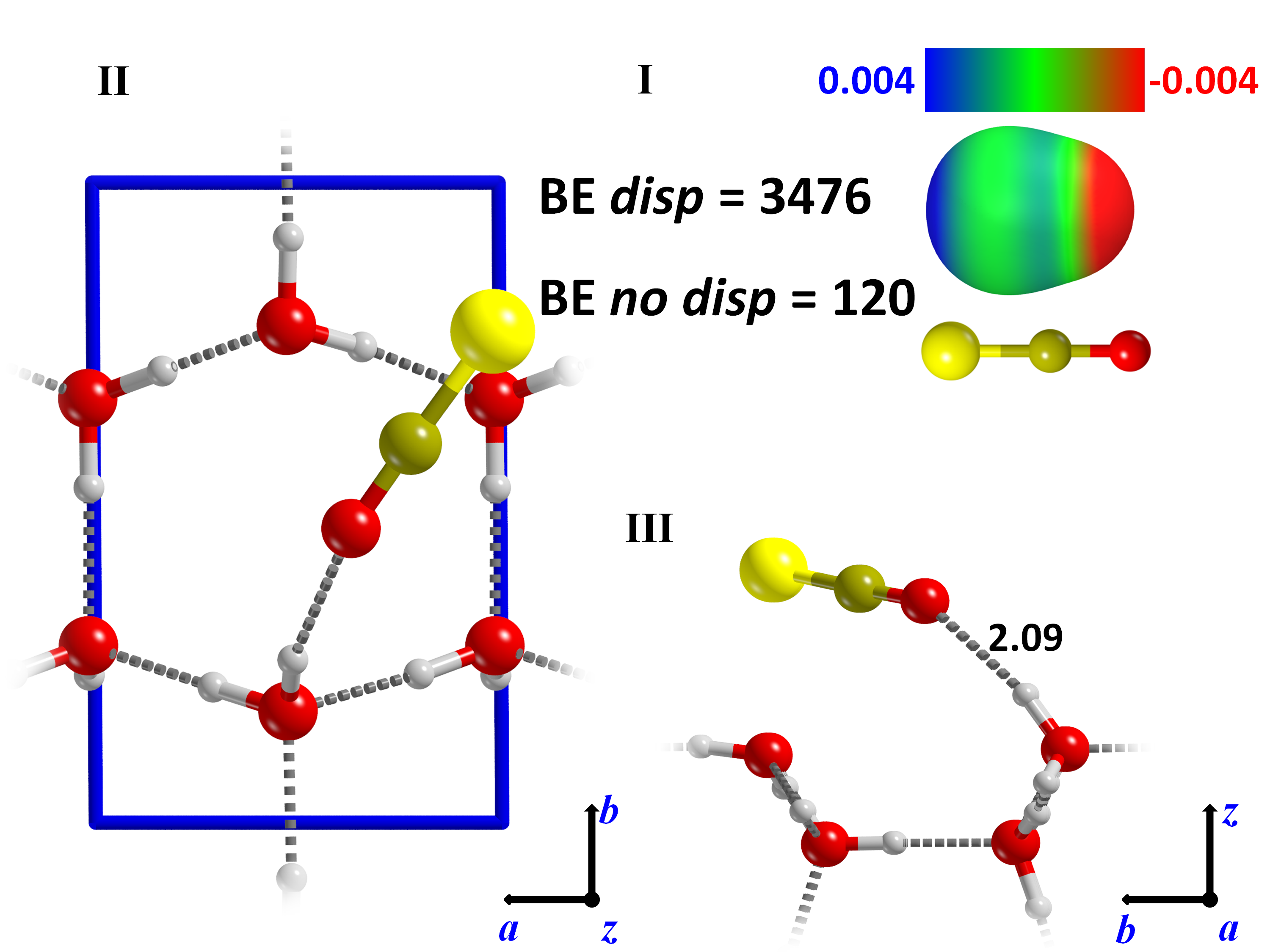}
\figsetgrpnote{I) Representation of the carbon sulphide (OCS) molecule along with its ESP surface, BSSE-corrected binding energies (\emph{BE}) with (\emph{disp}) and without (\emph{no disp}) dispersion. II) Top view of OCS-(010) P-ice surface interaction (unit cell highlighted in blue). III) Detail of the side view.}
\label{fig:ocs}
\figsetgrpend    

\figsetgrpstart
\figsetgrpnum{16.7}
\figsetgrptitle{HCN}
\figsetplot{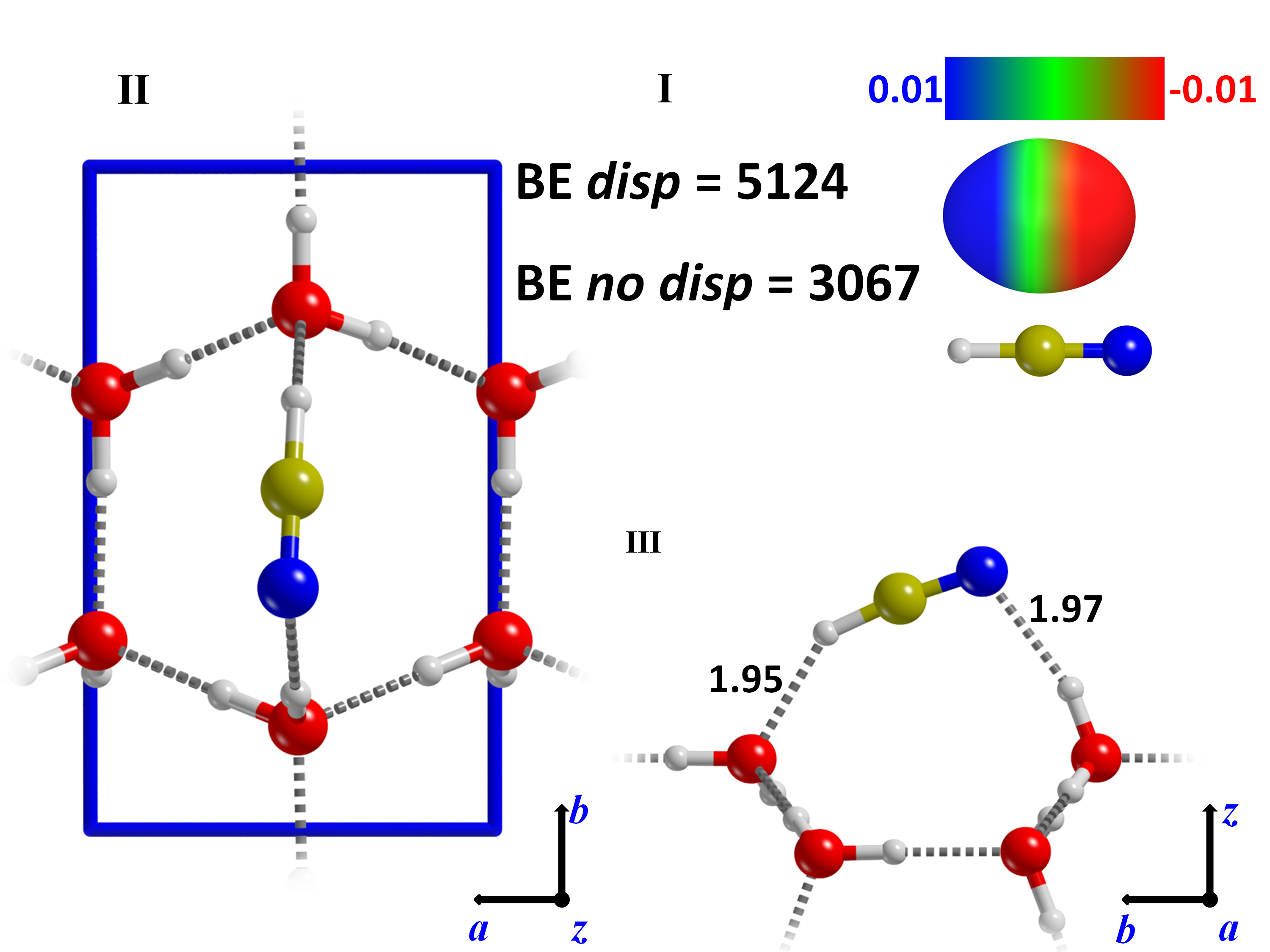}
\figsetgrpnote{I) Representation of the hydrogen cyanide (HCN) molecule along with its ESP surface, BSSE-corrected binding energies (\emph{BE}) with (\emph{disp}) and without (\emph{no disp}) dispersion. II) Top view of HCN-(010) P-ice surface interaction (unit cell highlighted in blue). III) Detail of the side view.}
\label{fig:hcn}
\figsetgrpend

\figsetgrpstart
\figsetgrpnum{16.8}
\figsetgrptitle{$H_2$O}
\figsetplot{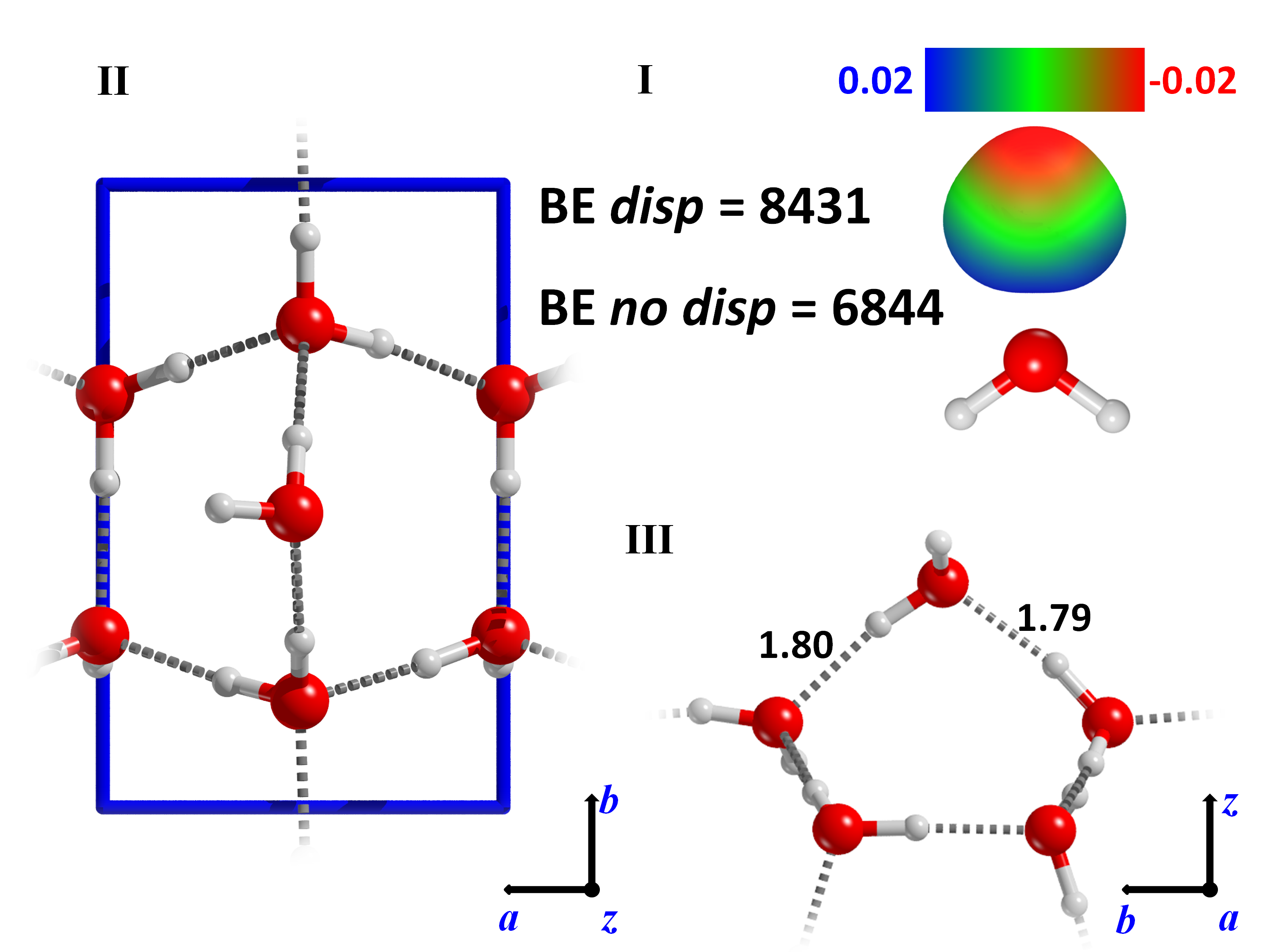}
\figsetgrpnote{I) Representation of the water (H\textsubscript{2}O) molecule along with its ESP surface, BSSE-corrected binding energies (\emph{BE}) with (\emph{disp}) and without (\emph{no disp}) dispersion. II) Top view of H\textsubscript{2}O-(010) P-ice surface interaction (unit cell highlighted in blue). III) Detail of the side view.}
\label{fig:h2o}
\figsetgrpend 

\figsetgrpstart
\figsetgrpnum{16.9}
\figsetgrptitle{NH\textsubscript{3}}
\figsetplot{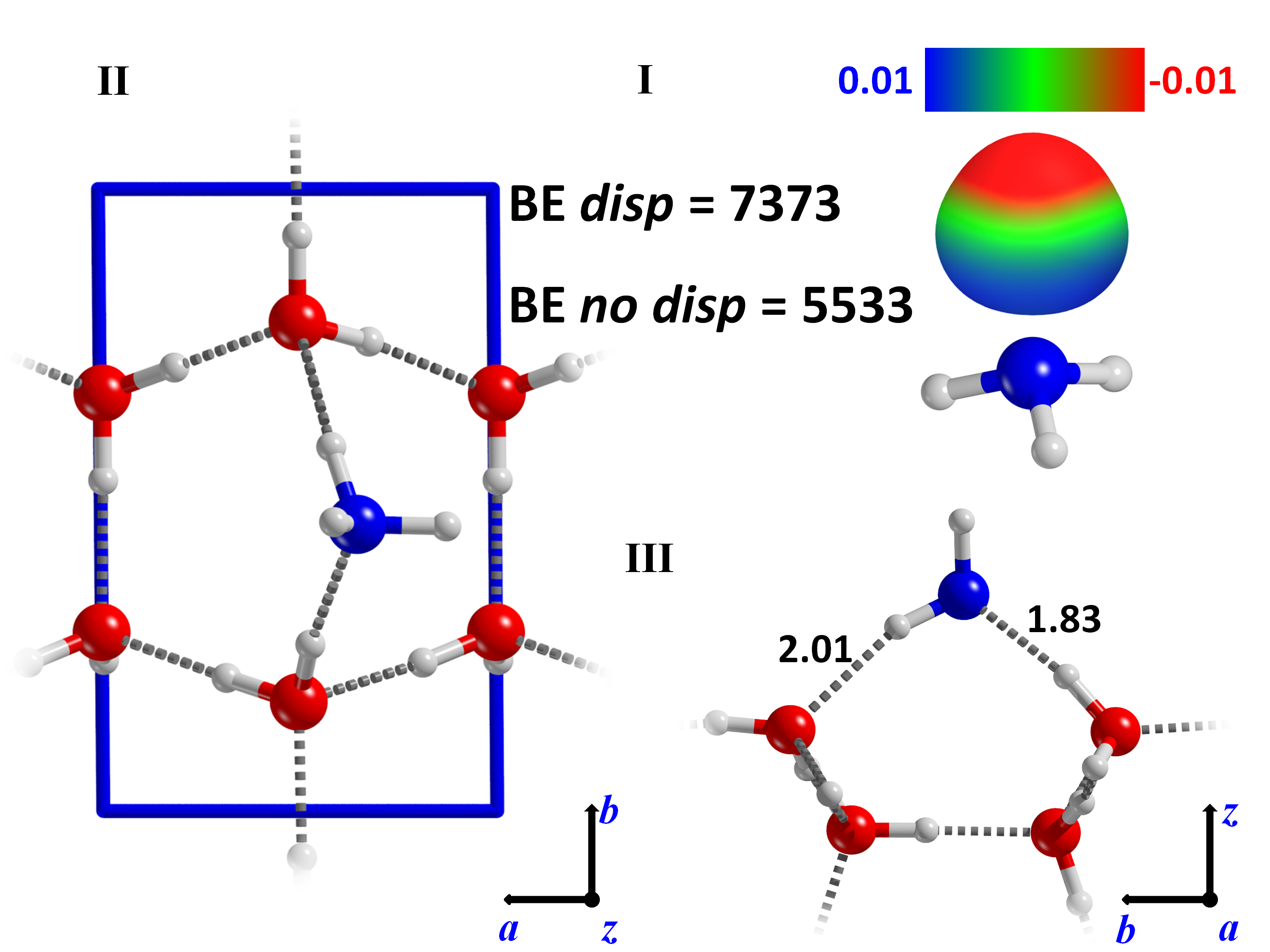}
\figsetgrpnote{I) Representation of the ammonia (NH\textsubscript{3}) molecule along with its ESP surface, BSSE-corrected binding energies (\emph{BE}) with (\emph{disp}) and without (\emph{no disp}) dispersion. II) Top view of NH\textsubscript{3}-(010) P-ice surface interaction (unit cell highlighted in blue). III) Detail of the side view.}
\label{fig:nh3}
\figsetgrpend

\figsetgrpstart
\figsetgrpnum{16.10}
\figsetgrptitle{H\textsubscript{2}S}
\figsetplot{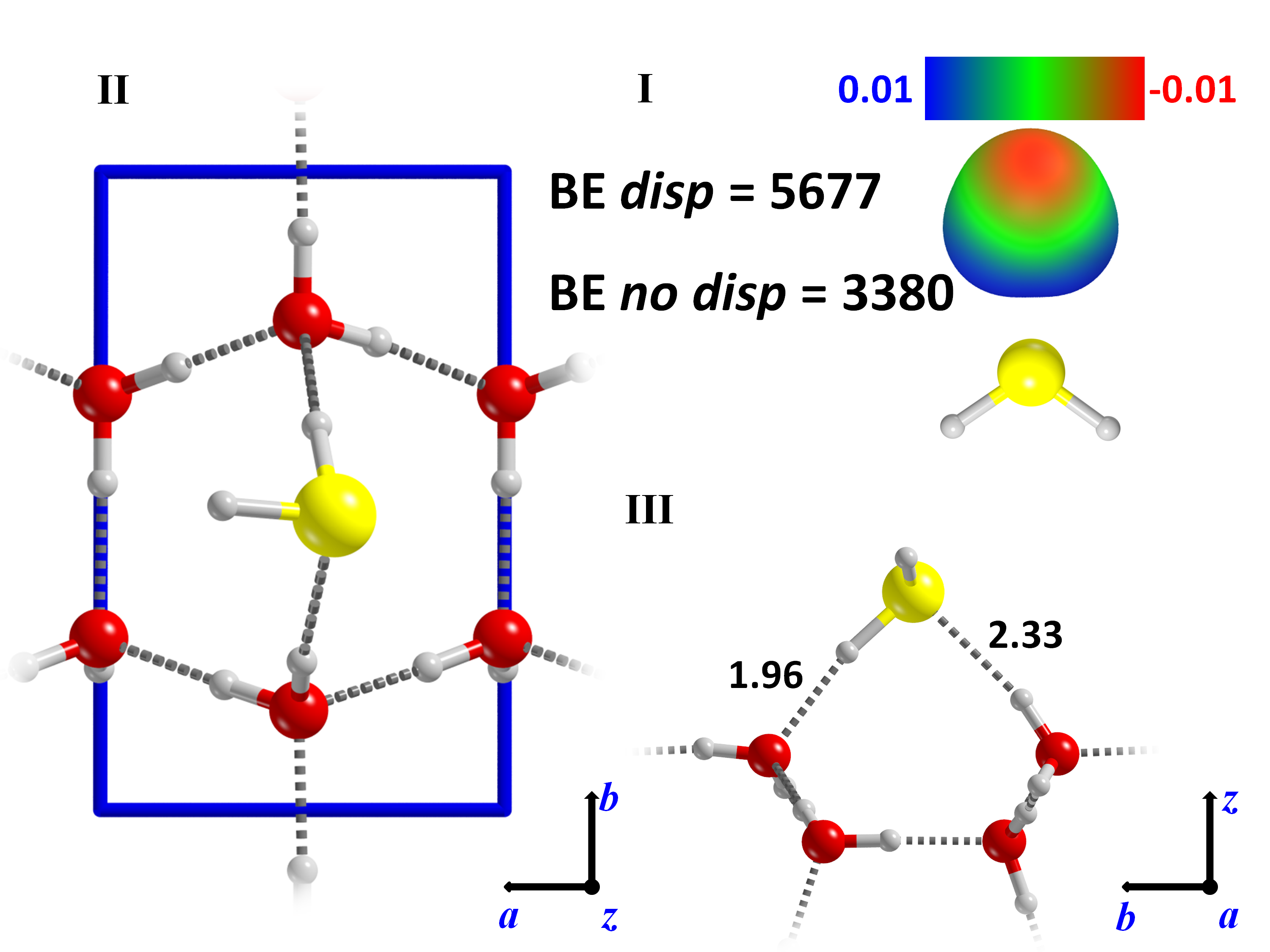}
\figsetgrpnote{I) Representation of the hydrogen sulphide (H\textsubscript{2}S) molecule along with its ESP surface, BSSE-corrected binding energies (\emph{BE}) with (\emph{disp}) and without (\emph{no disp}) dispersion. II) Top view of H\textsubscript{2}S-(010) P-ice surface interaction (unit cell highlighted in blue). III) Detail of the side view.}
\label{fig:h2s}
\figsetgrpend 

\figsetgrpstart
\figsetgrpnum{16.11}
\figsetgrptitle{HCl}
\figsetplot{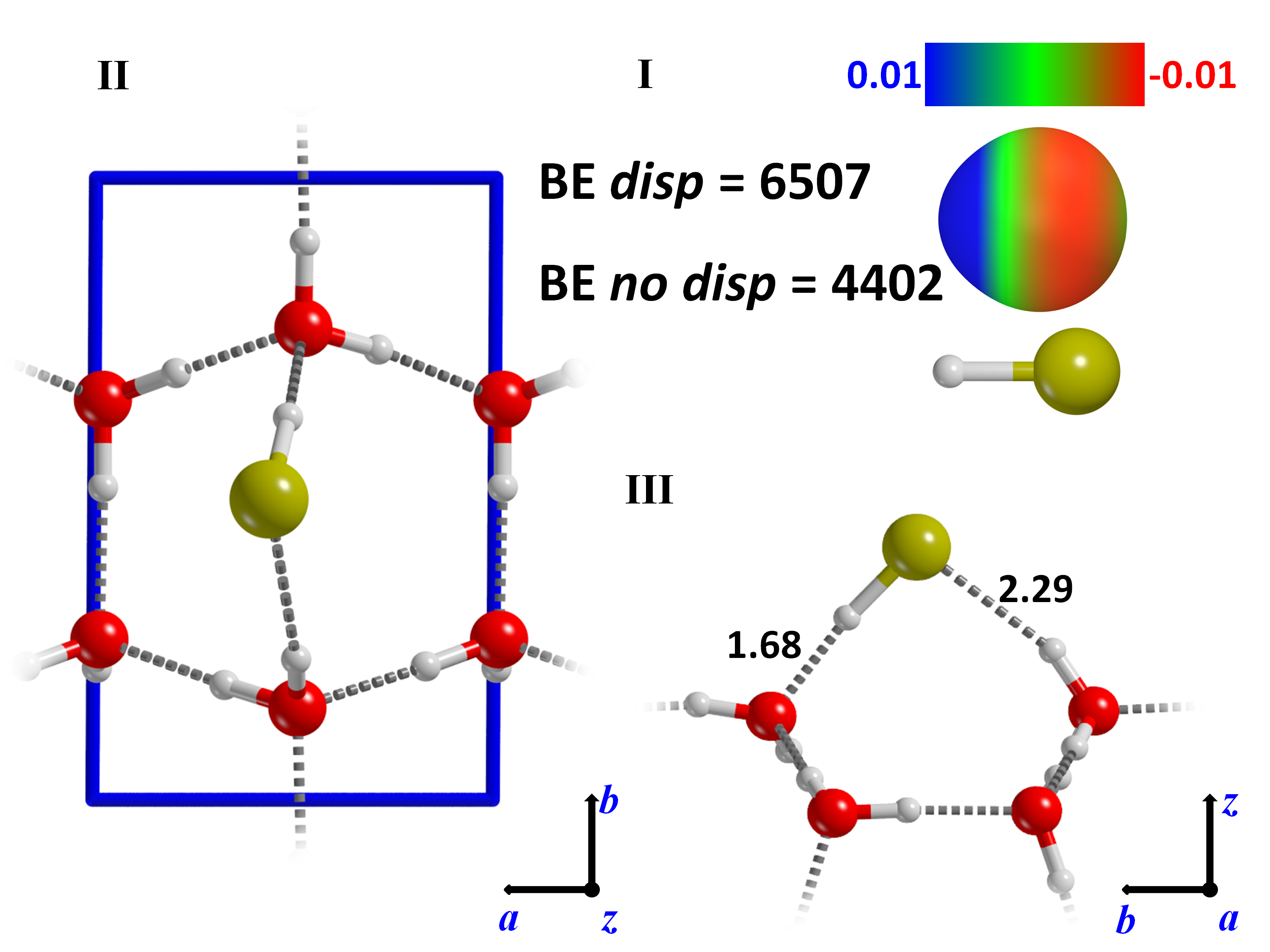}
\figsetgrpnote{I) Representation of the hydrogen cloride (HCl) molecule along with its ESP surface, BSSE-corrected binding energies (\emph{BE}) with (\emph{disp}) and without (\emph{no disp}) dispersion. II) Top view of HCl-(010) P-ice surface interaction (unit cell highlighted in blue). III) Detail of the side view.}
\label{fig:hcl}
\figsetgrpend     
    
\figsetgrpstart
\figsetgrpnum{16.12}
\figsetgrptitle{CH\textsubscript{3}OH}
\figsetplot{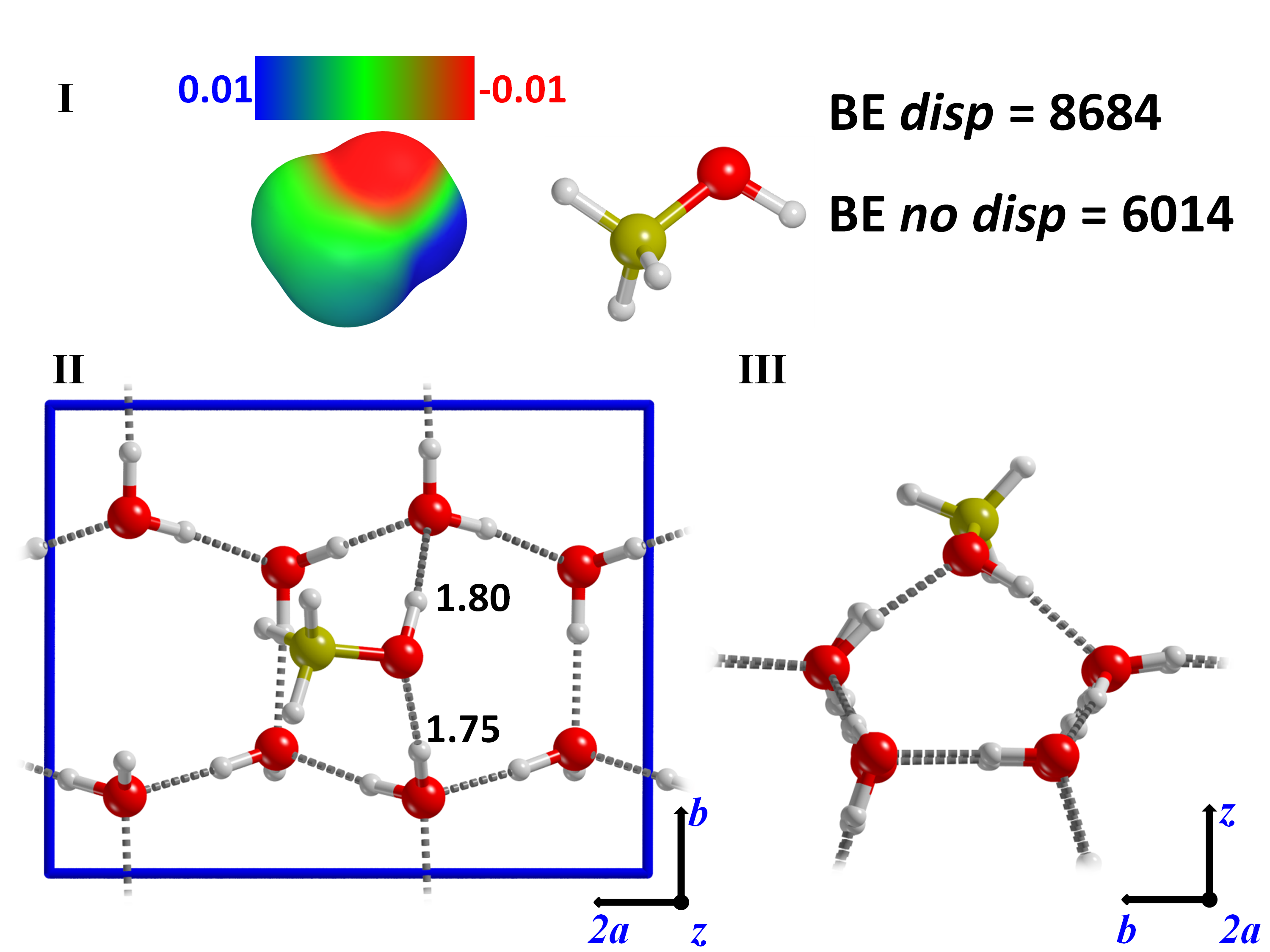}
\figsetgrpnote{I) Representation of the methanol (CH\textsubscript{3}OH) molecule along with its ESP surface, BSSE-corrected binding energies (\emph{BE}) with (\emph{disp}) and without (\emph{no disp}) dispersion. II) Top view of CH\textsubscript{3}OH-(010) P-ice 2x1 supercell surface interaction (unit cell highlighted in blue). III) Detail of the side view.}    
\label{fig:ch3oh}
\figsetgrpend

\figsetgrpstart
\figsetgrpnum{16.13}
\figsetgrptitle{CH\textsubscript{3}OH}
\figsetplot{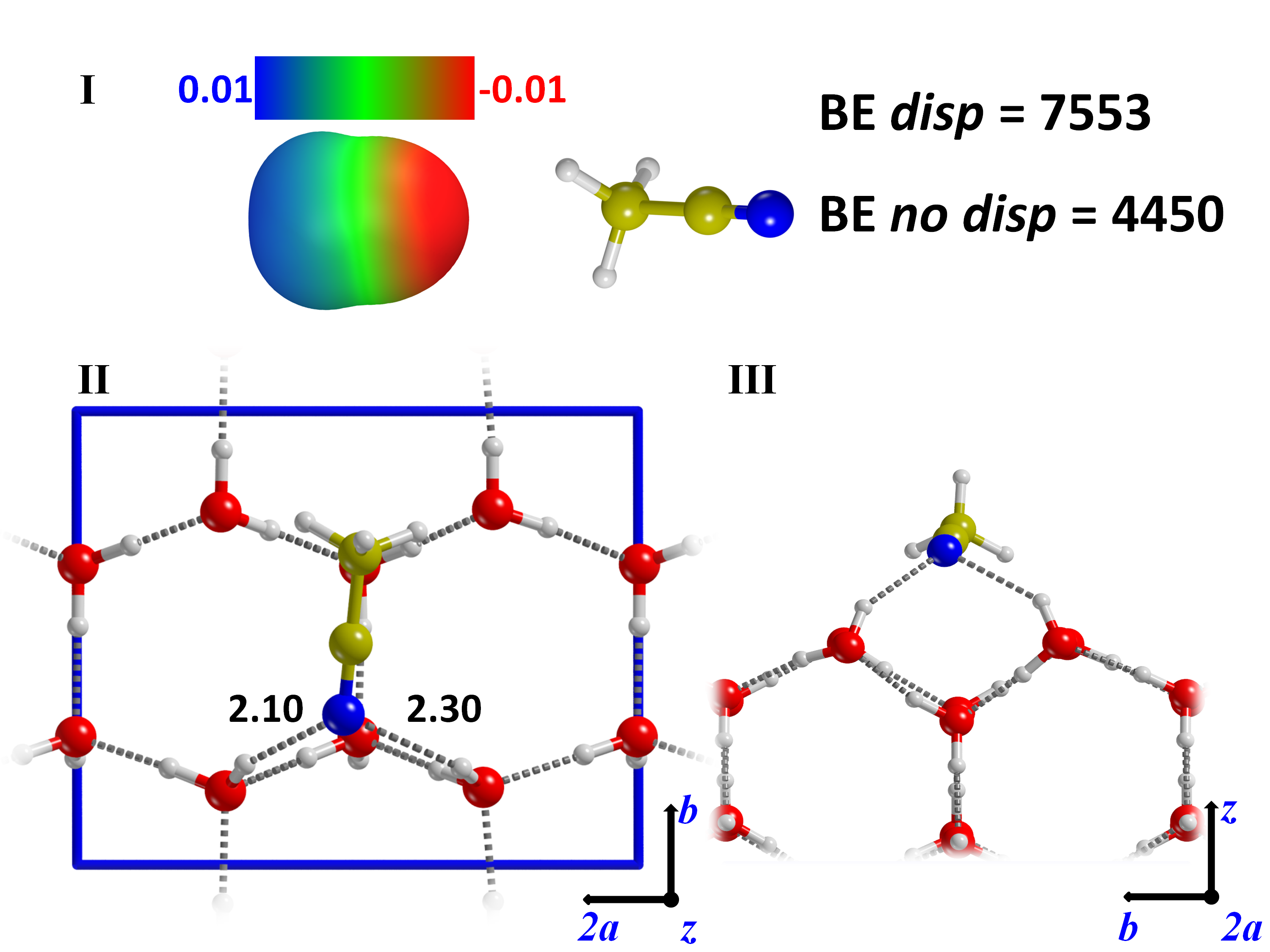}
\figsetgrpnote{I) Representation of the methanol (CH\textsubscript{3}OH) molecule along with its ESP surface, BSSE-corrected binding energies (\emph{BE}) with (\emph{disp}) and without (\emph{no disp}) dispersion. II) Top view of CH\textsubscript{3}OH-(010) P-ice 2x1 supercell surface interaction (unit cell highlighted in blue). III) Detail of the side view.}
\label{fig:ch3cn}
\figsetgrpend

\figsetgrpstart
\figsetgrpnum{16.14}
\figsetgrptitle{CH\textsubscript{3}OH}
\figsetplot{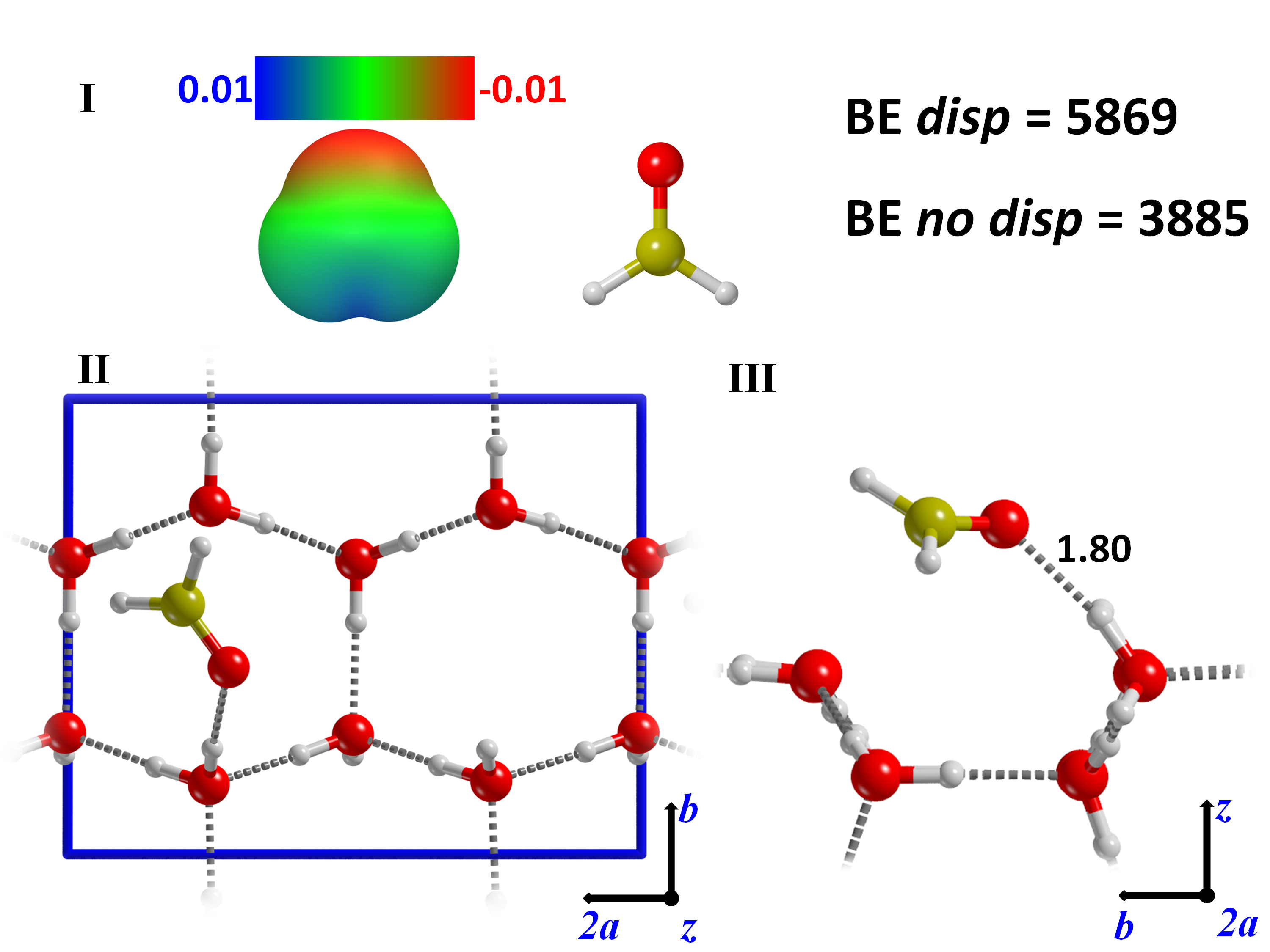}
\figsetgrpnote{I) Representation of the methanol (CH\textsubscript{3}OH) molecule along with its ESP surface, BSSE-corrected binding energies (\emph{BE}) with (\emph{disp}) and without (\emph{no disp}) dispersion. II) Top view of CH\textsubscript{3}OH-(010) P-ice 2x1 supercell surface interaction (unit cell highlighted in blue). III) Detail of the side view.}
\label{fig:h2co-sc1}
\figsetgrpend    

\figsetgrpstart
\figsetgrpnum{16.15}
\figsetgrptitle{CH\textsubscript{3}OH}
\figsetplot{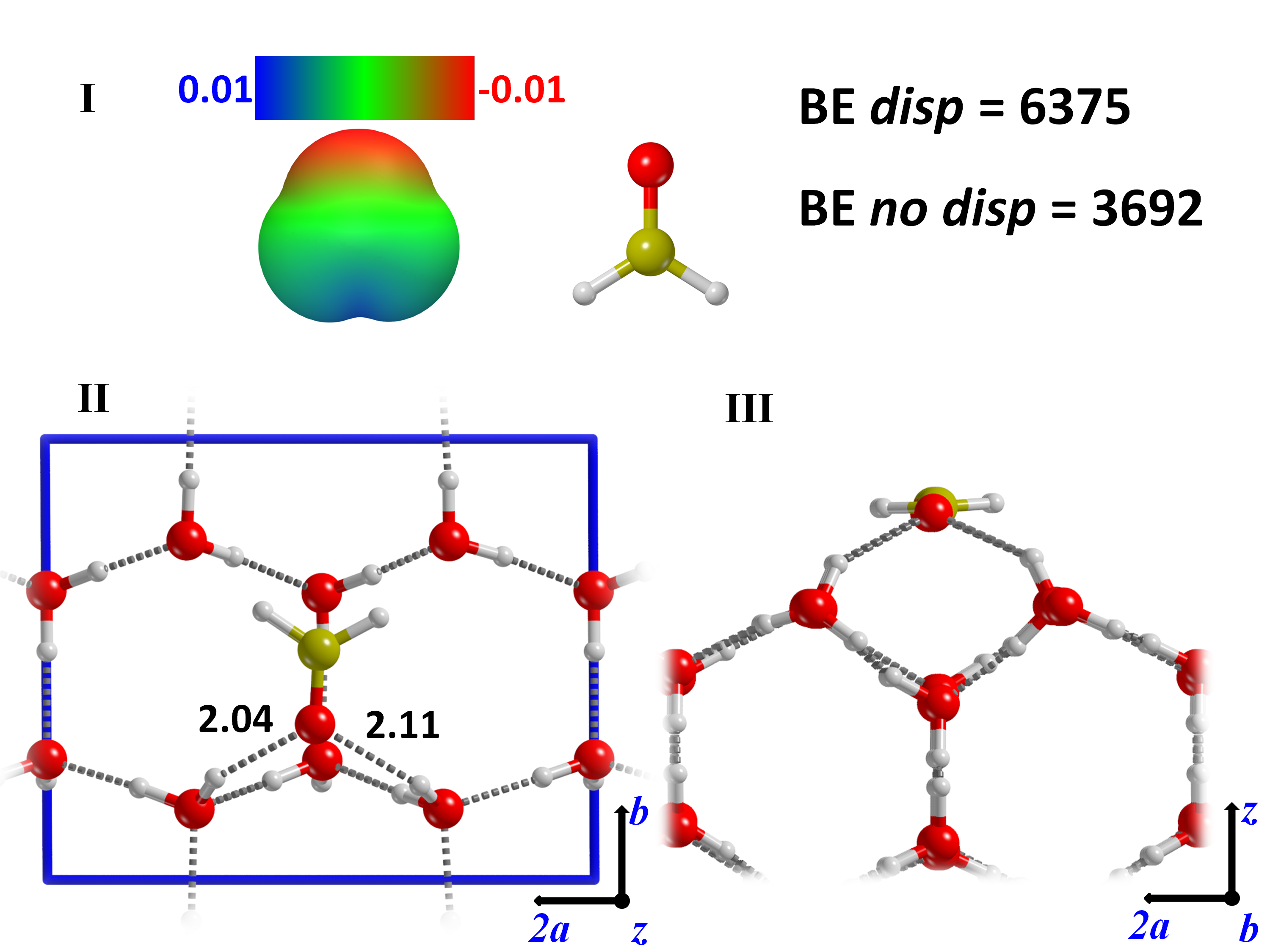}
\figsetgrpnote{I) Representation of the methanol (CH\textsubscript{3}OH) molecule along with its ESP surface, BSSE-corrected binding energies (\emph{BE}) with (\emph{disp}) and without (\emph{no disp}) dispersion. II) Top view of CH\textsubscript{3}OH-(010) P-ice 2x1 supercell surface interaction (unit cell highlighted in blue). III) Detail of the side view.}    
\label{fig:h2co-sc2}
\figsetgrpend

\figsetgrpstart
\figsetgrpnum{16.16}
\figsetgrptitle{CH\textsubscript{3}OH}
\figsetplot{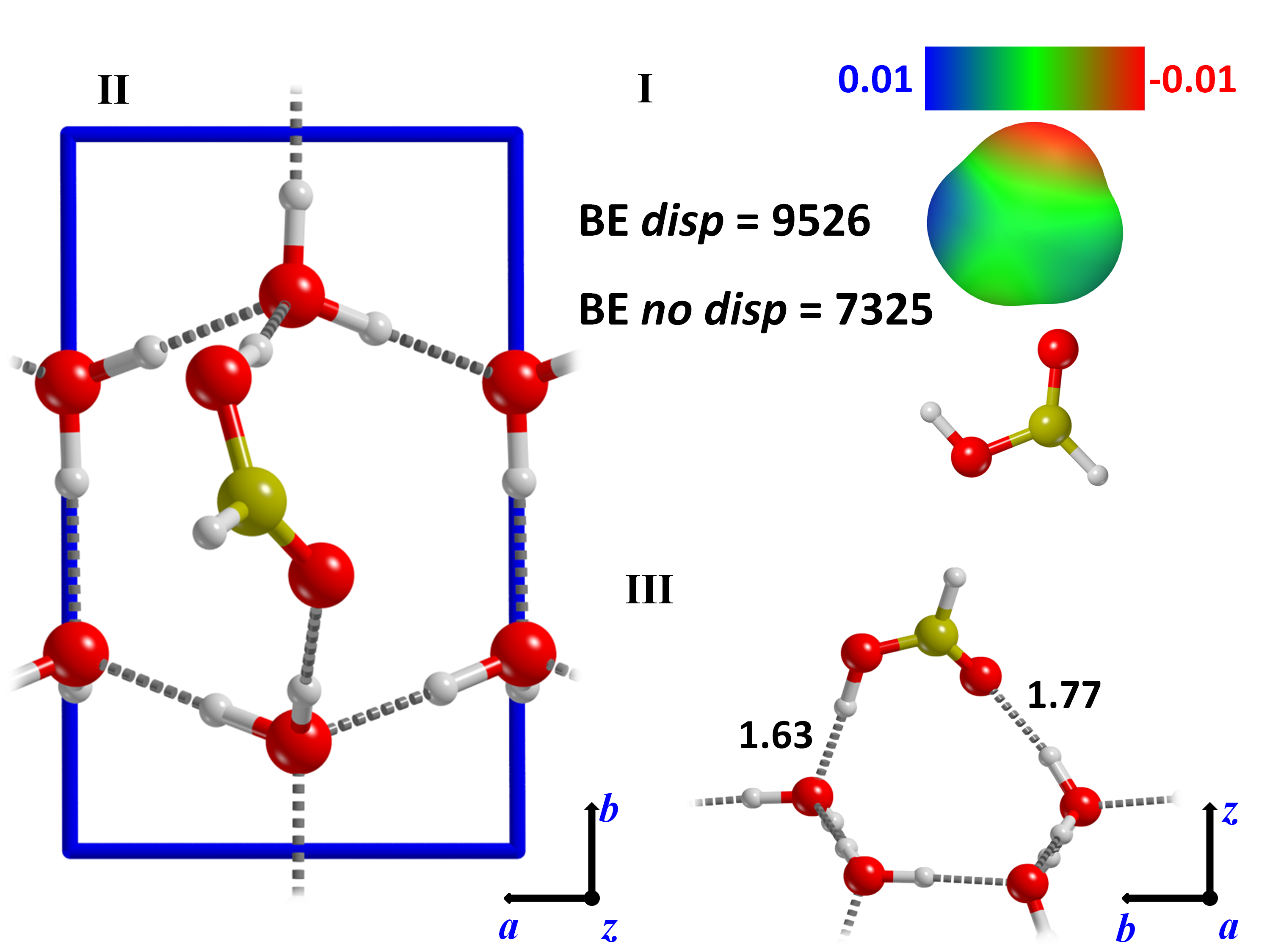}
\figsetgrpnote{I) Representation of the methanol (CH\textsubscript{3}OH) molecule along with its ESP surface, BSSE-corrected binding energies (\emph{BE}) with (\emph{disp}) and without (\emph{no disp}) dispersion. II) Top view of CH\textsubscript{3}OH-(010) P-ice 2x1 supercell surface interaction (unit cell highlighted in blue). III) Detail of the side view.}
\label{fig:HCOOH}
\figsetgrpend

\figsetgrpstart
\figsetgrpnum{16.17}
\figsetgrptitle{CH\textsubscript{3}OH}
\figsetplot{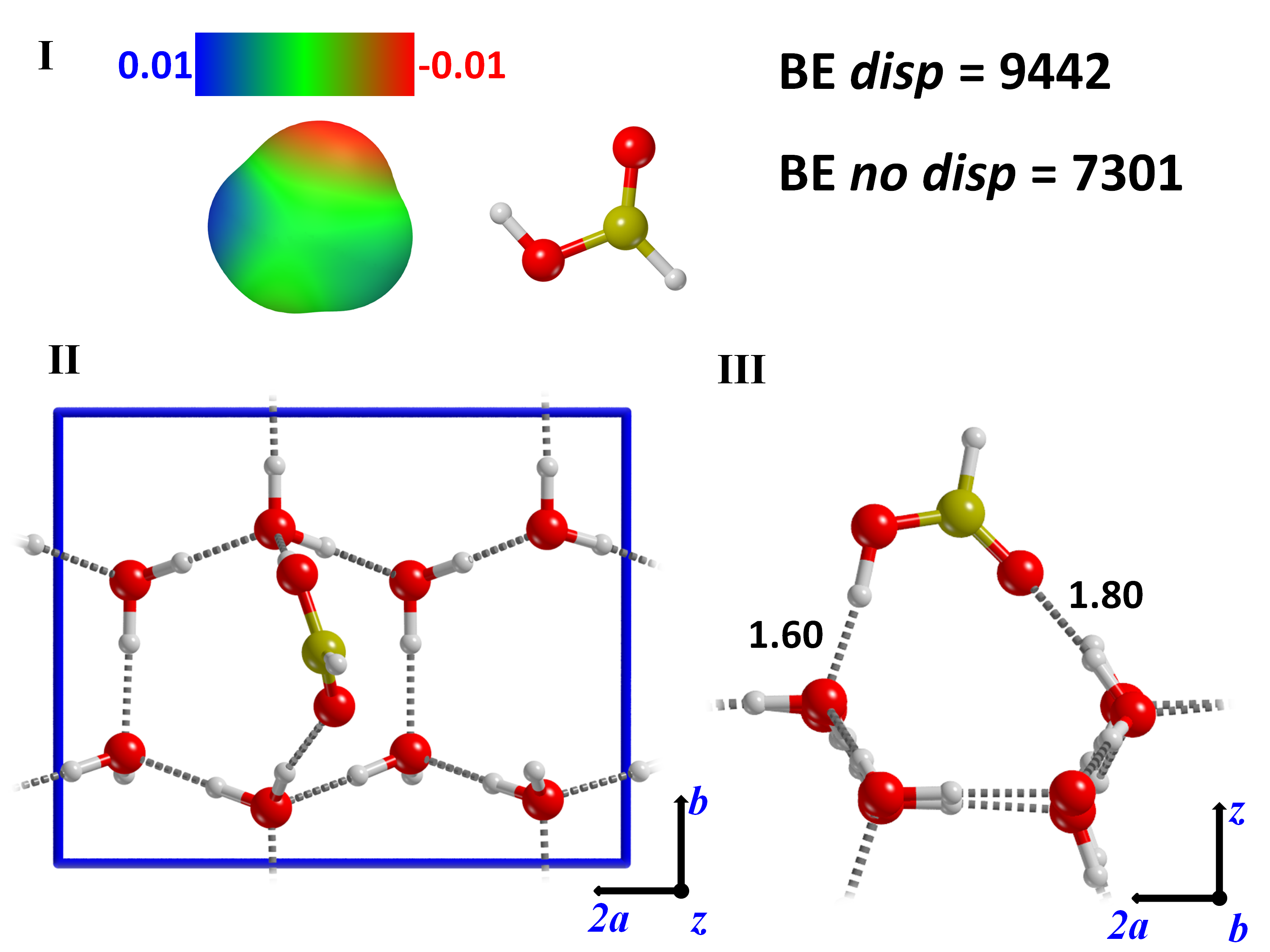}
\figsetgrpnote{I) Representation of the methanol (CH\textsubscript{3}OH) molecule along with its ESP surface, BSSE-corrected binding energies (\emph{BE}) with (\emph{disp}) and without (\emph{no disp}) dispersion. II) Top view of CH\textsubscript{3}OH-(010) P-ice 2x1 supercell surface interaction (unit cell highlighted in blue). III) Detail of the side view.}
\label{fig:HCOOH-sc}
\figsetgrpend
    
\figsetgrpstart
\figsetgrpnum{16.18}
\figsetgrptitle{CH\textsubscript{3}OH}
\figsetplot{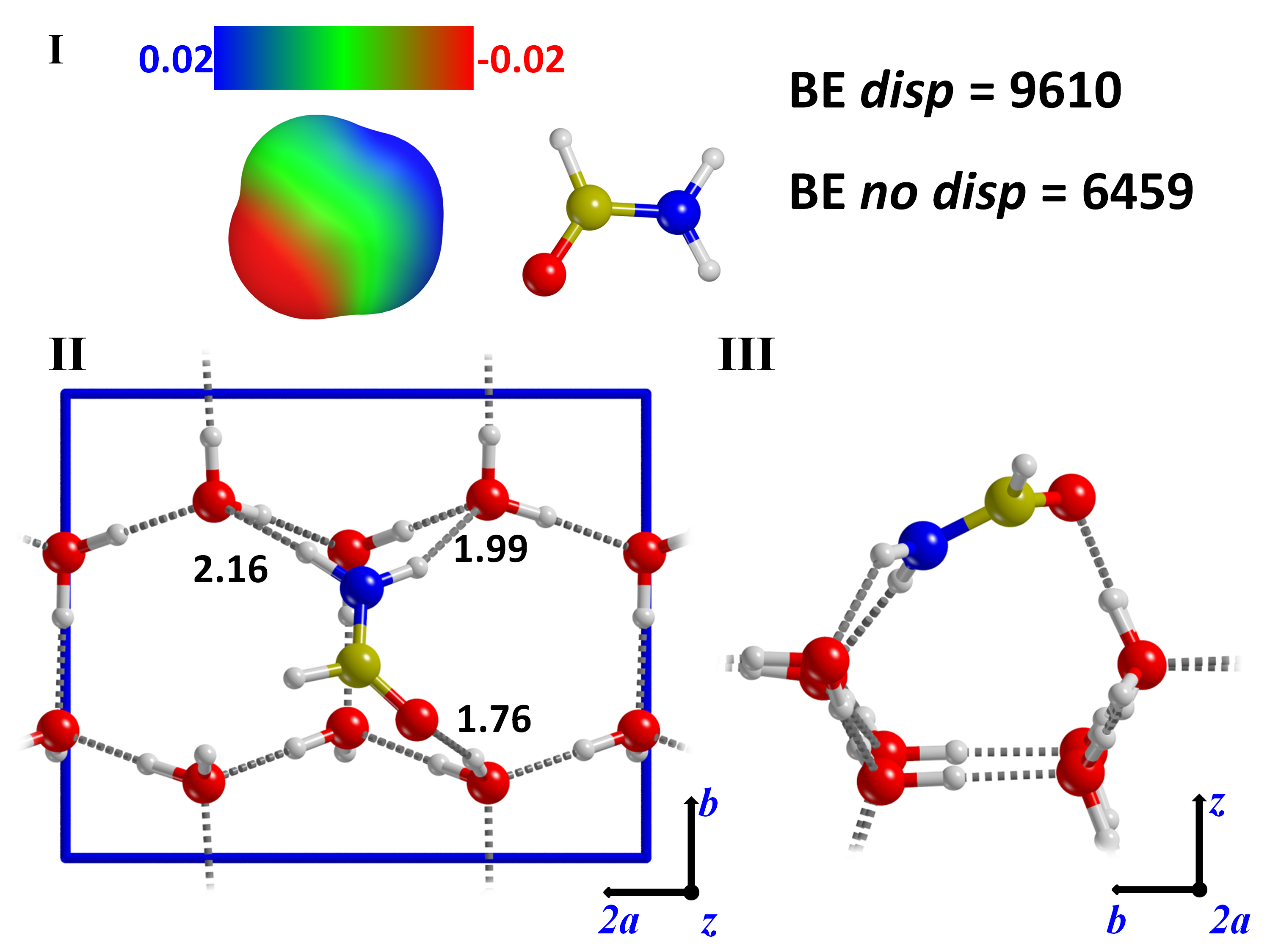}
\figsetgrpnote{I) Representation of the methanol (CH\textsubscript{3}OH) molecule along with its ESP surface, BSSE-corrected binding energies (\emph{BE}) with (\emph{disp}) and without (\emph{no disp}) dispersion. II) Top view of CH\textsubscript{3}OH-(010) P-ice 2x1 supercell surface interaction (unit cell highlighted in blue). III) Detail of the side view.}    
\label{fig:hconh2-sc1}
 \figsetgrpend
   
\figsetgrpstart
\figsetgrpnum{16.19}
\figsetgrptitle{CH\textsubscript{3}OH}
\figsetplot{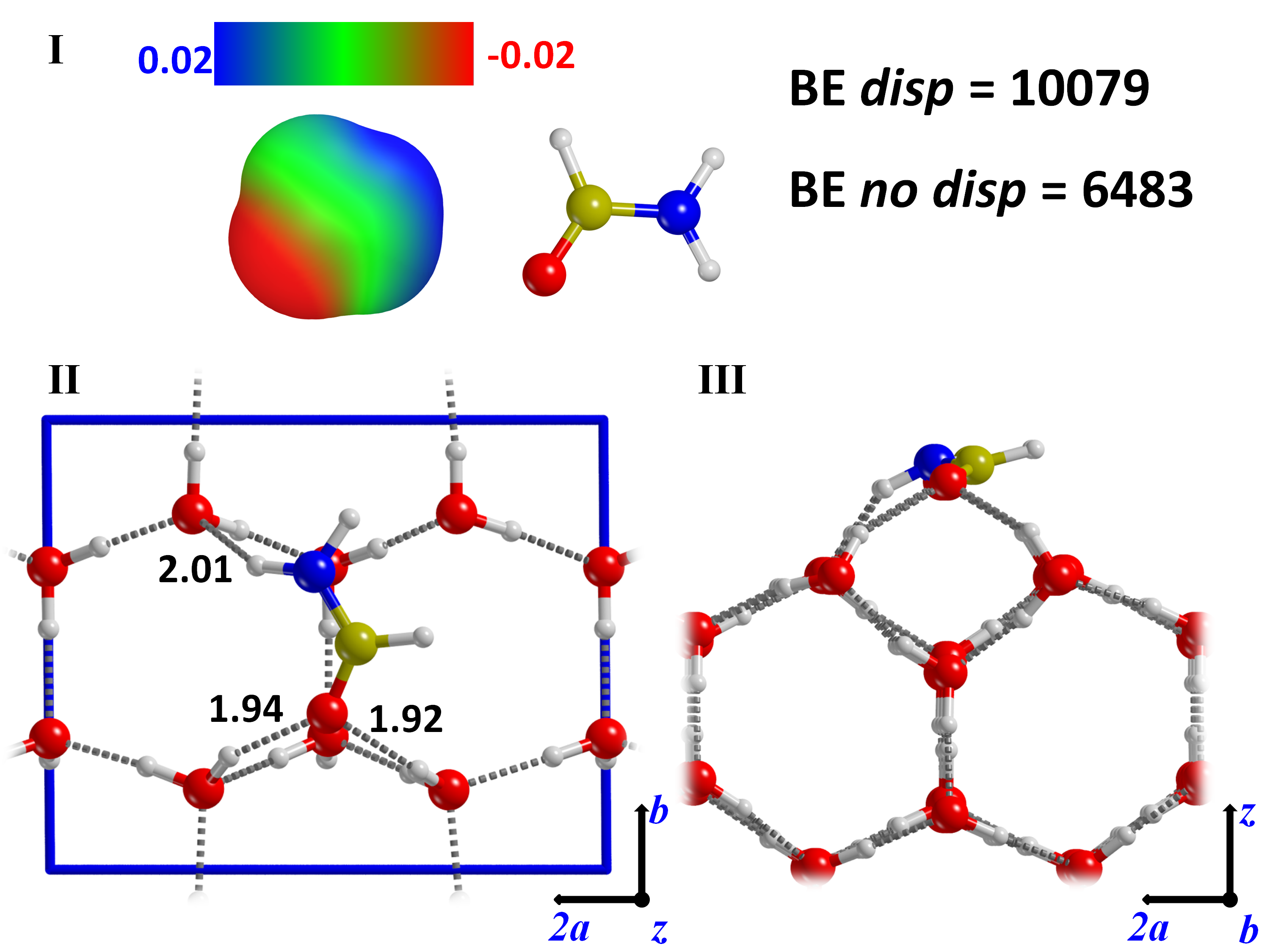}
\figsetgrpnote{I) Representation of the methanol (CH\textsubscript{3}OH) molecule along with its ESP surface, BSSE-corrected binding energies (\emph{BE}) with (\emph{disp}) and without (\emph{no disp}) dispersion. II) Top view of CH\textsubscript{3}OH-(010) P-ice 2x1 supercell surface interaction (unit cell highlighted in blue). III) Detail of the side view.}    
\label{fig:hconh2-sc2}
\figsetgrpend

\figsetgrpstart
\figsetgrpnum{16.20}
\figsetgrptitle{CH\textsubscript{3}OH}
\figsetplot{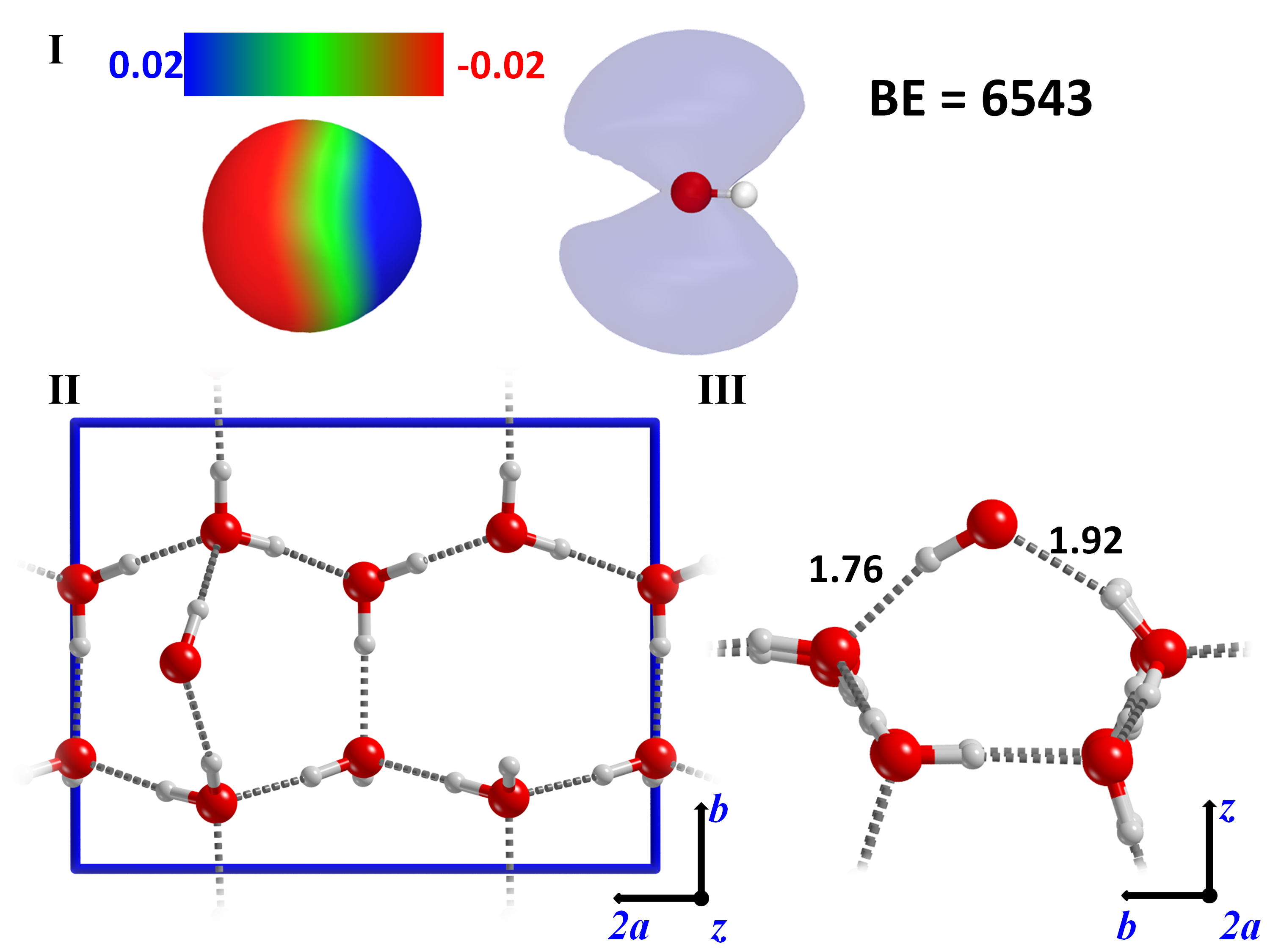}
\figsetgrpnote{I) Representation of the methanol (CH\textsubscript{3}OH) molecule along with its ESP surface, BSSE-corrected binding energies (\emph{BE}) with (\emph{disp}) and without (\emph{no disp}) dispersion. II) Top view of CH\textsubscript{3}OH-(010) P-ice 2x1 supercell surface interaction (unit cell highlighted in blue). III) Detail of the side view.}
\label{fig:oh}
 \figsetgrpend
 
\figsetgrpstart
\figsetgrpnum{16.21}
\figsetgrptitle{CH\textsubscript{3}OH}
\figsetplot{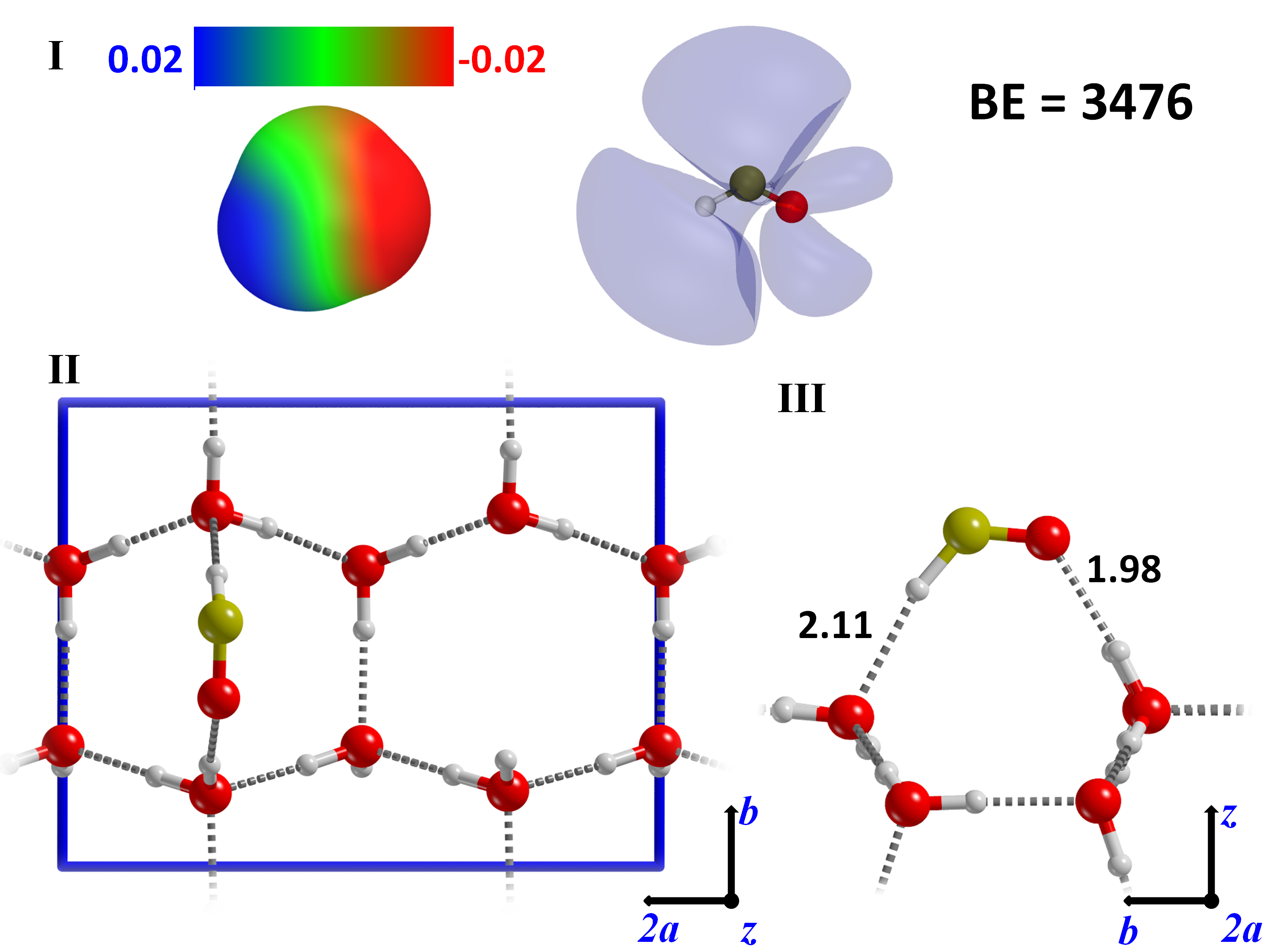}
\figsetgrpnote{I) Representation of the methanol (CH\textsubscript{3}OH) molecule along with its ESP surface, BSSE-corrected binding energies (\emph{BE}) with (\emph{disp}) and without (\emph{no disp}) dispersion. II) Top view of CH\textsubscript{3}OH-(010) P-ice 2x1 supercell surface interaction (unit cell highlighted in blue). III) Detail of the side view.}
\label{fig:hco}
\figsetgrpend

\figsetgrpstart
\figsetgrpnum{16.22}
\figsetgrptitle{CH\textsubscript{3}OH}
\figsetplot{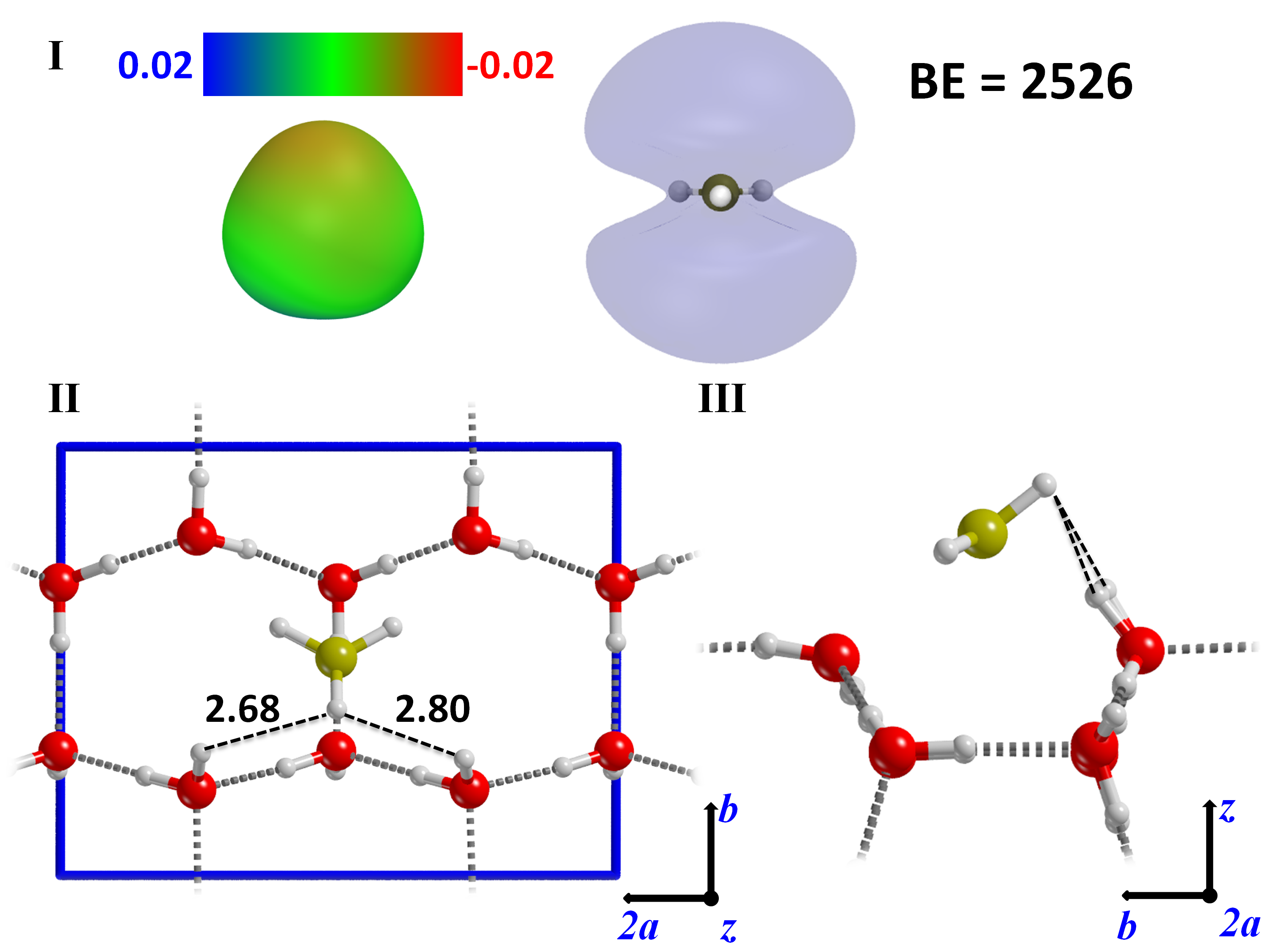}
\figsetgrpnote{I) Representation of the methanol (CH\textsubscript{3}OH) molecule along with its ESP surface, BSSE-corrected binding energies (\emph{BE}) with (\emph{disp}) and without (\emph{no disp}) dispersion. II) Top view of CH\textsubscript{3}OH-(010) P-ice 2x1 supercell surface interaction (unit cell highlighted in blue). III) Detail of the side view.}
\label{fig:ch3}
\figsetgrpend
    
\figsetgrpstart
\figsetgrpnum{16.23}
\figsetgrptitle{CH\textsubscript{3}OH}
\figsetplot{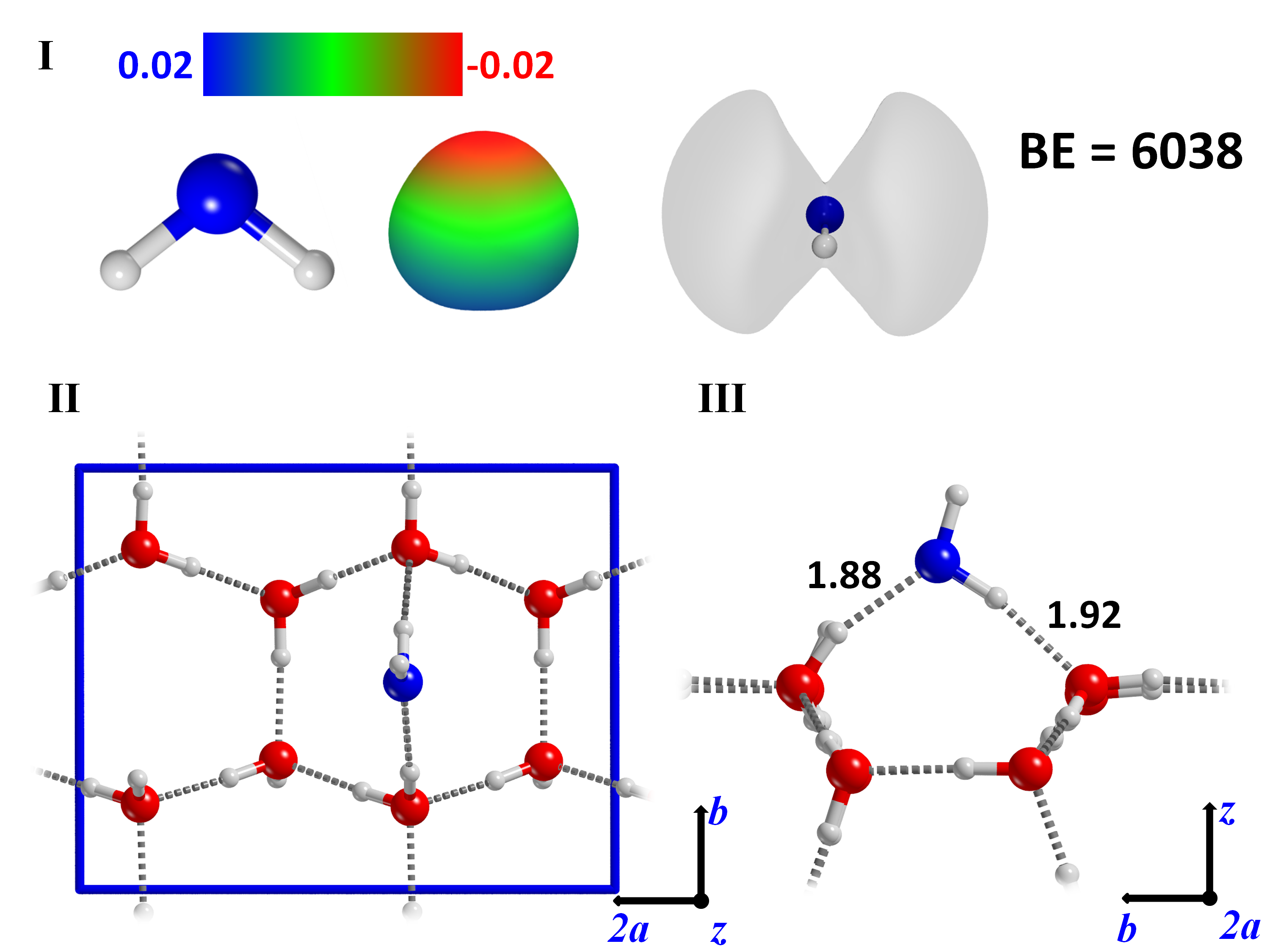}
\figsetgrpnote{I) Representation of the methanol (CH\textsubscript{3}OH) molecule along with its ESP surface, BSSE-corrected binding energies (\emph{BE}) with (\emph{disp}) and without (\emph{no disp}) dispersion. II) Top view of CH\textsubscript{3}OH-(010) P-ice 2x1 supercell surface interaction (unit cell highlighted in blue). III) Detail of the side view.}    
\label{fig:nh2}
\figsetgrpend
\figsetend
%%%%%%%%%%%%%%%%%%%%%%%%%%%%%%%%%%%%%%%%%%%%%%%%%%%%%%%%%%%%%%
%%%%%%%%%%%%%%%%%%%%%%%%%%%%%%%%%%%%%%%%%%%%%%%%%%%%%%%%%%%%%%
%%%%%%%%%%%%%%%%%%%%%%%%%%%%%%%%%%%%%%%%%%%%%%%%%%%%%%%%%%%%%%
\section{Surface distortion energy on ASW}\label{aoo-be-aws}

Hereafter we present the distortion energy contributions calculated on the amorphous slab will be compared with the same contributions calculated on the (010) P-ice model for some notable cases.

\begin{table}[h!]
    \centering
    \caption{Comparison of the distortion energy contributions (in Kelvin) calculated on the (010) P-ice slab and the mean value of these contributions calculated on the different adsorption cases on the amorphous ice model}
    \begin{tabular}{c c c}
    \toprule
    Species & $\delta$E\textsubscript{s} crystalline ice & $\langle$$\delta$ E\textsubscript{s} $\rangle$ amorphous ice \\
     \toprule
     CH\textsubscript{3}OH  & 662 & 2790 \\
     H\textsubscript{2}O  &  565 & 3211 \\
     HCN  &  589 &  1299 \\
     HCOOH  & 878& 1660\\
     NH\textsubscript{3}  &  734 & 1082 \\
  \bottomrule 
   \end{tabular}
\end{table}

\section{Basis set A-VTZ*}
\label{basis set}

\begin{table}[htbp]
  \centering
  \caption{Basis set for the H atom in CRYSTAL17 format used in this work}
    \begin{tabular}{rrrr}
    \toprule
    \multicolumn{1}{p{4.215em}}{\textbf{H}} & \multicolumn{1}{c}{0} &       &  \\
    \midrule
    \multicolumn{1}{p{4.215em}}{S} & 3     & 1     &  \\
    \multirow{3}[0]{*}{} & \multirow{3}[0]{*}{} & 34.06134 & 6.03E-03 \\
          &       & 5.123575 & 4.50E-02 \\
          &       & 1.164663 & 0.201897 \\
    \multicolumn{1}{p{4.215em}}{S} & 1     &       &  \\
          &       & 0.32723 & 1 \\
    \multicolumn{1}{p{4.215em}}{S} & 1     &       &  \\
          &       & 0.103072 & 1 \\
    \multicolumn{1}{p{4.215em}}{P} & 1     &       &  \\
          &       & 0.8   & 1 \\
    \bottomrule
    \end{tabular}%
  \label{tab:basis_H}%
\end{table}%

\begin{table}[htbp]
  \centering
  \caption{Basis set for the C atom in CRYSTAL17 format used in this work}
    \begin{tabular}{rrrr}
    \toprule
    \multicolumn{1}{p{4.215em}}{\textbf{C}} & \multicolumn{1}{c}{0} &       &  \\
    \midrule
    \multicolumn{1}{p{4.215em}}{S} & 5     & 1     &  \\
    \multirow{5}[0]{*}{} & \multirow{5}[0]{*}{} & 8506.038 & 5.34E-04 \\
          &       & 1275.733 & 4.13E-03 \\
          &       & 290.3119 & 2.12E-02 \\
          &       & 82.0562 & 8.24E-02 \\
          &       & 26.47964 & 0.240129 \\
    \multicolumn{1}{p{4.215em}}{S} & 1     &       &  \\
          &       & 9.241459 & 1 \\
    \multicolumn{1}{p{4.215em}}{S} & 1     &       &  \\
          &       & 3.364353 & 1 \\
    \multicolumn{1}{p{4.215em}}{S} & 1     &       &  \\
          &       & 0.871742 & 1 \\
    \multicolumn{1}{p{4.215em}}{S} & 1     &       &  \\
          &       & 0.363524 & 1 \\
    \multicolumn{1}{p{4.215em}}{S} & 1     &       &  \\
          &       & 0.128731 & 1 \\
    \multicolumn{1}{p{4.215em}}{P} & 4     &       &  \\
    \multirow{4}[0]{*}{} & \multirow{4}[0]{*}{} & 34.7095 & 5.33E-03 \\
          &       & 7.959088 & 3.59E-02 \\
          &       & 2.378697 & 0.142003 \\
          &       & 0.815401 & 0.342031 \\
    \multicolumn{1}{p{4.215em}}{P} & 1     &       &  \\
          &       & 0.289538 & 1 \\
    \multicolumn{1}{p{4.215em}}{P} & 1     &       &  \\
          &       & 0.100848 & 1 \\
    \multicolumn{1}{p{4.215em}}{D} & 1     &       &  \\
          &       & 1.6   & 1 \\
    \multicolumn{1}{p{4.215em}}{D} & 1     &       &  \\
          &       & 0.4   & 1 \\
    \bottomrule
    \end{tabular}%
  \label{tab:basis_C}%
\end{table}%

\begin{table}[htbp]
  \centering
  \caption{Basis set for the N atom in CRYSTAL17 format used in this work}
    \begin{tabular}{rrrr}
    \toprule
    \multicolumn{1}{p{4.215em}}{\textbf{N}} & \multicolumn{1}{c}{0} &       &  \\
    \midrule
    \multicolumn{1}{p{4.215em}}{S} & 5     & 1     &  \\
    \multirow{5}[0]{*}{} & \multirow{5}[0]{*}{} & 11913.42 & -5.23E-04 \\
          &       & 1786.721 & -4.04E-03 \\
          &       & 406.5901 & -2.08E-02 \\
          &       & 114.9253 & -8.12E-02 \\
          &       & 37.10588 & -0.23871 \\
    \multicolumn{1}{p{4.215em}}{S} & 1     &       &  \\
          &       & 12.97168 & 1 \\
    \multicolumn{1}{p{4.215em}}{S} & 1     &       &  \\
          &       & 4.730229 & 1 \\
    \multicolumn{1}{p{4.215em}}{S} & 1     &       &  \\
          &       & 1.252518 & 1 \\
    \multicolumn{1}{p{4.215em}}{S} & 1     &       &  \\
          &       & 0.512601 & 1 \\
    \multicolumn{1}{p{4.215em}}{S} & 1     &       &  \\
          &       & 0.179397 & 1 \\
    \multicolumn{1}{p{4.215em}}{P} & 4     &       &  \\
    \multirow{4}[0]{*}{} & \multirow{4}[0]{*}{} & 49.21876 & 5.55E-03 \\
          &       & 11.34894 & 3.81E-02 \\
          &       & 3.428509 & 0.149414 \\
          &       & 1.179951 & 0.348982 \\
    \multicolumn{1}{p{4.215em}}{P} & 1     &       &  \\
          &       & 0.417261 & 1 \\
    \multicolumn{1}{p{4.215em}}{P} & 1     &       &  \\
          &       & 0.142951 & 1 \\
    \multicolumn{1}{p{4.215em}}{D} & 1     &       &  \\
          &       & 2     & 1 \\
    \multicolumn{1}{p{4.215em}}{D} & 1     &       &  \\
          &       & 0.5   & 1 \\
    \bottomrule
    \end{tabular}%
  \label{tab:basis_N}%
\end{table}%

\begin{table}[htbp]
  \centering
  \caption{Basis set for the O atom in CRYSTAL17 format used in this work}
    \begin{tabular}{rrrr}
    \toprule
    \multicolumn{1}{p{4.215em}}{\textbf{O}} & \multicolumn{1}{c}{0} &       &  \\
    \midrule
    \multicolumn{1}{p{4.215em}}{S} & 5     & 1     &  \\
    \multirow{5}[0]{*}{} & \multirow{5}[0]{*}{} & 15902.65 & 5.15E-04 \\
          &       & 2384.954 & 3.98E-03 \\
          &       & 542.7196 & 2.05E-02 \\
          &       & 153.4041 & 8.03E-02 \\
          &       & 49.54572 & 0.237668 \\
    \multicolumn{1}{p{4.215em}}{S} & 1     &       &  \\
          &       & 17.33965 & 1 \\
    \multicolumn{1}{p{4.215em}}{S} & 1     &       &  \\
          &       & 6.330336 & 1 \\
    \multicolumn{1}{p{4.215em}}{S} & 1     &       &  \\
          &       & 1.699588 & 1 \\
    \multicolumn{1}{p{4.215em}}{S} & 1     &       &  \\
          &       & 0.689545 & 1 \\
    \multicolumn{1}{p{4.215em}}{S} & 1     &       &  \\
          &       & 0.23936 & 1 \\
    \multicolumn{1}{p{4.215em}}{P} & 4     &       &  \\
    \multirow{4}[0]{*}{} & \multirow{4}[0]{*}{} & 63.27052 & 6.07E-03 \\
          &       & 14.62331 & 4.19E-02 \\
          &       & 4.448952 & 0.161569 \\
          &       & 1.528151 & 0.356828 \\
    \multicolumn{1}{p{4.215em}}{P} & 1     &       &  \\
          &       & 0.529973 & 1 \\
    \multicolumn{1}{p{4.215em}}{P} & 1     &       &  \\
          &       & 0.175094 & 1 \\
    \multicolumn{1}{p{4.215em}}{D} & 1     &       &  \\
          &       & 2.4   & 1 \\
    \multicolumn{1}{p{4.215em}}{D} & 1     &       &  \\
          &       & 0.6   & 1 \\
    \bottomrule
    \end{tabular}%
  \label{tab:basis_O}%
\end{table}%

\begin{table}[htbp]
  \centering
  \caption{Basis set for the S atom in CRYSTAL17 format used in this work}
    \begin{tabular}{rrrr}
    \toprule
    \multicolumn{1}{p{4.215em}}{\textbf{S}} & \multicolumn{1}{c}{0} &       &  \\
    \midrule
    \multicolumn{1}{p{4.215em}}{S} & 5     &       &  \\
    \multirow{5}[0]{*}{} & \multirow{5}[0]{*}{} & 103954 & 2.47E-04 \\
          &       & 15583.79 & 1.92E-03 \\
          &       & 3546.129 & 9.96E-03 \\
          &       & 1002.681 & 4.04E-02 \\
          &       & 324.9029 & 0.130675 \\
    \multicolumn{1}{p{4.215em}}{S} & 1     &       &  \\
          &       & 115.5123 & 1 \\
    \multicolumn{1}{p{4.215em}}{S} & 2     &       &  \\
    \multirow{2}[0]{*}{} & \multirow{2}[0]{*}{} & 44.52821 & 0.504036 \\
          &       & 18.39789 & 0.230681 \\
    \multicolumn{1}{p{4.215em}}{S} & 1     &       &  \\
          &       & 5.510068 & 1 \\
    \multicolumn{1}{p{4.215em}}{S} & 1     &       &  \\
          &       & 2.125987 & 1 \\
    \multicolumn{1}{p{4.215em}}{S} & 1     &       &  \\
          &       & 0.436919 & 1 \\
    \multicolumn{1}{p{4.215em}}{S} & 1     &       &  \\
          &       & 0.157309 & 1 \\
    \multicolumn{1}{p{4.215em}}{P} & 5     &       &  \\
    \multirow{5}[0]{*}{} & \multirow{5}[0]{*}{} & 606.6937 & 2.32E-03 \\
          &       & 143.507 & 1.86E-02 \\
          &       & 45.74616 & 8.60E-02 \\
          &       & 16.87291 & 0.252484 \\
          &       & 6.63992 & 0.446327 \\
    \multicolumn{1}{p{4.215em}}{P} & 1     &       &  \\
          &       & 2.672714 & 1 \\
    \multicolumn{1}{p{4.215em}}{P} & 1     &       &  \\
          &       & 1.000009 & 1 \\
    \multicolumn{1}{p{4.215em}}{P} & 1     &       &  \\
          &       & 0.354389 & 1 \\
    \multicolumn{1}{p{4.215em}}{P} & 1     &       &  \\
          &       & 0.116713 & 1 \\
    \multicolumn{1}{p{4.215em}}{D} & 1     &       &  \\
          &       & 1.1   & 1 \\
    \multicolumn{1}{p{4.215em}}{D} & 1     &       &  \\
          &       & 0.275 & 1 \\
    \bottomrule
    \end{tabular}%
  \label{tab:basis_S}%
\end{table}%

\begin{table}[htbp]
  \centering
  \caption{Basis set for the Cl atom in CRYSTAL17 format used in this work}
    \begin{tabular}{rrrr}
    \toprule
    \multicolumn{1}{p{4.215em}}{\textbf{Cl}} & \multicolumn{1}{c}{0} &       &  \\
    \midrule
    \multicolumn{1}{p{4.215em}}{S} & 5     &       &  \\
    \multirow{5}[0]{*}{} & \multirow{5}[0]{*}{} & 117805.8 & 2.42E-04 \\
          &       & 17660.27 & 1.87E-03 \\
          &       & 4018.597 & 9.74E-03 \\
          &       & 1136.223 & 3.95E-02 \\
          &       & 368.1206 & 0.127972 \\
    \multicolumn{1}{p{4.215em}}{S} & 1     &       &  \\
          &       & 130.8615 & 1 \\
    \multicolumn{1}{p{4.215em}}{S} & 2     &       &  \\
    \multirow{2}[0]{*}{} & \multirow{2}[0]{*}{} & 50.47901 & 0.428741 \\
          &       & 20.91681 & 0.196685 \\
    \multicolumn{1}{p{4.215em}}{S} & 1     &       &  \\
          &       & 6.353139 & 1 \\
    \multicolumn{1}{p{4.215em}}{S} & 1     &       &  \\
          &       & 2.494801 & 1 \\
    \multicolumn{1}{p{4.215em}}{S} & 1     &       &  \\
          &       & 0.543359 & 1 \\
    \multicolumn{1}{p{4.215em}}{S} & 1     &       &  \\
          &       & 0.194344 & 1 \\
    \multicolumn{1}{p{4.215em}}{P} & 5     &       &  \\
    \multirow{5}[0]{*}{} & \multirow{5}[0]{*}{} & 681.0688 & 2.37E-03 \\
          &       & 161.1136 & 1.89E-02 \\
          &       & 51.38664 & 8.78E-02 \\
          &       & 18.95851 & 0.257074 \\
          &       & 3.003516 & 0.371524 \\
    \multicolumn{1}{p{4.215em}}{P} & 1     &       &  \\
          &       & 7.456529 & 1 \\
    \multicolumn{1}{p{4.215em}}{P} & 1     &       &  \\
          &       & 1.060936 & 1 \\
    \multicolumn{1}{p{4.215em}}{P} & 1     &       &  \\
          &       & 0.39452 & 1 \\
    \multicolumn{1}{p{4.215em}}{P} & 1     &       &  \\
          &       & 0.133233 & 1 \\
    \multicolumn{1}{p{4.215em}}{D} & 1     &       &  \\
          &       & 1.3   & 1 \\
    \multicolumn{1}{p{4.215em}}{D} & 1     &       &  \\
          &       & 0.325 & 1 \\
    \bottomrule
    \end{tabular}%
  \label{tab:basis_Cl}%
\end{table}%

\end{document}